\documentclass[twocolumn]{aastex63} 

\usepackage[utf8]{inputenc}
\usepackage[caption=false]{subfig}
\usepackage{natbib}
\usepackage{comment} 
\usepackage{csvsimple, array, longtable, booktabs, amsmath, amssymb, bm, lineno, color, soul, hyphenat} 
\usepackage{hyperref}
\usepackage[all]{hypcap} 
\hypersetup{ 
                colorlinks=true, %
                breaklinks=true, %
                pdfauthor={Maurice L. Wilson}, %
                pdftitle={CME Heating and Energy Budget}, %
                citecolor=blue, %
                linkcolor=blue %
                }

\usepackage{etoolbox}
\usepackage[normalem]{ulem}

\newcommand{\affilHarvardCfA}{Center for Astrophysics, Harvard \& Smithsonian, 60 Garden St, Cambridge, MA 02138, USA}

\newcommand{\rsun}{\ensuremath{ R_\odot }}
\newcommand{\Ang}{\ensuremath{ \mathrm{\AA} }}
\newcommand{\unitn}{cm\ensuremath{ ^{-3} }}
\newcommand{\unitv}{km~s\ensuremath{ ^{-1} }}
\newcommand{\ionn}[2]{#1~{\footnotesize #2}}
\newcommand{\clump}[1]{\emph{#1}} 
\newcommand{\scinot}[2]{#1\ensuremath{\times}10\ensuremath{^{#2}}}
\newbool{egshow}\setbool{egshow}{true}\ifbool{egshow}{ \newcommand{\egcite}{e.g.,} }{ \newcommand{\egcite}{} }
\newcommand{\rOVI}{\ensuremath{ \frac{ \rm{O\ VI}\ 1032 }{ \rm{O\ VI}\ 1038 } }} 
\newcommand{\rLyman}{\ensuremath{ \frac{ \rm{H\ I}\ 1216 }{ \rm{H\ I}\ 1026 } }}  
\newcommand{\rOV}{\ensuremath{ \frac{ \rm{O\ V}\ 1214 }{ \rm{O\ V}\ 1218 } }}  
\newcommand{\rCIII}{\ensuremath{ \frac{ \rm{C\ III}\ 977 }{ \rm{O\ VI}\ 1032 } }}
\newcommand{\rOVbr}{\ensuremath{ \frac{ \rm{O\ V}\ 1218 }{ \rm{O\ VI}\ 1032 } }}
\newcommand{\rhialpha}{\ensuremath{ \frac{ \rm{H\ I}\ 1216 }{ \rm{O\ VI}\ 1032 } }}
\newcommand{\rhibeta}{\ensuremath{ \frac{ \rm{H\ I}\ 1026 }{ \rm{O\ VI}\ 1032 } }}
\newcommand{\rNIII}{\ensuremath{ \frac{ \rm{N\ III}\ 992 }{ \rm{N\ III}\ 999 } }}

\newcommand{\vsp}{7pt}

\begin{document}

\title{ Constraining the CME Core Heating and Energy Budget with SOHO/UVCS }
\shorttitle{CME Heating and Energy Budget}

\shortauthors{Wilson et al.}

\author[0000-0003-1928-0578]{Maurice L. Wilson}
\affiliation{\affilHarvardCfA}

\author[0000-0002-7868-1622]{John C. Raymond} 
\affiliation{\affilHarvardCfA}

\author[0000-0003-1611-227X]{Susan T. Lepri} 
\affiliation{ Department of Climate and Space Sciences and Engineering, University of Michigan, Ann Arbor, MI 48109, USA }

\author[0000-0001-9231-045X]{Roberto Lionello} 
\affiliation{\affilHarvardCfA}
\affiliation{Predictive Science Incorporated, 9990 Mesa Rim Rd. Suite 170, San Diego, CA 92121, USA}

\author[0000-0001-6628-8033]{Nicholas A. Murphy} 
\affiliation{\affilHarvardCfA}

\author[0000-0002-6903-6832]{Katharine K. Reeves} 
\affiliation{\affilHarvardCfA}

\author[0000-0002-9258-4490]{Chengcai Shen} 
\affiliation{\affilHarvardCfA}

\correspondingauthor{Maurice L. Wilson}
\email{maurice.wilson@cfa.harvard.edu}



\begin{abstract}

We describe the energy budget of a coronal mass ejection (CME) observed on 1999 May 17 with the Ultraviolet Coronagraph Spectrometer (UVCS).  We constrain the physical properties of the CME's core material as a function of height along the corona by using the spectra taken by the single-slit coronagraph spectrometer at heliocentric distances of 2.6 and 3.1 solar radii.  We use plasma diagnostics from intensity ratios, such as the \ionn{O}{VI} doublet lines, to determine the velocity, density, temperature, and non-equilibrium ionization states.  We find that the CME core's velocity is approximately 250~km/s, and its cumulative heating energy is comparable to its kinetic energy for all of the plasma heating parameterizations that we investigated.  Therefore, the CME's unknown heating mechanisms have the energy to significantly affect the CME's eruption and evolution.  To understand which parameters might influence the unknown heating mechanism, we constrain our model heating rates with the observed data and compare them to the rate of heating generated within a similar CME that was constructed by the MAS code's 3D MHD simulation.  The rate of heating from the simulated CME agrees with our observationally constrained heating rates when we assume a quadratic power law to describe a self-similar CME expansion. Furthermore, the heating rates agree when we apply a heating parameterization that accounts for the CME flux rope's magnetic energy being converted directly into thermal energy.  This UVCS analysis serves as a case study for the importance of multi-slit coronagraph spectrometers for CME studies.

 
\textit{Online materials}: color figures

\end{abstract}

\section{Introduction}

In 2021, we acknowledge the 50th anniversary of coronal mass ejection (CME) observations along with the advent of a privatized billionaire space race.  A decade into the very first space race, observations of bright plasma from a CME were recorded for the first time \citep{Hansen.1971, Tousey.1973, Gosling.1974}.  Now, CMEs are understood to be magnetized plasma clouds originating from long filament or prominence loops of relatively cool plasma.  Stored magnetic energy is abruptly released with the cool plasma, which subsequently expands while travelling through the corona and interplanetary medium.  
The physical mechanisms that launch and continuously drive the behavior seen in CMEs are still ambiguous.  This ambiguity results in a broad range of physical interpretations being considered to explain the initiation, the morphology, the composition, and the total energy budget of CMEs.  

Many of the observationally-supported interpretations suggest CMEs consist of a bright outer shell that leads a faint flux rope which surrounds a dense core of plasma \citep{Illing.1985}. At supersonic speeds, the leading edge is preceded by a shock front of gas that often correlates with solar energetic particles (SEPs) that can disrupt satellite communications \citep[\egcite][]{Kahler.1994, Laming.2013}.  At any speed, the leading edge is the simplest feature to track in white light images as the CME propagates through the corona.  It contains bright coronal gas that is initially compressed by the eruption \citep[\egcite][]{Ma.2011, Howard.2018}.  The flux rope is often referred to as the void because of how it appears in images and spectra due to the flux rope's dim brightness.  Instead, measurements of its ionization states are gathered \textit{in situ} near 1~AU for interplanetary coronal mass ejections (ICMEs) \citep[\egcite][]{Lepri.2001, Rivera.2019.April}.  
The dense core of the CME contains a large mass of plasma spanning a wide range of ionization states.  This plasma originates from a (filament or) prominence loop that can extend above a current sheet as the prominence material erupts from the solar surface \citep[\egcite][]{Liewer.2009, Reeves.2015}.  Overall, these observed features form the canonical three-part CME that consists of a leading edge, a flux rope, and a core.  Upon eruption, the energy budget of this CME might be influenced by an accompanying solar flare.

For hundreds of simultaneous flare-CME events, many terms in the energy budget are within the range of 10$^{29}$---10$^{32}$~erg 
\citep[\egcite][]{Aschwanden.2014,Aschwanden.2017, Emslie.2005}.
When there is no accompanying flare, the energy budget of the CME is often found for only one or two components of the three-part CME.  Due to their frequently imaged bright features, the core and leading edge are the two most convenient components of the CME to study when determining the energy budget; although, the magnetic energy requires measurements from the flux rope.  

Compared to the rest of the total energy budget, the magnetic energy of a CME is difficult to measure.  Serendipitous measurements of the magnetic energy are usually gathered \textit{in situ} near 1~AU if an ICME bombards a spacecraft \citep[\egcite][]{Davies.2020, Scolini.2020},
while targeted measurements are typically acquired through remote observations of pre-CME prominences and filaments on the solar disk \citep[\egcite][]{Leroy.1983, Solanki.2003}.  Upon eruption, most of the CME's magnetic energy is concentrated in the flux rope.  It is difficult to track this magnetic energy after the eruption due to the faint emission within this component of the CME.  Attempts have been made to bridge the gap between the measurements of the CME magnetic field at 1~\rsun\ (remotely) and 1~AU (\textit{in situ}).  Magnetohydrodynamic (MHD) models have been used to gain insight on the morphology of the magnetic field structure by extrapolating from solar disk measurements, extrapolating from \textit{in situ} measurements, or interpolating between both measurements \cite[\egcite][]{Usmanov.1995, Feng.2010}.  However, the mechanisms that transform the complex, coronal flux rope into a relaxed, interplanetary plasma cloud are still largely unconfirmed.  This creates much uncertainty for magnetic energy estimates of CMEs seen traveling through the corona, frequently via white light images.

It is much more feasible to measure and continuously track the kinetic and potential energy components of the CME energy budget.  Both forms of energy require a value for the CME's mass, which can be estimated directly from white light coronagraph images.  The images show features along the plane of sky (POS) and capture the light scattered by free electrons; and, the information inferred from the features is averaged along the line-of-sight (LOS) within the optically thin coronal medium.  Such information can be misinterpreted due to projection effects.  Frequently, geometric assumptions are made to mitigate misunderstandings caused by projection effects when determining the mass or three-dimensional structure of CMEs \citep[\egcite][]{Ciaravella.2003, Emslie.2004, Vourlidas.2010}.  For the kinetic and potential energy, the uncertainty due to errors in the mass estimate can be avoided if only the specific energy (i.e., quantities of energy per mass) is used to compare and contrast the energy budgets of various CMEs, which may have masses that are evaluated with distinct techniques and sources of uncertainty.  

The heating energy is another component of the CME energy budget that is often plagued by uncertainties.  This is because the physical mechanisms responsible for continuously generating thermal energy are not understood.  Processes that cool the plasma or redistribute its thermal energy can occur while the plasma is being heating even though observations may sometimes indicate minor changes in the plasma temperature.  Evidence for the extended, post-eruption heating has been found through observations of erupting prominence material observed as absorption features that are later seen as emission features, presumably due to its temperature increasing \citep[][]{Filippov.2002, Lee.2017}.  Additionally, \textit{in situ} measurements at 1~AU have indicated the need for CME heating until the ionization states are frozen-in \citep[][]{Rakowski.2007}, i.e. until the plasma density is low enough or velocity is fast enough for the local environment's ionization and recombination processes to no longer alter the CME's ionization states.  Quantifying the energy of the heating process may provide clues for its underlying physical mechanisms.  Several studies have quantitatively assessed the cumulative heating energy component of the energy budget and found it to be comparable to the kinetic energy \citep[\egcite][]{Akmal.2001, Murphy.2011}.  It is clear that the heating is an important process that can improve our understanding of the CME's evolution during and after the initial eruption. 

In this paper, we provide constraints on the heating energy of localized plasma within a CME by using fortuitous spectroscopic measurements of a CME crossing the (single) slit of a coronagraph spectrometer at multiple heights in the corona.  Our work with this unique dataset is supported by measurements from solar disk photometry and white light coronagraph images of the CME.  This paper is organized as follows.

In Section~\ref{sect: observations}, we describe the data acquired from three instruments of the \textit{Solar and Heliospheric Observatory} (\textit{SOHO}).  We have photometry from the Extreme ultraviolet Imaging Telescope (EIT), white light images from the Large Angle Spectroscopic Coronagraph (LASCO), and spectra from the Ultraviolet Coronagraph Spectrometer (UVCS) to study the CME that erupted in 1999 on May 17.  In Section~\ref{sect: data analysis}, we interpret the features seen within the data to distinguish between a variety of structures within the CME core.  In Section~\ref{sect: plasma diagnostics}, we discuss how plasma diagnostics inferred from the spectra constrain the plasma parameters.  The constraints provide upper and lower limits on the physical properties that we find from our 1D numerical models and non-equilibrium ionization (NEI) calculations, which we explain in Section~\ref{sect: numerical model}.  The constrained 1D models are compared to the 3D MHD model of a slow CME.  This CME's evolution is simulated by the Magnetohydrodynamic Algorithm outside a Sphere (MAS) code and we discuss this in Section~\ref{sect: mas model}.  Our energy budget results and heating rates for the observed CME are given in Section~\ref{sect: results}.  We demonstrate our methodology
through the detailed analysis presented for one heating parameterization.  The analyses for our other parameterizations are given in the Appendices.  Lastly, in Section~\ref{sect: summary} we summarize our work and give closing remarks about the current dearth of coronagraph spectrometers, which is an issue that will be resolved by the UVSC Pathfinder and LOCKYER missions. 

\section{Observations of the CME}\label{sect: observations}

We study observations taken of a CME that occurred on 17 May 1999.  Three instruments on board the \textit{SOHO} spacecraft clearly captured the CME: EIT between 00:48 and 03:12 UTC, LASCO C2 camera between 00:49 and 5:25 UTC, and UVCS between 03:08 and 04:38 UTC.  We used EIT and LASCO to confirm the CME detection and obtain rough estimates of the CME's velocity.  We use spectra from UVCS to analyze the evolution of its physical properties.

\subsection{EIT Photometry}\label{sect: eit}

\begin{figure}
    \centering
    \includegraphics[width=\linewidth, trim=1.4in 0.6in 1.5in 0.45in, clip=True]{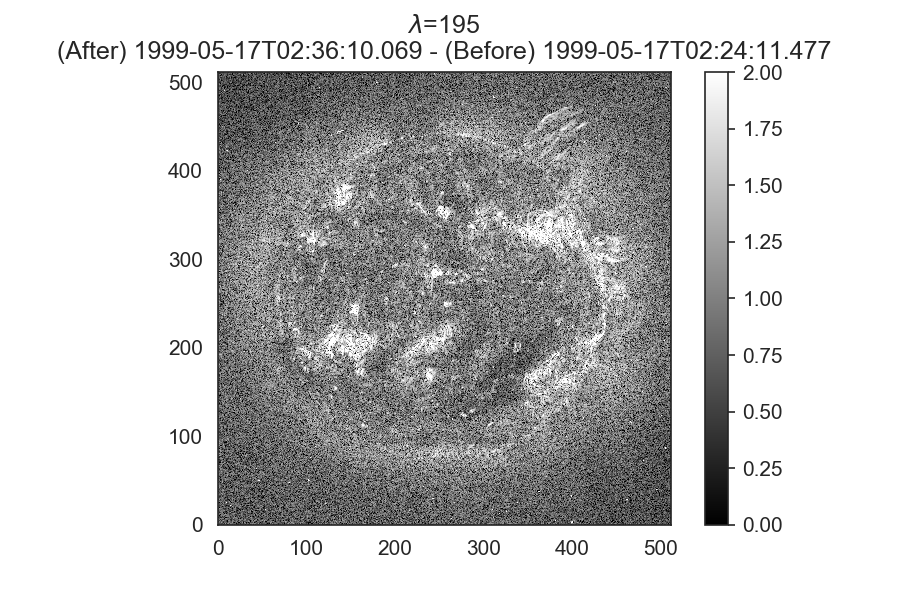}
    \caption{EIT difference image from observations taken at 02:24 and 02:36 UTC in the 195~\Ang\ bandpass.  In the top right, plumes of elongated CME material continuously and slowly erupt off the limb between 00:48 and 03:12 UTC.}
    \label{fig: eit}
\end{figure}

The EIT \citep{Delaboudiniere.1995} observations show filamentary structures erupting near the northwest limb of the Sun.  This is most evident in the 195~\Ang\ bandpass with images taken at a 12 minute cadence and an exposure time of 4.5 seconds.  These structures can be seen in the difference image given in Figure~\ref{fig: eit}.  Multiple filamentary structures are near the position angle (PA) of 315$^\circ$ (counter-clockwise from north pole).  They elongate and travel radially outward between times 00:48 and 3:00~UTC.  
It is not clear where the launch site was on the Sun given that only one image of these structures was captured by the 304~\Ang\ bandpass.
Taken at 1:18~UTC with an exposure time of 32 seconds, the 304~\Ang\ image shows many towering prominence loops that extend downward to footpoints that do not reside in the foreground.  
These observations suggest that, before 00:48~UTC, the CME either has yet to launch or is traveling behind the solar disk; and, beyond 3:00~UTC, the CME material has traveled beyond the field of view or is no longer emitting radiation within the bandpass.  Images in the 171~\Ang\ and 284~\Ang\ bandpasses were taken at times outside of this time window and therefore did not provide relevant information.  Based on the time window, the structures imaged by EIT begin to erupt at least two hours before the UVCS observations capture the CME at a heliocentric distance of 1.4~\rsun\ along SOHO's POS.  Assuming the EIT structures continued to travel radially outward at a constant speed, the observation times suggest a speed of $\sim$80~\unitv\ along the POS for CME material traveling from the limb to a heliocentric distance of 1.4~\rsun.  However, we later discuss the importance of confirming observations of the same, specific structures at multiple heights when attempting to deduce the velocity of CME material.


\subsection{LASCO Photometry}\label{sect: lasco}

\begin{figure}
    \centering
    \includegraphics[width=\linewidth, trim=2.2in 1.3in 1.55in 0.55in, clip=True]{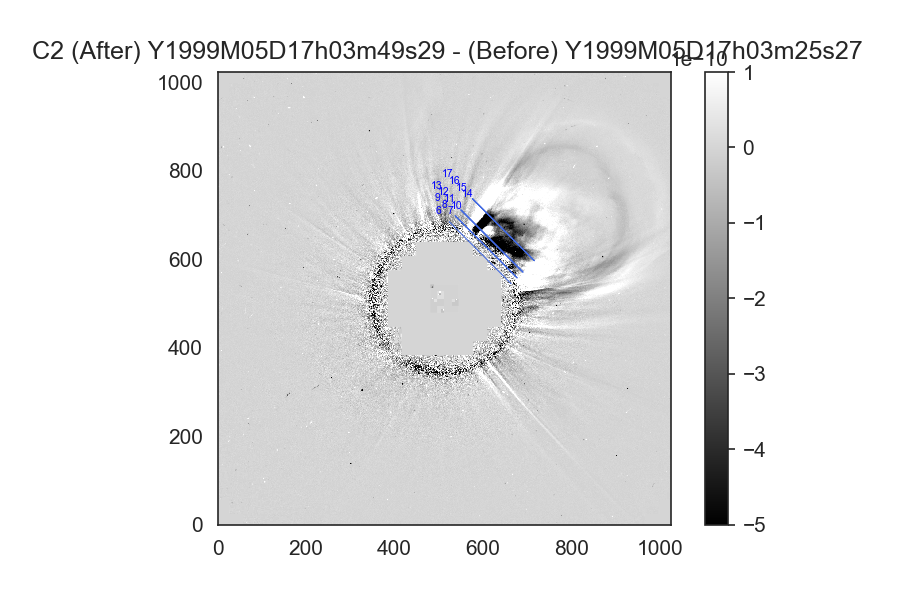}
    \caption{LASCO difference image from observations taken at 03:25 and 03:49 UTC show the CME's amorphous core and shell-like leading edge.  The blue lines represent the UVCS slit aperture at times corresponding to the observations IDs (cf. Table~\ref{table: observations}).}
    \label{fig: slits on lasco}
\end{figure}

White light photometry of LASCO \citep{Brueckner.1995} can be obtained from any of its three cameras: C1, C2, and C3.  The C1 camera however was no longer operational after 1998; therefore, data on the 1999 CME is available  only from the C2 and C3 camera.  They provide a combined field of view covering 2.5 to 30~\rsun.  Among the LASCO images that capture the CME, we primarily consider the images that occur near the UVCS observation times.  This is limited to the C2 images taken from 2:49 to 4:49~UTC each with an exposure time of 25 seconds.  An example of when UVCS observations coincide with LASCO is given in Figure~\ref{fig: slits on lasco}.  At distinct times, the single slit aperture of the UVCS instrument monitors the corona at distinct heliocentric distances ($d_H$).  In our example, we overlay the slit (illustrated as a blue line) onto a difference image of LASCO white light photometry only if the UVCS slit-image is taken at a time within $\pm$20 minutes of a single LASCO C2 image.  Within this time interval, Figure~\ref{fig: slits on lasco} shows the UVCS slit's center at 1.7, 1.9, 2.1, and 2.6~\rsun\ at different times.  The UVCS observations that have an assigned identification (ID) ranging from 6 to 17 have (blue) slits that are overlaid onto the difference image.  These UVCS observation IDs and times are given in Table~\ref{table: observations}.  

The positions of the slit during this CME event suggest that UVCS primarily observed the bright, dense core of the CME.  Throughout all of the observations listed in Table~\ref{table: observations}, neither the CME's current sheet, the faint void, nor the leading edge are discernable within the UVCS data.  According to the CDAW CME Catalog \citep{Gopalswamy.2009}, the leading edge seen in LASCO's C2 and C3 images travels at 500~\unitv\ beyond 3~\rsun\ with a 5~m~s$^{-2}$ acceleration on average along the POS.  The core of the CME is seen in the same white light photometry and contains amorphous features that significantly alter in appearance from one image to another.  This makes it difficult to determine the core's speed when using only white light imagery.  Based on the C2 images we use, this problem is exacerbated with core material at lower heights in the corona.  As the material expands and travels higher in the corona, some features of the CME core are more clear in their discernible shape in one image than another image, and they also become fainter.  Consequently, we depend on the UVCS information for velocity estimates of features seen within the CME core.  However, the leading edge velocity from LASCO does serve as an upper limit for the core's velocity along the POS.

\capstartfalse
\begin{table}
    \begin{longtable}{lccc}
        \label{table: observations}
        \\\multicolumn{4}{c}{ {\scshape \tablename\ \thetable}:  UVCS Observations\iffalse of 17 May 1999 CME\fi } \\\midrule\midrule
        ID & Time (UTC) & $d_H\ (R_\odot)$ & $\Delta t$ (s) \\\midrule
        0 & 03:07:50 & 1.4 & 200 \\
        1 & 03:11:14 & 1.4 & 200 \\
        2 & 03:15:13 & 1.5 & 200 \\
        3 & 03:18:39 & 1.5 & 200 \\
        4 & 03:22:36 & 1.7 & 180 \\
        5 & 03:26:04 & 1.7 & 180 \\
        6 & 03:29:33 & 1.7 & 180 \\
        7 & 03:33:34 & 1.9 & 180 \\
        8 & 03:37:04 & 1.9 & 180 \\
        9 & 03:40:33 & 1.9 & 180 \\
        10 & 03:44:32 & 2.1 & 180 \\
        11 & 03:48:03 & 2.1 & 180 \\
        12 & 03:51:32 & 2.1 & 180 \\
        13 & 03:55:02 & 2.1 & 180 \\
        14 & 03:58:47 & 2.6 & 180 \\
        15 & 04:02:18 & 2.6 & 180 \\
        16 & 04:05:46 & 2.6 & 180 \\
        17 & 04:09:16 & 2.6 & 180 \\
        18 & 04:12:47 & 2.6 & 180 \\
        19 & 04:16:49 & 3.1 & 200 \\
        20 & 04:20:14 & 3.1 & 180 \\
        21 & 04:23:47 & 3.1 & 200 \\
        22 & 04:27:13 & 3.1 & 180 \\
        23 & 04:30:45 & 3.1 & 200 \\
        24 & 04:34:10 & 3.1 & 180 \\
        25 & 04:37:43 & 3.1 & 180 \\\midrule
        \multicolumn{4}{ p{0.6\linewidth} }{ \textbf{Notes.} The POS heliocentric distance, $d_H$, corresponds to the slit's central pixel. } 
    \end{longtable}
\end{table}
\capstarttrue

\vspace{30pt}

\subsection{UVCS Observations of CME Core}

Effective ultraviolet coronograph spectrometers attempt to minimize the contamination of bright solar disk emission while maximizing the signal of relatively weak coronal emission lines.  Many spectrometers constructed for this purpose are modeled after the design originally introduced by \citet{Kohl.1978}.  As one of such instruments, UVCS was designed to detect coronal emission  covering the 940--1360~\Ang\ wavelength range as a means for studying the physical conditions of coronal plasma from $d_H \approx 1.5\ \rsun$ out to $d_H \approx 10\ \rsun$ away from the center of the solar disk in the plane of sky~\citep{Kohl.1995,Kohl.2006, Gardner.1996,Gardner.2000,Gardner.2002}.  UVCS consists of two spectrometers~\citep{Pernechele.1997}: the Lyman-$\alpha$ channel can cover the 1145--1285~\Ang\ range but is optimized for the \ionn{H}{I} Lyman-$\alpha$ line at 1216~$\mathrm{\AA}$ while the \ionn{O}{VI} channel can cover the 940--1125~\Ang\ range but is optimized for the \ionn{O}{VI} 1032 and 1038~\Ang\ doublet lines.  In this work, we only use data from the \ionn{O}{VI} channel.  We analyze data from both the ``primary" light path and the ``redundant" light path, albeit both paths lead to the \ionn{O}{VI} detector.  The redundant mirror provides the spectral coverage needed to monitor \ionn{H}{I} Lyman-$\alpha$ emission without using the Lyman-$\alpha$ channel.

On 17 May 1999, UVCS was staring near the northwest limb of the Sun at a position angle of 315$^\circ$ with the slit positioned at heliocentric distances ranging from 1.42 to 3.10~$R_\odot$.  The core of a CME passes through the field of view and there are 26 images taken with exposure times of either 180 seconds or 200 seconds.  This is tabulated in Table~\ref{table: observations} and all of these images capture features of the CME core at the same position angle.  There are no observations that occur immediately before or immediately after the CME event at this position angle.  
The spatial binning along the slit is 3 pixels (21'').    

\begin{figure*}\label{fig: all lines}
    \centering
    \includegraphics[width=0.8\linewidth, trim=0cm 0cm 0cm 0cm, clip]{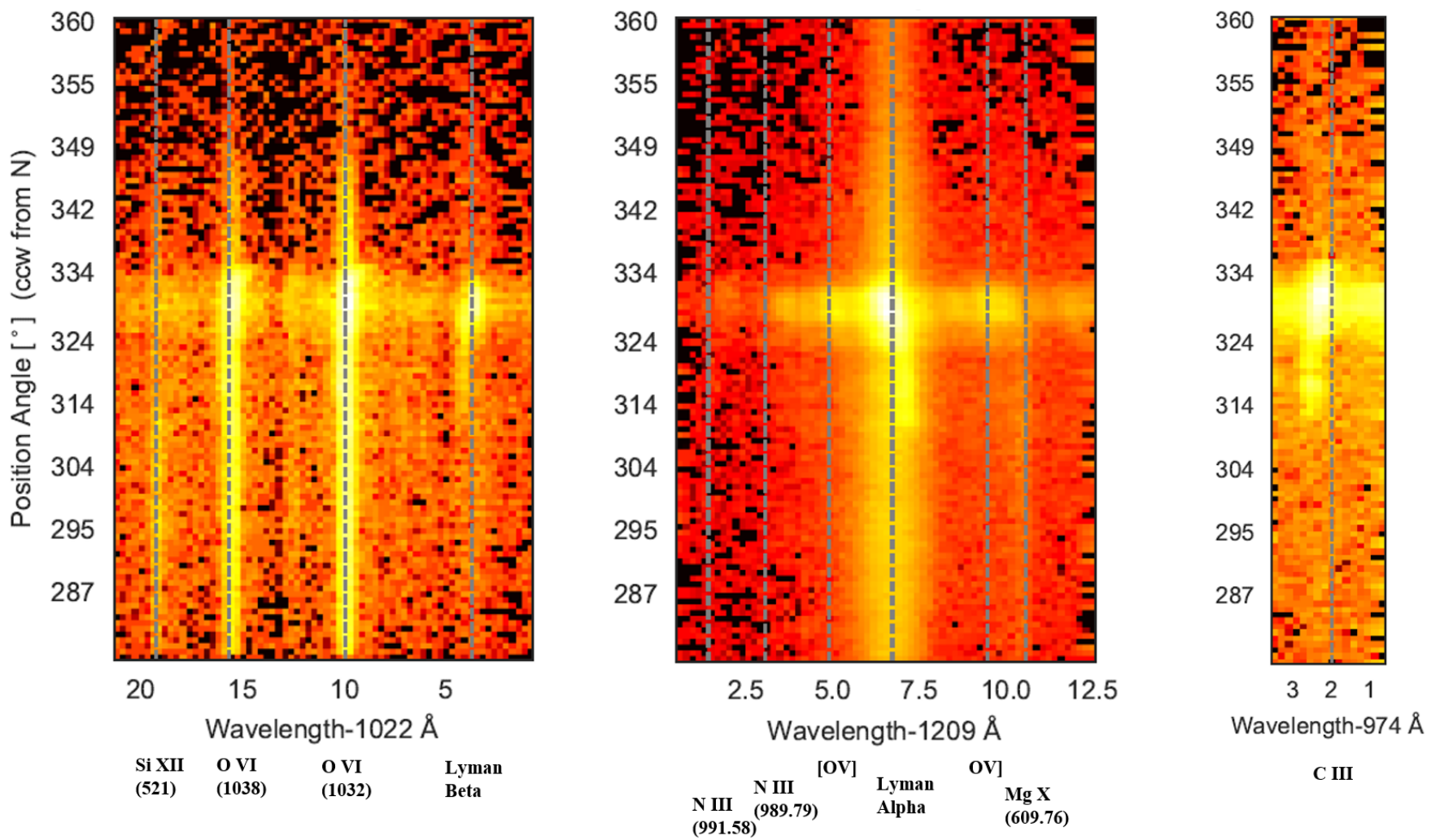} 
    \caption{UVCS data taken when slit aperture is positioned to $d_H=1.4$~\rsun\  at 03:08 UTC.  From top to bottom, the spectral lines in this example show the background corona, a very bright clump of CME core material, and diffuse CME core material.}
    
\end{figure*}

Due to the limitations of telemetry, three distinct panels within the \ionn{O}{VI} channel's spectral coverage were stored.  As shown in the example of Figure~\ref{fig: all lines}, the three panels were stored with a binning of 3, 2, and 2 pixels in the dispersion direction which corresponds to -0.298, -0.199, and -0.199~\Ang\ respectively for the primary light path and 0.274, 0.183, and 0.183 respectively for the redundant light path.  The negative dispersion indicates that the wavelengths will increase in the opposite direction (as seen in Figure~\ref{fig: all lines}).  The three panels have wavelength ranges respectively corresponding to 1023--1043, 979--993, and 975--978~\Ang\ for the primary path.  For the redundant path, the wavelength ranges are 1163--1182, 1209--1222, 1223--1226~\Ang\ respectively.  See Table~\ref{table: spectral lines} for the most prominent spectral lines identified along with their peak ion formation temperature under ionization equilibrium.  The wavelength calibration, flux calibration, and corrections in detector distortions and flat fielding are processed via the UVCS Data Analysis Software version 5.1 (DAS51). 
\vspace{\vsp}

\newcommand{\tablefsize}{\normalsize}
\capstartfalse
\begin{table*}
    \begin{longtable}{lllclc}
        \label{table: spectral lines}
        
        \\\multicolumn{6}{c}{\normalsize {\scshape \tablename\ \thetable}:   Prominent lines detected by UVCS during CME} \\\midrule\midrule
         \tablefsize Wavelength ($\mathrm{\AA}$) & \tablefsize Ion  & \multicolumn{3}{c}{ \tablefsize Transition} & \tablefsize log $T_m$ \\\midrule
         \tablefsize 1215.67           & \tablefsize  H I  &  \multicolumn{3}{c}{\tablefsize Lyman-$\alpha$} & \tablefsize 4.5 \\ 
         \tablefsize 1025.72           & \tablefsize H I  &  \multicolumn{3}{c}{\tablefsize Lyman-$\beta$}  & \tablefsize 4.5 \\
         \tablefsize 977.02            & \tablefsize  C III  &\tablefsize  $2\rm{s}^2\ \ ^1\rm{S}_0$ &-&\tablefsize $2\rm{s}2\rm{p}\ \ ^1\rm{P}_1$ & \tablefsize 4.8 \\
         \tablefsize 989.79, 991.58           & \tablefsize  N III &\tablefsize $2\rm{s}^2 2\rm{p}\ \ ^2\rm{P}_{1/2,3/2}$ &-&\tablefsize $2\rm{s}2\rm{p}^2\ \ ^2\rm{D}_{3/2,5/2}$ & \tablefsize 4.9 \\
         \tablefsize 1213.85           & \tablefsize [OV]  &\tablefsize  $2\rm{s}^2\ \ ^1\rm{S}_0$ &-&\tablefsize $2\rm{s}2\rm{p}\ \ ^3\rm{P}_2$ & \tablefsize  5.4 \\
         \tablefsize 1218.39           & \tablefsize O V]  &\tablefsize  $2\rm{s}^2\ \ ^1\rm{S}_0$ &-&\tablefsize $2\rm{s}2\rm{p}\ \ ^3\rm{P}_1$  & \tablefsize  5.4 \\
         \tablefsize 1031.91, 1037.61  & \tablefsize O VI  &\tablefsize  $2\rm{s}\ \ ^2\rm{S}_{1/2}$ &-&\tablefsize $2\rm{p}\ \ ^2\rm{P}_{3/2,1/2}$  & \tablefsize  5.5  \\
         \tablefsize 609.76, 624.93    & \tablefsize Mg X  &\tablefsize $2\rm{s}\ \ ^2\rm{S}_{1/2}$ &-&\tablefsize $2\rm{p}\ \ ^2\rm{P}_{3/2,1/2}$  & \tablefsize  6.1  \\
         \tablefsize 499.37, 520.66    & \tablefsize Si XII  &\tablefsize $2\rm{s}\ \ ^2\rm{S}_{1/2}$ &-&\tablefsize $2\rm{p}\ \ ^2\rm{P}_{3/2,1/2}$ & \tablefsize 6.4  \\\midrule
         \multicolumn{6}{l}{ \textbf{Notes.} The Mg X and Si XII spectral lines are seen in their second spectral order.}
         
    \end{longtable}
\end{table*}
\capstarttrue

\begin{figure*}\label{fig: all spectra}
    \centering
    \captionsetup[subfigure]{position=top, labelfont=bf,textfont=normalfont,singlelinecheck=off,justification=raggedright} 
    \subfloat[\label{fig: all spectra lower}]{%
    \includegraphics[page=1, scale=0.5, trim=4.0cm 4cm 22.5cm 2.05in, clip]{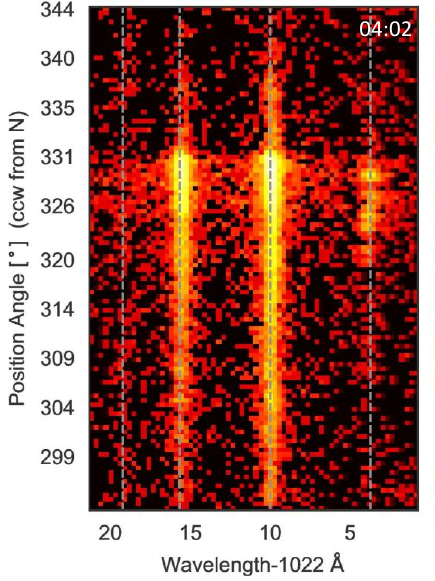}
    \includegraphics[page=2, scale=0.5, trim=4.5cm 4cm 22.5cm 2.05in, clip]{allspectra_ABC.pdf}
    \includegraphics[page=3, scale=0.5, trim=4.5cm 4cm 22.5cm 2.05in, clip]{allspectra_ABC.pdf}
    \includegraphics[page=4, scale=0.5, trim=4.5cm 4cm 22.5cm 2.05in, clip]{allspectra_ABC.pdf}}
    \captionsetup[subfigure]{position=top, labelfont=bf,textfont=normalfont,singlelinecheck=off,justification=raggedright} 
    \subfloat[\label{fig: all spectra higher}]{%
    \includegraphics[page=5, scale=0.5, trim=4.0cm 4cm 22.5cm 2.05in, clip]{allspectra_ABC.pdf}
    \includegraphics[page=6, scale=0.5, trim=4.5cm 4cm 22.5cm 2.05in, clip]{allspectra_ABC.pdf}
    \includegraphics[page=7, scale=0.5, trim=4.5cm 4cm 22.5cm 2.05in, clip]{allspectra_ABC.pdf}
    \includegraphics[page=8, scale=0.5, trim=4.5cm 4cm 22.5cm 2.05in, clip]{allspectra_ABC.pdf}}
    \caption{ For four exposures, the UVCS panel showing the \ionn{O}{VI} doublet lines is presented.  \textbf{(a)} This row corresponds to the last four exposures with the slit at $d_H = 2.6$~\rsun.  \textbf{(b)} This row corresponds to the last four exposures with the slit at $d_H = 3.1$~\rsun.}
\end{figure*}

\section{Data Analysis}\label{sect: data analysis}

As shown in Figure~\ref{fig: all lines}, the three UVCS panels have information from the ultraviolet spectral lines listed in Table~\ref{table: spectral lines}.  In the leftmost panel, the spectral lines of interest come from the primary optical path.  The \ionn{Si}{XII} line at 521~\Ang\ is present in its second spectral order at 1042~\Ang.  Between the \ionn{O}{VI} doublet lines (1032~\Ang\ and 1038~\Ang) is a \ionn{H}{I} Lyman-$\alpha$ instrumental ghost due to imperfect spacing of grating grooves.  In the middle panel, the spectral lines come from the primary and redundant optical paths.  The two \ionn{N}{III} lines are from the primary path and the rest are from the redundant path.  Similar to the \ionn{Si}{XII} line, the \ionn{Mg}{X} line at 610~\Ang\ is detected in its second spectral order at 1220~\Ang.  The rightmost panel only contains the \ionn{C}{III} emission at 977~\Ang\ coming from the primary path.  

At various position angles along the slit, material from the CME or background corona can be Doppler shifted away from the gray dashed line in Figure~\ref{fig: all lines}.  For example, in most of the spectral lines, there is an abnormally bright bulge at PA~=~330$^\circ$.  This CME material extends to lower position angles along the slit.  Near PA~=~315$^\circ$, the material is clearly redshifted.

From one image to another, there are only slight changes to the position angles and Doppler shifts of the CME material.  Per spectral line, prominent features within the CME material can be tracked from one height to another by monitoring the consistency of the structure's position angle, Doppler shift, brightness relative to other spectral lines, and spatial size (along the slit).  As the CME evolves, these characteristics should change; but, we expect only minor changes over the time intervals required for UVCS observations to shift from height to another.  These shifts often occur at intervals of about ten minutes (cf. Table~\ref{table: observations}).

Tracking specific structures seen along the slit from one height to another becomes difficult if a structure becomes too faint or becomes visually mingled with another structure.  Tracking can also be difficult if a structure observed at one height has similar characteristics as a different structure observed at a higher height.  If one confuses distinct structures as the same structure, this will lead directly to a miscalculation of the velocity.  An example of this occurs when distinct parts of the same elongated, filamentary structure are observed at distinct heights along the POS.  

Tracking specific structures should not depend solely on observations of plasma that share similar characteristics at multiple heights.  For the sake of accurately assessing the evolution of the specific clump of plasma, it is necessary to consider the reasons why the clump might be bright in one image and faint in the next image or be found at a different PA or different Doppler shift in subsequent images.  In such scenarios, the continuity of information between images for a single clump of plasma becomes ambiguous.  Therefore, we use three approaches to confirm that the single-slit UVCS has observed the same structure over multiple images at distinct coronal heights: we consider the spatial position, the brightness, and the velocity of clumps at each height.

\subsection{Confirmation from spatial position}\label{sect: spatial confirmation}

The automated programming for UVCS operations was set to take 2, 2, 3, 3, 4, 5, and 7 exposures when observing heights 1.4, 1.5, 1.7, 1.9, 2.1, 2.6, and 3.1~\rsun\ respectively.  The most images come from heights 2.6 and 3.1~\rsun\ and thus those images would provide the best chance at discerning the same material at multiple heights.  The last four images taken at 2.6~\rsun\ are shown in Figure~\ref{fig: all spectra lower} and the last four images taken at 3.1~\rsun\ are shown in Figure~\ref{fig: all spectra higher}.  Each image only shows the panel of the detector that contains the \ionn{O}{VI} doublet lines, and the visual contrast of each image is arbitrarily set to best emphasize the \ionn{O}{VI} features.  Therefore, some \ionn{H}{I} Lyman-$\beta$ features are present but difficult to see; and, the brightness of one image should not be compared to that of another. 

At 2.6~\rsun\ (cf. Figure~\ref{fig: all spectra lower}), the first image contains a bright clump of CME core material at 330$^\circ$ in the \ionn{O}{VI} lines.  This material extends to lower positions on the slit.  The Lyman-$\beta$ emission shows distinct clumps of \ionn{H}{I} material near the same position angle of 330$^\circ$.  In the second image (\clump{A}), the brightest clump of \ionn{O}{VI} material is slightly higher than before.  There is now a clump at the lower position of 322$^\circ$, which has a slightly wider spectral width than the material at the same position angle in the previous image taken $\sim$3 minutes prior.  The white arrow points at this new clump for the 1038~\Ang\ line, although the same phenomenon occurs at 1032 and 1026~\Ang\ and in other UVCS panels at 1216 and 977~\Ang\ that are not shown.  The third image (\clump{B}) shows the highest clump at a slightly higher position angle than 3 minutes prior, and the white arrow is higher to show that the lower clump's position is higher as well.  Now, the \ionn{H}{I} emission at 1026~\Ang\ (and at 1216~\Ang) is relatively faint at that lower position but still has a clump of \ionn{H}{I} material that remains bright at the higher position.  In the fourth and final image (\clump{C}), the white arrow is nearly overlapping with the highest clump to indicate that much of the lower material seen 3 minutes prior is now predominantly at this high position, although some of this clump's material is still seen at the lower position.  Therefore, between the four images taken at intervals of $\sim$3 minutes, there is a clump of plasma that seems to appear at image~\clump{A} and seems to be one portion of a filamentary structure that is seen again in image~\clump{B} and again in image~\clump{C}. The long filamentary structure, which is travelling outward at a near radial direction along the POS, must be oriented at a small angle from the slit.  This would cause the same strand of material to be imaged at gradually higher (or gradually lower) position angles as it passes by.  This is occurring while another bright strand of material is consistently seen in all four images in almost all spectral lines near PA~=~331$^\circ$.  Although not shown, these phenomena amongst the four images are evident in the \ionn{C}{III} emission as well.

At 3.1~\rsun\ (cf. Figure~\ref{fig: all spectra higher}), similar phenomena occur.  A bright bulge appears in image~\clump{A} (indicated by the white arrow) and its neighboring or connecting material (also indicated by the white arrow) seems to appear at slightly higher positions in image~\clump{B} and image~\clump{C}.  This qualitative assessment of the clumps' positions is evidence to support the hypothesis that the clumps of plasma observed in the last three exposures with the slit at 2.6~\rsun\ are the same clumps of plasma observed in the last three exposures with the slit at 3.1~\rsun.  Although this is clear for the clumps marked by the white arrows, it is likely true also for the consistently bright clumps at the higher position along the slit.  Clumps at the higher position angle seem to keep a similar size (along the slit) throughout images \clump{A}, \clump{B}, and \clump{C} at 2.6~\rsun; and at 3.1~\rsun, the clumps at the higher position angle also exhibit a consistent size throughout images \clump{A}, \clump{B}, and \clump{C}.  Thus, the pattern of behavior seen at the higher position angle remains the same between 2.6 and 3.1~\rsun.   For clumps at both position angles, the two \ionn{O}{VI} emission lines provide the best evidence to qualitatively confirm the hypothesis, but other spectral lines have their own features that follow similar patterns which support the hypothesis as well.

For the clumps observed at heights below 2.6~\rsun, such patterns of behavior are not clearly seen.  At each height below 2.6~\rsun, only two, three, or four images were taken and no distinguishable feature seemed to ``appear'' at multiple heights (like the lowest clump at 2.6~\rsun\ that appears in image \clump{A} and later appears at 3.1~\rsun\ again in image \clump{A}).  Ultimately, the spatial characteristics of the clumps observed below 2.6~\rsun\ do not confirm a multi-height detection.

\subsection{Confirmation from Brightness}\label{sect: brightness confirmation}

\begin{figure}
    \centering
    \includegraphics[width=\linewidth, trim=1.4cm 0.8cm 2.0cm 1.5cm, clip]{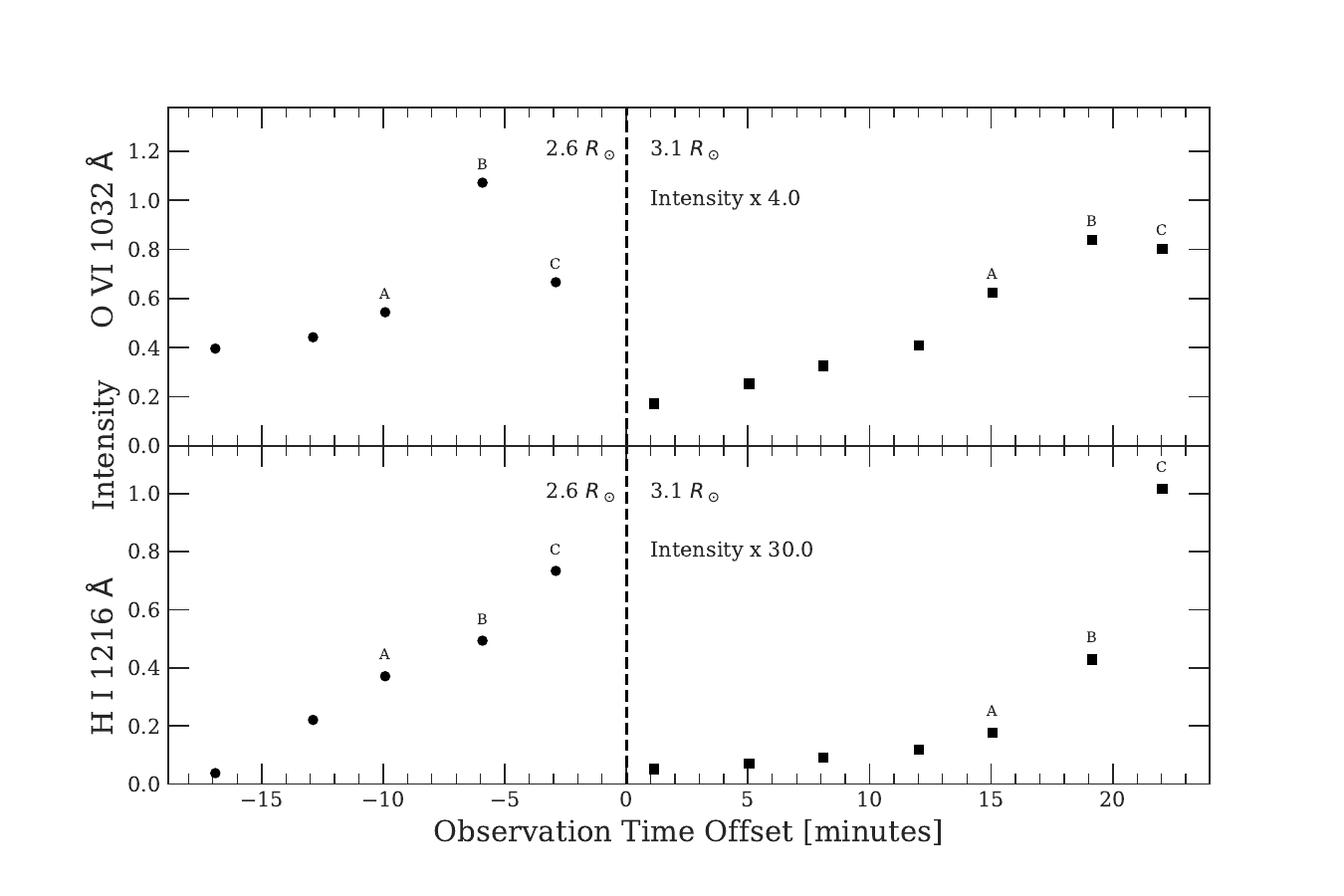}
    \caption{Light curve for composite clumps seen when the slit is positioned at 2.6~\rsun\ (left of vertical dashed line) and at 3.1~\rsun\ (right of vertical dashed line). Top: \ionn{O}{VI} 1032~\Ang\ light curves are shown.  Bottom: \ionn{H}{I} 1216~\Ang\ light curves are shown.  See text in \S\ref{sect: brightness confirmation} for further details.}
    \label{fig: light curve}
\end{figure}

A more quantitative confirmation comes from the clumps' brightness at each height and is summarized in Figure~\ref{fig: light curve}.  We record the total intensity for each spectral line after subtracting out the background corona.  None of the UVCS exposures occur immediately before or after the CME event.  Therefore, we use the relatively faint regions near the top and bottom of the detector (e.g., PA~$\sim$ 340$^\circ$ or 300$^\circ$ for data in Figure~\ref{fig: all spectra}) to determine the average background coronal flux for each spectral line and subtract it from the regions of CME material.  The light curve shows the total intensity amongst all clumps within a given spectral line.  
Although we can clearly distinguish one clump from another along the slit aperture, the changes in position angle and brightness introduce uncertainties in defining a consistent size for each individual clump. 
This is exacerbated as multiple clumps become very close to one another along the slit aperture in a given image.  
Therefore, we maintain consistency by tracking the total brightness of all of the CME material along a given spectral line as one composite clump, instead of tracking the brightness of each individual clump.

Figure~\ref{fig: light curve} shows light curves for the \ionn{O}{VI} 1032~\Ang\ emission (in units of $10^{10}$~photons~steradian$^{-1}$~cm$^{-2}$~s$^{-1}$) and the \ionn{H}{I} 1216~\Ang\ emission (in units of $10^{12}$~photons~steradian$^{-1}$~cm$^{-2}$~s$^{-1}$) for slit positions of 2.6 and 3.1~\rsun.  The vertical dashed line visually separates the data taken at 2.6~\rsun\ from the data taken at 3.1~\rsun.   The intensities at 3.1~\rsun\ are arbitrarily amplified by a factor four for 1032~\Ang\ and thirty for 1216~\Ang\ in order to visually place the light curves on the same scale.  

As previously mentioned, the most images per height are taken at 2.6 and 3.1~\rsun, which give enough information to make useful height-to-height comparisons.  In the light curves for 1032~\Ang, the final three images (\clump{A}, \clump{B}, and \clump{C}) at both heights yield a pattern where the composite clump intensity is brightest for image~\clump{B}, second-brightest for image~\clump{C}, and third-brightest for image~\clump{A}.  This suggests that the composite clumps observed at 2.6~\rsun\ are the same as the composite clumps observed at 3.1~\rsun.  The same can be said about the light curves for 1216~\Ang, which have their own pattern of monotonically increasing over time.  Although this height-to-height similarity may also be true for the images taken prior to seeing composite clump~\clump{A}, we focus on the last three images since they provide the best signal to noise ratio.  

We find these two forms of confirmation despite the composite clumps' decrease in brightness as they travel from 2.6 to 3.1~\rsun.  The \ionn{O}{VI} emission at 1032~\Ang\ drops by a factor four and the \ionn{H}{I} emission at 1216~\Ang\ drops by a factor of thirty.  The difference in factors might be attributed to the \ionn{H}{I} being in a cooler region of the CME core that is separate along the LOS from the \ionn{O}{VI}.  The general decrease in brightness can occur for many different reasons.  Some of the decrease may come from a change in density and temperature as the material expands between 2.6 and 3.1~\rsun.  The decrease in brightness for distinct spectral lines can be due to changes in ionization states within the emitting plasma.  Ultimately, the degeneracy amongst parameters that affect the CME's brightness obscures the specific underlying mechanisms that are responsible for the specific decreases in brightness observed by UVCS.

\subsection{Multi-height velocity}\label{sect: multi-height velocity}

Since composite clumps \clump{A}, \clump{B}, and \clump{C} seem to appear in the UVCS observations at both 2.6 and 3.1~\rsun, we can estimate their total velocities.  We determine the POS velocity by accounting for the two distinct times each clump is observed at two distinct heights (and position angles) in the corona.  At both heights, we find the intensity-weighted centroid of each composite clump for each spectral line.  Using the centroid positions and observation times, we estimate an average velocity in the POS for the composite clump between 2.6 and 3.1~\rsun.  The centroid positions indicate a direction for the POS velocity vector that is almost radially outward.  This is due to the composite clumps being found at nearly the same position angles at 2.6~\rsun\ (with centroid PA~$\sim$~327$^\circ$) and 3.1~\rsun\ (with centroid PA~$\sim$~324$^\circ$).  When all of these factors are considered, each composite clump within its respective spectral line yields a multi-height, average velocity in the POS equal to $\sim$250~\unitv.

To give an example, we determine a centroid position for each spectral line in which composite clump~\clump{B} is found.  For this clump, the average of the centroid PAs is 327.4$^\circ$ when the slit's center is at 2.6~\rsun\ and 323.7$^\circ$ when the slit's center is at 3.1~\rsun.  The distance between the centroids is 0.55~\rsun\ with a difference of 25 minutes in observation times.  This corresponds to a speed of 255.1~\unitv\ along the POS.  This is applied to the clump's LOS velocity at 2.6~\rsun\ and its LOS velocity at 3.1~\rsun.  As a source of uncertainty, the estimated difference in times of observing clump~\clump{B} may be erroneous due to the 3-minute exposures.  Considering this, an observation time difference of 28 minutes makes the POS velocity 227.8~\unitv\ and a difference of 22 minutes yields 289.9~\unitv, which suggests a $\sim$30~\unitv\ uncertainty about the POS estimate of 255.1~\unitv. 

We determine the instantaneous LOS velocity component from Doppler shifts of each spectral line.  %
As an example, the spectral lines emitted by clump~\clump{B} exhibit Doppler shifts that average to 54.3~\unitv\ as a blueshift at 2.6~\rsun\ and 56.1~\unitv\ as a redshift at 3.1~\rsun.  The transition from blueshift to redshift could be evidence of helical motion; but, there are not enough observations of clump~\clump{B} (at multiple heights) to confirm periodicity in its Doppler shifts and thus helicity in its motion.  

For our final velocities, if a composite clump has a POS estimate of 250~\unitv\ and a LOS estimate (for a given ion and spectral line) of 50~\unitv, this altogether yields a total velocity magnitude of 255~\unitv\ with a direction oriented 11$^\circ$ out of the POS.  To account for unknown sources of error, we conservatively adopt an upper limit of 300~\unitv\ and a lower limit of 200~\unitv\ for each composite clump.  This multi-height, average velocity is used in \S\ref{sect: velocity confirmation} to obtain the aforementioned velocity-based confirmation of composite clumps \clump{A}, \clump{B}, and \clump{C}.

We do not attempt the velocity-based confirmation for composite clumps found at heights below 2.6~\rsun\ (i.e., $d_H=$1.4, 1.5, 1.7, 1.9, and 2.1).  Considering their positions along the slit (as in \S\ref{sect: spatial confirmation}) and their light curves (as in \S\ref{sect: brightness confirmation}), there are not enough images taken at these heights to confirm that a clump captured at one height was also captured at another height.  Without either of these forms of confirmation, two distinct heights and observation times cannot be used to determine the multi-height velocity estimate of any of these clumps.  Therefore, we exclude these clumps from the velocity-based confirmation test in \S\ref{sect: velocity confirmation}.  The lack of various forms of confirmation implies that each of these clumps were likely observed at only a single height, which is typical for observations by single-slit coronagraph spectrometers.  Therefore, we do not use these clumps when constraining the CME core's physical properties as a function of height.

\vspace{\vsp}
\begin{figure}
    \centering
    \includegraphics[width=\linewidth, trim=0.3cm 0.3cm 0.3cm 0.88cm, clip]{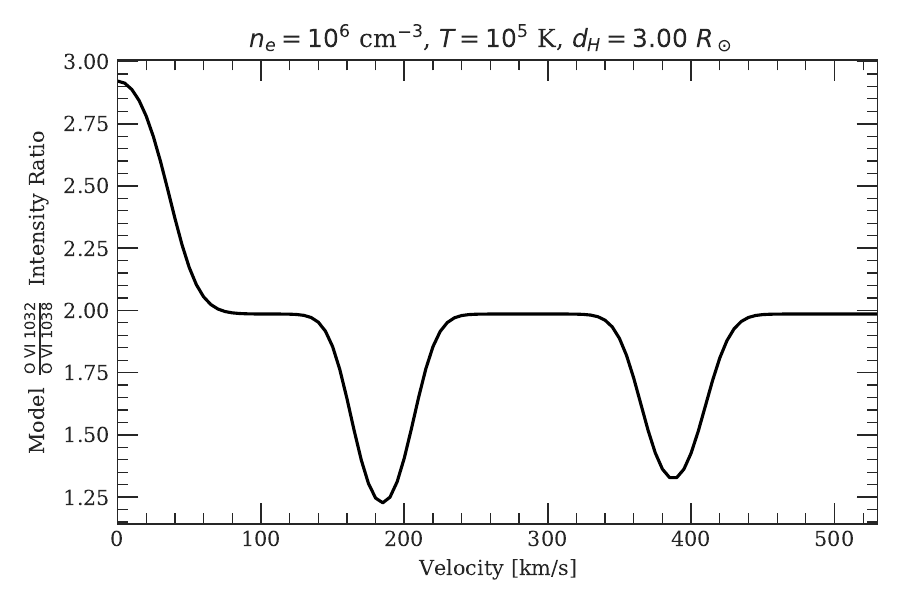}
    \caption{Model intensity ratios, $I_{1032}/I_{1038}$, when the scattering plasma is at a heliocentric distance $r=3.0$~\rsun\ under a temperature of $10^5$~K and a density of $10^6$~\unitn\ while travelling radially outward from the solar surface. } 
    \label{fig: model OVI}
\end{figure}

\begin{figure}
    \centering
    \includegraphics[width=\linewidth, trim=0.3cm 0.3cm 0cm 0.2cm, clip]{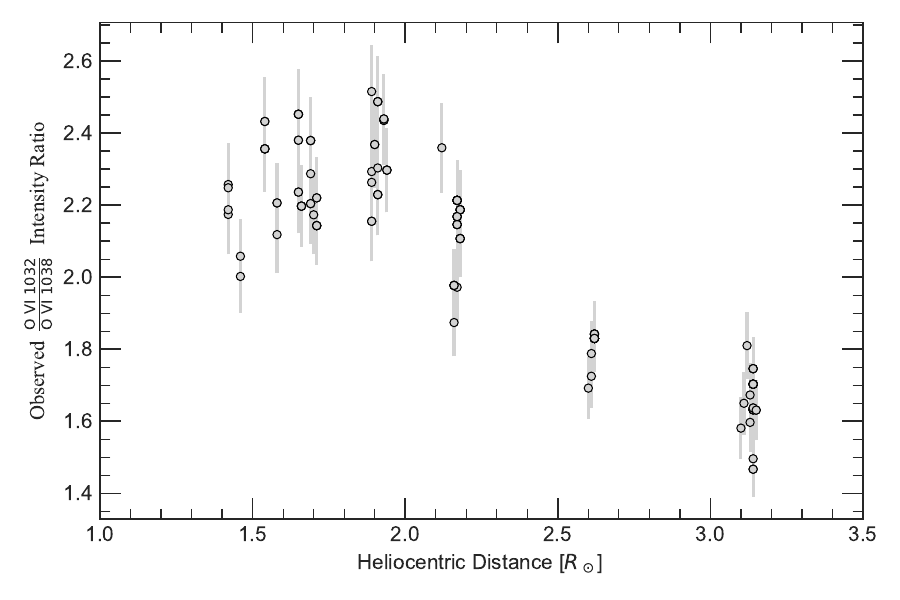}
    \caption{Intensity ratios of individual clumps observed in the \ionn{O}{VI} doublet lines.}
    \label{fig: uvcs OVI}
\end{figure}

\capstartfalse
\begin{table*}
\begin{longtable}{rcccccc}

\label{table: uvcs ratios}
\\\multicolumn{7}{c}{ {\scshape \tablename\ \thetable}:  UVCS composite clump intensity ratios} \\\midrule\midrule
Ratio\ \ \ \  &  \clump{A1}  &  \clump{A2}  &  \clump{B1}  &  \clump{B2}  &  \clump{C1}  &  \clump{C2} \\\midrule
\csvreader[
        before filter=\ifnumless{\thecsvinputline}{4}{\csvfilteraccept}{
        \ifnumgreater{\thecsvinputline}{5}{%
        \ifnumequal{\thecsvinputline}{8}{\csvfilterreject}{%
        \ifnumequal{\thecsvinputline}{10}{\csvfilterreject}{%
        \ifnumequal{\thecsvinputline}{11}{\csvfilterreject}{ \csvfilteraccept }%
        }%
        }%
        }{\csvfilterreject}%
        },
        late after line=\\
        ]{latex_intensityratios.csv}{}{
        \ifthenelse{ \equal{\csvcoli}{OVI} }{\rOVI}{}%
        \ifthenelse{ \equal{\csvcoli}{Lyman} }{\rLyman}{}%
        \ifthenelse{ \equal{\csvcoli}{CIII} }{\rCIII}{}%
        \ifthenelse{ \equal{\csvcoli}{hi1216} }{\rhialpha}{}%
        \ifthenelse{ \equal{\csvcoli}{hi1025.72} }{\rhibeta}{}%
        \ifthenelse{ \equal{\csvcoli}{ov1218} }{\rOVbr}{}%
        & \csvcolii & \csvcoliii & \csvcoliv & \csvcolv & \csvcolvi & \csvcolvii  
        }
        \midrule
        \multicolumn{7}{ p{\linewidth} }{ \textbf{Notes.}  The relevant intensity ratios for three composite clumps (\clump{A}, \clump{B}, and \clump{C}) are given.  Each clump of CME material is labeled with a \clump{1} to represent its observation at the first coronal height, 2.6~\rsun, and a \clump{2} to represent the second coronal height, 3.1~\rsun.  As elaborated in the text of \S\ref{sect: constraints}, the intensity ratio uncertainties we adopt are based on both the calibration of UVCS data and the uncertainties within atomic models that describe each transition line.  }
\end{longtable}
\end{table*}
\capstarttrue


\section{Plasma Diagnostics} 
\label{sect: plasma diagnostics}
We can deduce the physical properties of the observed plasma by decomposing the components of the UV radiation observed.  We use atomic models to determine the contribution from collisional excitation or radiative excitation of the emitting ions.  Assuming the \textit{coronal model approximation}, ions are excited from their ground state primarily by free electron collisions or photo-absorption and subsequently are de-excited primarily through spontaneous radiative decay.  For excited ions in metastable states, the radiative decay rate is much slower and the collisional de-excitation rate is no longer negligible.  \cite{Akmal.2001} exploited this fact with the [\ionn{O}{V}] 1214~\Ang\ and \ionn{O}{V}] 1218~\Ang\ lines (cf.~Table~\ref{table: spectral lines}).  Because of the collisional de-excitation, the intensity ratio of the \ionn{O}{V} lines became a useful density diagnostic for their CME analysis.  The intensity ratio between the \ionn{N}{III} 990 and 992~\Ang\ lines can serve as a density diagnostic as well.  

Unfortunately, for three of these lines, clumps \clump{A}, \clump{B}, and \clump{C} are too faint to clearly distinguish them from the grating-scattered light from bright Lyman-$\alpha$ emission and the background corona.  For the last three exposures taken at both 2.6 and 3.1~\rsun, only the \ionn{O}{V}] 1218~\Ang\ line is bright enough.  Therefore we do not make use of the other three lines in our plasma diagnostics.  Also, the three clumps are not seen in the second-order \ionn{Mg}{X} and \ionn{Si}{XII} lines.  We therefore discard these spectral lines from our analysis as well.

\subsection{Two Components of Emissivity} \label{sect: emissivity}

The \ionn{O}{VI} doublet can serve as both a velocity and density diagnostic if we consider the aforementioned two processes of plasma excitation in the corona.  For emission at wavelength $\lambda$, the two processes are responsible for the two components of emissivity.  This yields a total local intensity $I_{\lambda}$, in units of photons~cm$^{-2}$~s$^{-1}$~steradian$^{-1}$, that can be summarized as the following: 
\begin{equation}\label{eq: intensity}
    \begin{split}
        I_{\lambda} = &\ \frac{1}{4\pi} \int_{\rm{LOS}} ( \epsilon_{c,\lambda} + \epsilon_{r,\lambda} )ds , \\
        \epsilon_{c,\lambda} = &\ n_{Z,z} \cdot n_e q_{ex, \lambda}(Z,z,T) , \\
        \epsilon_{r,\lambda} = &\ n_{Z,z} \cdot I_\odot(\lambda_i + \delta\lambda_i) \sigma_\lambda \mathcal{W}(r) . \\
    \end{split}
\end{equation}
The collisional excitation component $\epsilon_{c,\lambda}$ includes collisional excitation rate coefficient $q_{ex,\lambda}$ in units of photons~cm$^3$~s$^{-1}$.  
The radiative excitation component $\epsilon_{r,\lambda}$ includes the dilution factor $\mathcal{W}(r)$ which is the solid angle,  $2\pi \Big( 1- \sqrt{ 1 - (R_\odot /r)^2 } \Big)$, subtended by the solar disk with respect to the scattering plasma that is at a heliocentric distance $r$ away: a distance that is not confined to the POS like $d_H$.  The effective cross section for scattering radiation of a given wavelength is $\sigma_\lambda$ and the $I_\odot$ is the intensity from the solar disk radiation that is to be scattered.  The incident radiation from the disk emits at $\lambda_i$ and is Doppler shifted by $\delta\lambda_i$ with respect to the velocity of the plasma.  The free electron density $n_e$ only directly affects the collisional excitations.  The ion density $n_{Z,z}$ directly affects the ion's collisional and radiative excitations.  It can be characterized as $n_{Z,z} = n_H A_{Z} f_{z}$, where the number density of ion $z$ of element $Z$ is the product of the hydrogen density ($n_H$), the $Z$ element's abundance ($A_{Z}$) relative to hydrogen, and the fraction ($f_{z}$) of all ions $z$ of element $Z$.  Values for $q_{ex,\lambda}$ are given by the CHIANTI atomic database \citep{Dere.1997, Dere.2019}.  
Values for $\sigma_\lambda$  are based on oscillator strengths from CHIANTI, and values for $I_\odot$ are based on observations from \cite{Vernazza.1978} taken near solar minimum.  We multiply their $I_\odot$($\lambda_i$~=~\ionn{H}{I}~1216) values by a factor of 1.37 and all other solar disk emission lines by 1.5 to account for the solar maximum activity in 1999 according to measurements from the \textit{UARS}/SOLSTICE instrument \citep{Rottman.2001}.

The ratio of the \ionn{O}{VI} doublet lines, $I_{1032}/I_{1038}$, can be a useful diagnostic when their collisional and radiative components are considered.  The ratio of their collisional components is $\epsilon_{c,1032}/\epsilon_{c,1038}$~=~2 due to the collision strengths of their atomic transitions.  Consequently, the ratio of total intensities becomes $R$~=~$I_{1032}/I_{1038}$~=~2.0 when the collisional components dominate.  When both radiative components dominate, the ratio becomes $R$~$>$~2 and indicates a slow velocity (i.e., small $\delta\lambda_i$) for the scattering plasma (cf. Figure~\ref{fig: model OVI}).  
However, at speeds greater than 100~\unitv\ the ratio of total intensities becomes $R$~$\leq$~2.0.  Figure~\ref{fig: model OVI} shows an example of how these characteristics of the \ionn{O}{VI} ratio can serve as a useful velocity diagnostic for the CME that we study.  Pumping of the 1038~\Ang\ line's radiative component brings the ratio to $R$~$<$~2.0.  This radiative pumping happens in Figure~\ref{fig: model OVI} due to the Doppler-shifted solar disk emission from \ionn{C}{II}: 1036.34~\Ang\ or 1037.02~\Ang\ following the transition $2\rm{s}^2 2\rm{p}\  ^2\rm{P}_{1/2\ \rm{or}\ 3/2}$~---~$2\rm{s} 2\rm{p}^2\ ^2\rm{S}_{1/2}$.  This is applicable for plasma traveling near 400 or 200~\unitv\ respectively.

Due to thermal broadening of the absorption line profiles, the range of $R<2$ velocities and $R>2$ velocities can broaden under hotter conditions.  These ranges are also affected by our assumption of 25~\unitv\ for any nonthermal broadening.  As long as the CME's velocity is roughly estimated and the observed ratio $R$ deviates from 2.0, the \ionn{O}{VI} doublet lines can provide tighter constraints on the range of original velocity estimates.  While the \ionn{O}{VI} doublet is radiatively pumped by Doppler-shifted emission, we can take the ratio between a line's collisional component and radiative component to evaluate an average electron density:

\begin{equation}\label{eq: OVI density}
    n_e = \frac{ I_{\odot}(\lambda_i = \rm{\ionn{C}{II}}\ 1037) \sigma_{1038} \mathcal{W}(r) }{ q_{ex, 1038} } \frac{R}{2-R} \ ,\ R<2, 
\end{equation}
which is useful for clumps \clump{A}, \clump{B}, and \clump{C} traveling at $\sim$250~\unitv.

Using the same concepts, we can also estimate a LOS thickness of the plasma cloud.  Once the plasma's intensity is measured, we can exercise a forward modelling procedure to estimate the two components of emissivity by using a model temperature, density (and ionization state), and velocity.  With $I_\lambda$ being measured and the emissivities being modelled, we can estimate a LOS length $s_{LOS}$ of the emitting plasma cloud.  To set an observational constraint on our model emissivities, we require that the estimated LOS length is greater than 10\% of the observed clump's POS size along the slit $s_{slit}$ and less than three times the clump's POS size: 
\begin{equation}\label{eq: length}
    0.1\times s_{slit}  <  s_{LOS}  <  3\times s_{slit},
\end{equation}
which can serve as useful upper and lower limits, especially when an observed clump is very faint and its $s_{slit}$ is difficult to determine.  
As an example of a typical size, clump~\clump{B}'s average size amongst its spectral lines is $s_{slit}= 0.57$~\rsun\ at 2.6~\rsun\ and $s_{slit}= 0.63$~\rsun\ at 3.1~\rsun.  This size is defined by the distance between positions of median-intensity on either side of the peak-intensity of the composite clump.  

All of the techniques used in Equations \ref{eq: intensity}, \ref{eq: OVI density}, \ref{eq: length} can exploit the Doppler dimming effect \citep{Hyder.1970} of the two-component emissivity observed from coronal, ultraviolet radiation.  Techniques like these can yield diagnostics on the emitting plasma and have been employed by many spectroscopic studies of CMEs \citep[\egcite][]{Ciaravella.2001, Raymond.2004, Bemporad.2006} and the solar wind \citep[\egcite][]{Kohl.1982, Noci.1987, Cranmer.2008, Strachan.2012, Gilly.2020}.  

\subsection{Confirmation from Velocity}\label{sect: velocity confirmation}

We use the \ionn{O}{VI} doublet ratio as a velocity diagnostic to confirm or reject the notion that the single-slit UVCS serendipitously captured images of the same unpredictable CME material at multiple heights in the corona.  We assign $\pm$5\% uncertainties to our observed \rOVI\ intensity ratios to account primarily for relatively minor uncertainties in the radiative components' solar disk line profiles and effective scattering cross sections.  The ratio for each individual clump within each image of the UVCS observations is plotted in Figure~\ref{fig: uvcs OVI}.  

Below 2.1~\rsun, the $R>2$ ratios imply that the clumps have speeds between 0 and 100~\unitv\ (cf.~\S\ref{sect: emissivity} and Figure~\ref{fig: model OVI}).  In this case, the greatest intensity ratio observed, $R$~=~2.52, corresponds to the slowest velocity estimate, which is roughly $\sim$45~\unitv\ depending on the density and temperature of the plasma.  However, as mentioned in \S\ref{sect: multi-height velocity}, the velocity is not confirmed by observations of the same clump at multiple heights. 

At 2.1~\rsun, the $R\approx 2$ ratios suggest that the \ionn{O}{VI} doublet alone is no longer a useful diagnostic.  
If the CME core material is accelerating, the instaneous velocities estimated from the $R>2$ ratios below 2.1~\rsun\ can act as a lower limit for the velocities determined at 2.1~\rsun\ with $R\approx 2$ ratios.  This still leaves much ambiguity in the clumps' velocities as well as a lack of confirmation in their multi-height detection.  Thus, these clumps are not included in our observational constraints of the CME core.

Above 2.1~\rsun, the observed $R<2$ ratios imply that the velocity can be determined if a different, independent estimate of the velocity is first acquired.  We elaborated in \S\ref{sect: multi-height velocity} how the three composite clumps observed at 2.6 and 3.1~\rsun\ have velocities of $\sim$250~\unitv.  Therefore, we constrain the model intensity ratios (like that of Figure~\ref{fig: model OVI}) to $R<2$ models that correspond to velocities between 200 and 300~\unitv.  Figure~\ref{fig: model OVI} is an example showing how there are $R<2$ models that reside within this velocity range.  Therefore, the physical conditions modelled can be constrained by the observed intensity ratios in a way that shows agreement between two independent velocity estimates: the instantaneous velocity estimates from Doppler dimming models and the average velocity estimates from multi-height observations of the plasma.

\subsection{Constraints on Physical Properties} \label{sect: constraints}

Confirming the precise details of the plasma's physical properties will require constraints from more than just the \ionn{O}{VI} ratio.  Other spectral line ratios can act as useful diagnostics for various properties, including temperature, density, velocity, and ionization states.  We present these observational constraints in Table~\ref{table: uvcs ratios}.  For reasons elaborated in \S\ref{sect: brightness confirmation}, we only consider the ratios of the total intensities for the three composite clumps, instead of the many individual clumps.  We conservatively assign uncertainties of 25\%, 30\%, 20\%, 30\%, and 30\% for the \rLyman, \rhialpha, \rhibeta, \rOVbr, and \rCIII\ ratios respectively which include uncertainties in model atomic rates, adopted solar disk emission line profiles, model scattering cross sections, and distinct UVCS calibrations for the primary and redundant optical paths.  

Using distinct ratios as observational constraints carries distinct assumptions about the plasma cloud's environment.  For example, multi-ion ratios like the \rhialpha\ ratio, require the models to assume an isothermal plasma cloud mixed with \ionn{H}{I} ions and \ionn{O}{VI} ions.  Such assumptions should be handled with care.  
Although \cite{Olsen.1994} and \cite{Allen.1998} suggested that the Lyman-$\alpha$ profile of the slow solar wind (from Olsen) and the fast solar wind (from Allen) acts as a good proxy for the free proton effective temperature, we do not assume this to be true for temperatures that are derived by intensity ratios that include Lyman-$\alpha$ emission.



\begin{figure*}\label{fig: Q-B2 profiles}
    \centering
    \captionsetup[subfigure]{position=top, labelfont=bf,textfont=normalfont,singlelinecheck=off,justification=raggedright} 
    \subfloat[\label{fig: density Q-B2}]{\includegraphics[page=1, width=0.49\linewidth, trim=0cm 0cm 0cm 0.65in, clip]{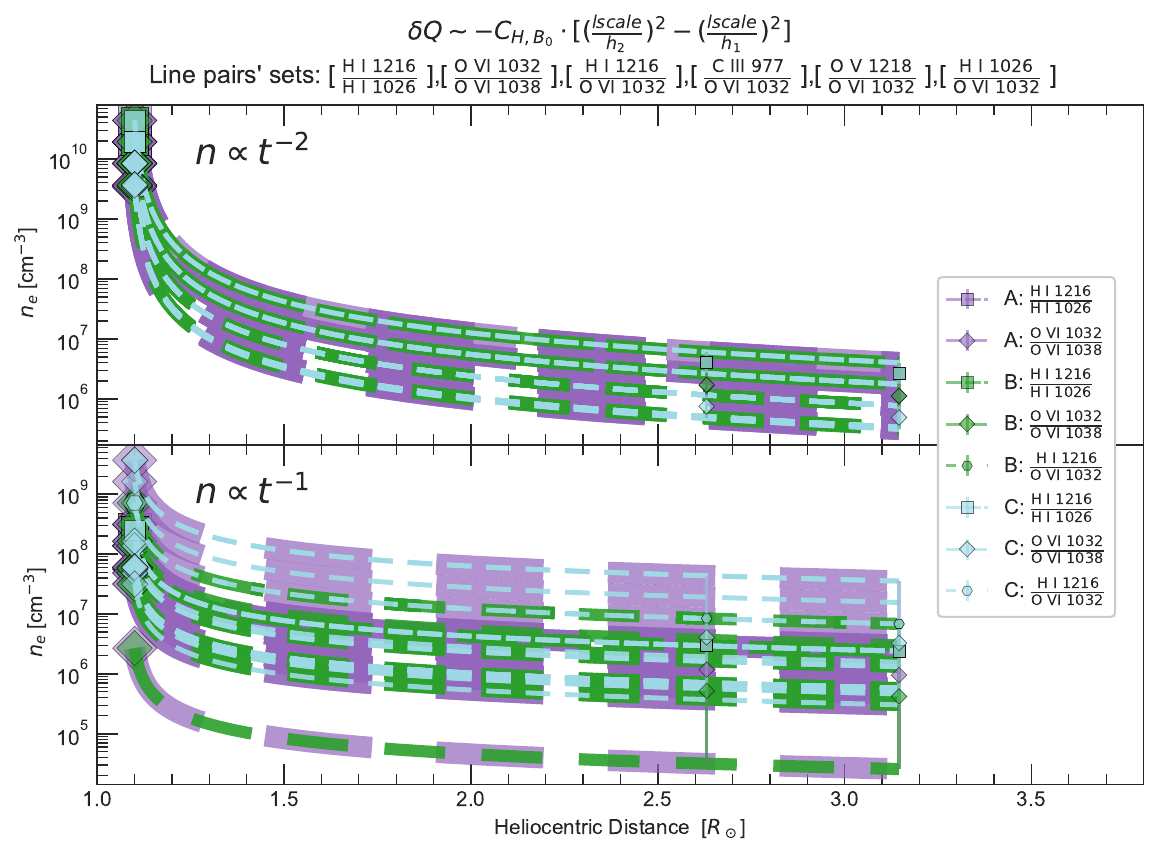}} \hfill  
    \subfloat[\label{fig: temp Q-B2}]{\includegraphics[page=1, width=0.49\linewidth, trim=0cm 0cm 0cm 0.65in, clip]{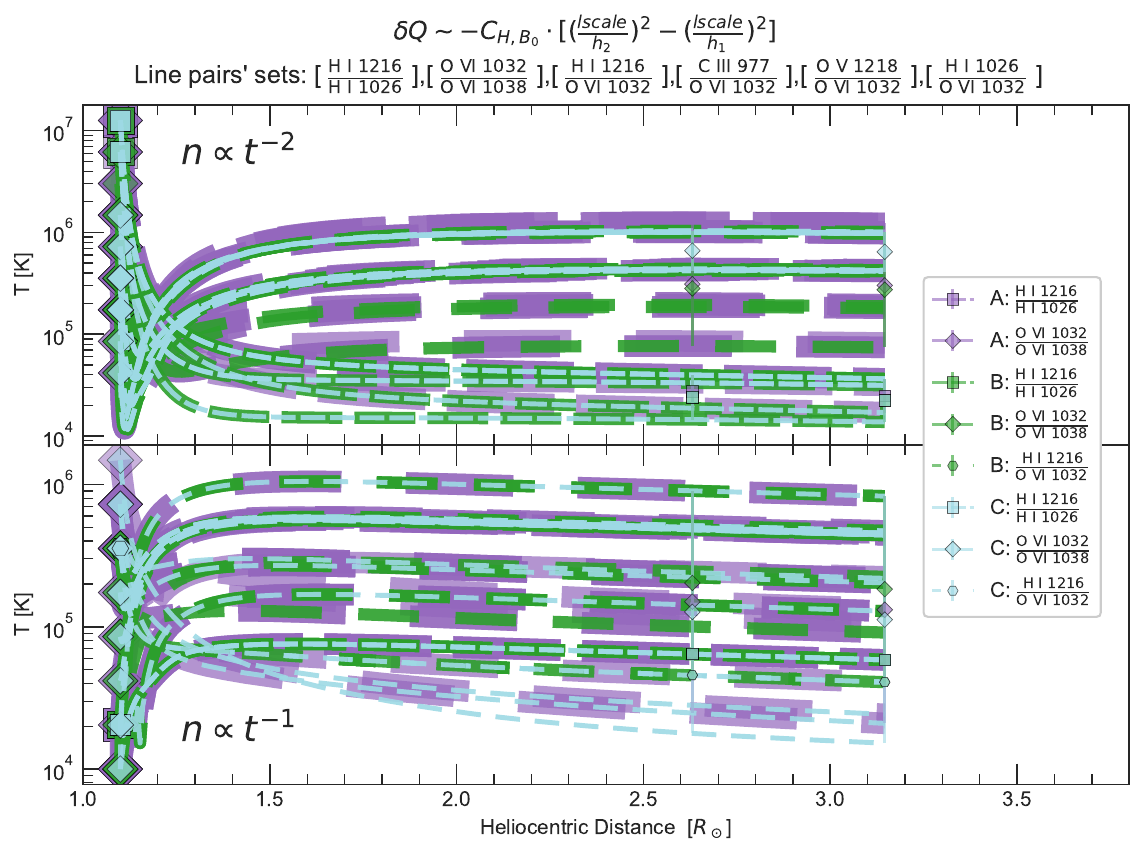}}
    \caption{ Resultant set of electron density and temperature profiles after using the $Q_{\rm{KR}}$ heating function where $\alpha_B=2$.  The square, diamond, and hexagon symbols located at 2.6 and 3.1~\rsun\ are arbitrarily positioned in the middle of their respective upper and lower limits for visual clarity.  The vertical position of this symbol is not statistically more significant than other values within its range.  \textbf{(a)} These model densities have corresponding temperatures, velocities, and ionization states that yield an intensity ratio congruent to the data. \textbf{(b)} These model temperatures have corresponding densities, velocities, and ionization states that yield an intensity ratio congruent to the data. }
\end{figure*}

\section{Numerical Models for Heating Rates}\label{sect: numerical model}

The physical properties of the observed plasma give insight on the rate of heating experienced by the plasma.  We determine these physical properties by modelling a parcel of plasma as it expands and travels radially away from the solar disk.  We primarily monitor its thermal energy as we follow the plasma with a Lagrangian approach and assume a self-similar expansion.  We allow the total density $n$ to monotonically decrease over a total time $t$ and we describe this expansion rate with the power law
\begin{equation}\label{eq: expand}
        \frac{ n_2 }{ n_1 } =  \bigg( \frac{t_2}{t_1} \bigg)^{-\alpha_t}
\end{equation}
where the density $n_1$ at time $t_1$ changes to the density $n_2$ at time $t_2$ at a rate that corresponds to the expansion index $\alpha_t$.  

We model the expansion rate with an index of 3.0 (cubic), 2.0 (quadratic), or 1.0 (linear) to act as an approximation for the dimensionality of the plasma's expansion.  The density is an observable quantity of the CME but the dimensionality of its expansion is unclear because observations of the three composite clumps do not provide much information about the CME's morphological properties.  To minimize our geometric assumptions, only the density parameter is directly affected by our assumed expansion rate, $n \propto t^{-\alpha_t}$.  However, the lack of morphological information can make the physical interpretation of $\alpha_t$ ambiguous.  For example, it is possible for a plasma to undergo a 3D expansion at a rate of $n \propto r^{-2} \propto t^{-2}$ (i.e., $\alpha_t$=2) under simplified conditions where a fluid experiences a steady state flow while its volume expands at a constant rate in each dimension.  Potential discrepancies like this in dimensionality are exacerbated when considering that a filamentary structure within a CME may be expanding in its long length faster than it expands in its short radius---thus, creating more ambiguity when defining a \textit{single} rate of expansion for CMEs through a power law.  

We use the expansion power law to drive one of our cooling terms for the plasma.  Any decrease in thermal energy via expansion is represented by $dE =  k_B \langle T/n \rangle dn$, where the quantity $T/n$ is averaged over a given time interval and used to express our total thermal expansion as
\begin{equation}\label{eq: energy expansion}
    E_{\delta n} = k_B T \frac{\delta n}{n}\quad \rm{erg\ particle}^{-1} ,  
\end{equation}
where $T$ is temperature, $\delta n$ is the change in density {\color{black}($n_2-n_1$)}, and $k_B$ is the Boltzmann constant.  This is a portion of the thermal energy $k_B T$ that is converted into work that expands the gas.  Due to the $\delta n$ dependence, a cubic expansion rate would allow for a far greater cooling than a linear expansion rate.  

The cooling is augmented by conversions of thermal energy into radiation that escapes the system.  The radiative cooling is expressed as
\begin{equation}\label{eq: rad cooling}
    P_{rad} =  n_e n_{Z,z} \Lambda_{Z,z}(T) \quad \ \rm{erg\ cm}^{-3}\ \rm{s}^{-1} ,
\end{equation}
where $n_e$ is the free electron density (in units of cm$^{-3}$), $n_{Z,z}$ is the density for ion $z$ of element $Z$, and $\Lambda_{Z,z}(T)$ is the cooling rate coefficient (in units of erg~cm$^{3}$~s$^{-1}$) that accounts for the emission line and continuum processes that can occur at a given temperature for a given ion.  We adopt the cooling rate coefficients computed by the CHIANTI atomic database \citep{Dere.1997, Dere.2019}.

We model these cooling mechanisms to reduce the thermal energy while an unknown heating mechanism augments the thermal energy by one of our five heating parameterizations.  The first two are not motivated by any known physical mechanisms of a CME.  
One is proportional to the density of the plasma and the other is proportional to the square of the density.  We characterize the former as $Q_n = C_H\cdot n$ and the latter as $Q_{n^2} = C_H\cdot n^2$, which each have a heating coefficient $C_H$.  
With the simple $Q_n$ function, we can test the effects of homogeneously generating a constant rate of thermal energy within a CME.  The utility of the $Q_{n^2}$ is in its square-density dependence.  We can gain insight on how the relatively high density environment near the solar surface may drastically affect the heating, and this would directly counteract the square-density dependent radiative cooling.

The third heating parameterization was adopted by \cite{Allen.1998,Allen.2000} to model the fast solar wind as the motion of neutral hydrogen, free protons, and free electrons are influenced by Alfv\'en waves.  It is described as 
\begin{equation}\label{eq: Q-ht}
    Q_{AHH} = C_H\cdot e^{ -\frac{ r_a }{ \mathcal{H} } } \quad \ \rm{erg\ cm}^{-3}\ \rm{s}^{-1} ,
\end{equation}
where $r_a$ is the altitude (equal to $r -1.0 \rsun$) and $\mathcal{H}$ is the scale height.  We adopt 0.7~\rsun\ as our scale height to remain consistent with \cite{Allen.1998}, as this was one of two model scale heights they considered.  Here, the heating rate coefficient, $C_H$, has units of $\rm{erg\ cm}^{-3}\ \rm{s}^{-1}$.  See also \cite{Withbroe.1988} and \cite{Lionello.2009} for additional implementations of this heating function.

Our last two parameterizations are expressed by one magnetic heating function.  Just as we used a power law to express the dimensionality of our self-similar expansion, we also present a power law to express the dimensionality of this magnetic heating:
\begin{equation}
    \label{eq: Q-B3}
    \begin{split}
        Q_{KR} & = C_{H,B_0}  \bigg[ \bigg( \frac{l}{ r_{a,1} } \bigg)^{\alpha_B} - \bigg( \frac{l}{ r_{a,2} } \bigg)^{\alpha_B}  \bigg] \ \rm{erg\ cm}^{-3} \\ 
               & = \frac{B^2_0}{8\pi}  \bigg[ \bigg( \frac{l}{r_{a,2}-\delta r_a} \bigg)^{\alpha_B} - \bigg( \frac{l}{ r_{a,2} } \bigg)^{\alpha_B}  \bigg]  ,
    \end{split}
\end{equation}
where the coefficient $C_{H,B_0}$ is a constant magnetic pressure (in units of erg~cm$^{-3}$) that includes an initial magnetic field strength $B_0$.  
This magnetic field strength is mostly applicable to a CME's flux rope, for which we assign a characteristic length scale $l$ and a magnetic expansion index $\alpha_B$.  We consider $l$ to be 0.1~\rsun, which is typical of a pre-CME flux rope; and, we consider $\alpha_B$ to be either 3.0 or 2.0, which represents a 3D or 2D expansion of the flux rope.  We test both choices of $\alpha_B$ and distinguish each choice as its own heating parameterization.  We use $\alpha_B$ to parameterize the unknown morphological properties of the magnetic flux rope.  Such properties could influence $\delta r_a$: the plasma's change in altitude ($r_{a,2} - r_{a,1}$) within a characteristic timescale while traveling between the two altitudes at some average, bulk speed within the corona.  

This magnetic heating function is inspired by the \cite{Kumar.1996} model for heating when the magnetic helicity is conserved \citep{Taylor.1974, Berger.1984} in a self-similarly expanding flux rope.  They analytically perform a dimensional analysis of magnetic helicity and suggest that it can follow the form $H_m \sim$ \textit{constant} $\sim l\cdot \int B^2 dV$, where $l$ is some characteristic length scale and $B^2$ is the magnetic energy that must decrease as the volume increases.  
In their model, a portion of the free magnetic energy is gradually converted to thermal energy as the flux rope extends to higher heights in the corona.  We mimic this by using a fraction (given as the total quantity in square brackets) of the magnetic energy to consistently heat the parcel of plasma.  
The nature of our specific $Q_{KR}$ function's magnetic heating is useful but it is not meant to explain any specific properties of the CME's (unobserved) flux rope.  Without knowing the morphology of the observed CME, we do not attempt to deduce the properties of its flux rope within this paper.  

\begin{figure}
    \centering
    \includegraphics[page=1, width=\linewidth, trim=0cm 0.2cm 0cm 0.65in, clip]{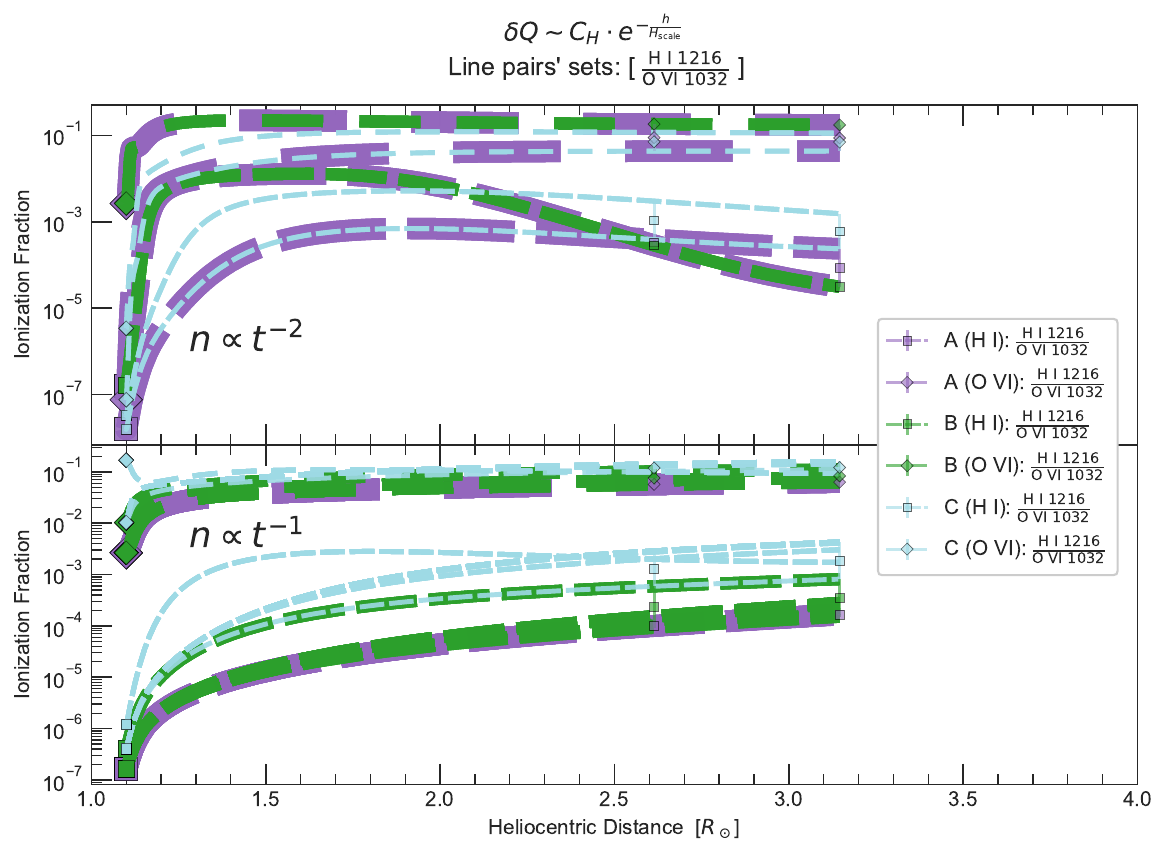}
    \caption{Resultant set of ionization fraction profiles after using the $Q_{\rm{AHH}}$ heating function and constraints from only the \rhialpha\ ratio.}
    \label{fig: ionization Q-ht}
\end{figure}


\subsection{Initial and Final Physical Conditions}

Our numerical modelling procedure yields the physical conditions of the plasma as a function of time and height in the corona.  We have the two cooling terms counteract one of the five heating parameterizations in order to change the internal thermal energy of the parcel of plasma: {\color{black} $U = \frac{3}{2} n k_B \Delta T$.  We apply the following formula,
\begin{equation}\label{eq: energy}
    U = Q - P_{rad}\cdot \Delta t + n\cdot E_{\delta n}, 
\end{equation}
for a given heating parameterization, $Q$, and a given self-similar expansion index, $\alpha_t$.  When using the magnetic heating parameterization, $Q_{\rm{KR}}$, we allow the expansion index for the entire CME, $\alpha_t$, to differ from the index $\alpha_B$ that we use for the rate of expansion caused by the unobserved flux rope's morphology.}

We start {\color{black}our procedure} by establishing a grid of initial conditions and allow each cell (or model) of the grid to evolve until the heights of clumps \clump{A},~\clump{B},~and~\clump{C} are reached.  The initial conditions we consider include densities of log$(n_{0}/\rm{\unitn})\in [6.0, 12.0]$ and temperatures of log$(T_{0}/\rm{K}) \in [4.0, 7.1]$ experienced by a plasma cloud in ionization equilibrium at $r=1.1$~\rsun.  The range of heating rate constants ($C_H$) we consider varies from one heating function to another.  Our initial conditions also include coronal elemental abundances from \cite{Feldman.1992} and ion populations in ionization equilibrium from the CHIANTI database.  %

After initiation, we reject models with temperatures that evolve beyond our temperature ceiling of $10^8$~K or below our temperature floor of $10^4$~K.  Once the heights of our observed clumps are reached, we use each model's instantaneous temperature, density, and velocity to determine the emissivities and intensity ratios for the spectral lines observed by UVCS.  %
At these heights, each model must meet the observational constraints established by the multi-height velocity limits, the LOS length limits, and the intensity ratio limits.  We assign each model a range of instantaneous velocities that lie within the multi-height velocity limits: $200 \leq v\ /\ \rm{\unitv} \leq300$.  When compared to the observed intensity of an emission line, the model's emissivities (derived from the model's temperature, density, and velocity) must yield an estimate for the clump's LOS length that is similar to the clump's POS size.  Subsequently, the model's emissivities for a pair of emission lines must yield an intensity ratio that agrees with observations.  Each cell-model within our grid that meets these criteria is included in our final evaluation of the energy budget.  The model's cumulative heating (specific) energy, HE$_{C}$, is compared to the sum of the kinetic (specific) energy, $\frac{1}{2} v^2$, and the gravitational potential (specific) energy, $GM_\odot ( 1/1.1\rsun - 1/ r )$.  The cumulative heating energy is the sum of the model's initial thermal energy and continuous production of thermal energy via the heating function.  Thus, each model will have a lower cumulative heating at the lower height of $r=$ 2.6~\rsun\ than at 3.1~\rsun.  This 1D numerical modelling procedure is a variation of the methods utilized by other UVCS CME heating analyses \citep[][]{Akmal.2001, Ciaravella.2001, Lee.2009, Landi.2010, Murphy.2011}.

\begin{figure*}\label{fig: Q-n}
    \centering
    
    \captionsetup[subfigure]{position=top, labelfont=bf,textfont=normalfont,singlelinecheck=off,justification=raggedright} 
    
    \subfloat[\label{fig: uvcs Q-n single-ratio}]{\includegraphics[page=1, width=0.49\linewidth, trim=0cm 0cm 0cm 0cm, clip=true]{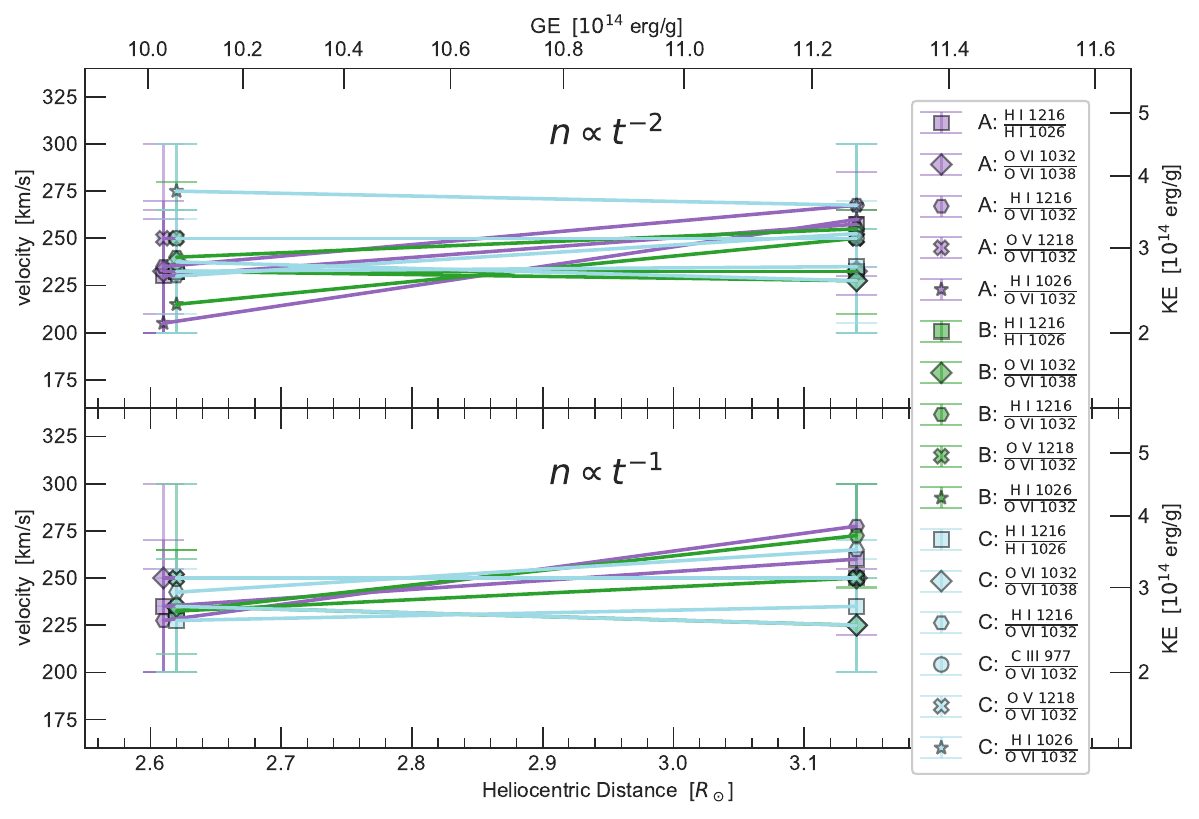}} \hfill  
    \subfloat[\label{fig: energy Q-n single-ratio}]{\includegraphics[page=1, width=0.49\linewidth, trim=0cm 0.1in 0cm 0.6in, clip=true]{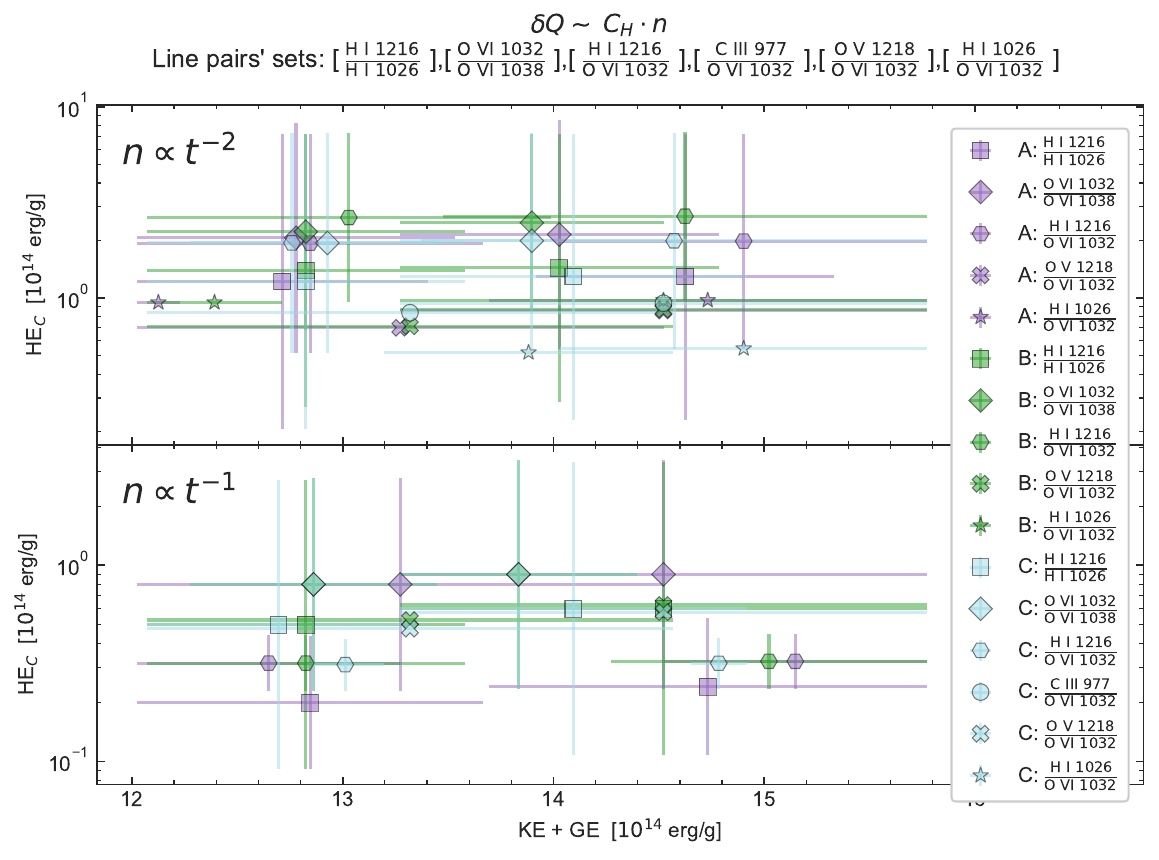}}
    
    \subfloat[\label{fig: uvcs Q-n multi-ratio}]{\includegraphics[page=1, width=0.49\linewidth, trim=0cm 0cm 0cm 0cm, clip=true]{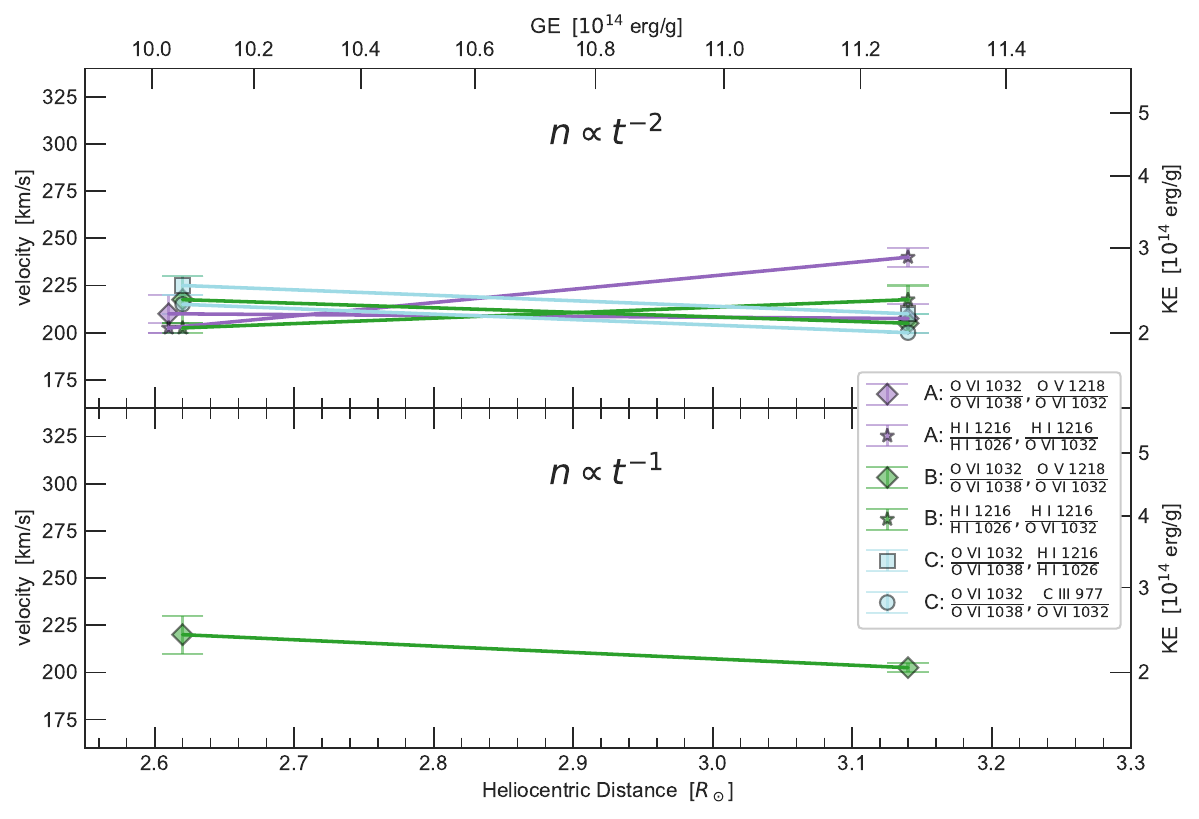}} \hfill  
    \subfloat[\label{fig: energy Q-n multi-ratio}]{\includegraphics[page=1, width=0.49\linewidth, trim=3cm 0.1in 4.5cm 0.6in, clip=true]{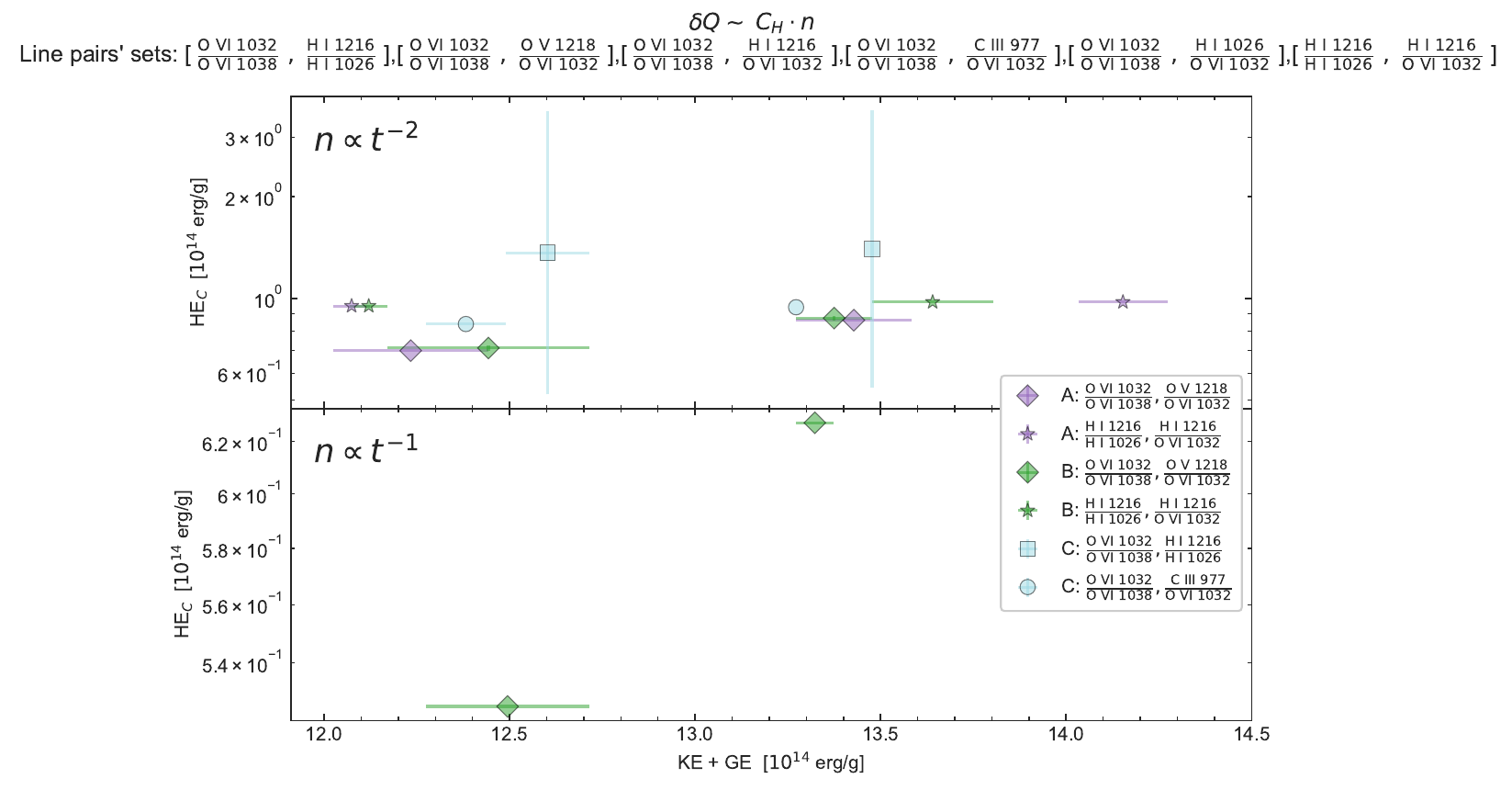}}
    
    \caption{ Observationally constrained models using the $Q_n$ heating parameterization.  Each symbol (i.e., each combination of color and shape) appears twice to represent each clump's observation at two distinct coronal heights, which each indicates a gravitational potential energy.  This is the potential energy overcome by the CME core as it travels from the solar surface to an observation site.  \textbf{(a,c)} The kinetic energy and gravitational potential energy are given.  \textbf{(b,d)} The cumulative heating energy is given.  The horizontal error bars derive directly from the range of model kinetic energies deduced by the constrained model velocities.}
\end{figure*}

\begin{figure*}\label{fig: Q-n conditions}
    \centering
    
    \captionsetup[subfigure]{position=top, labelfont=bf,textfont=normalfont,singlelinecheck=off,justification=raggedright} 
    
    \subfloat[\label{fig: density Q-n}]{\includegraphics[page=1, width=0.49\linewidth, trim=4cm 0cm 4cm 0.65in, clip]{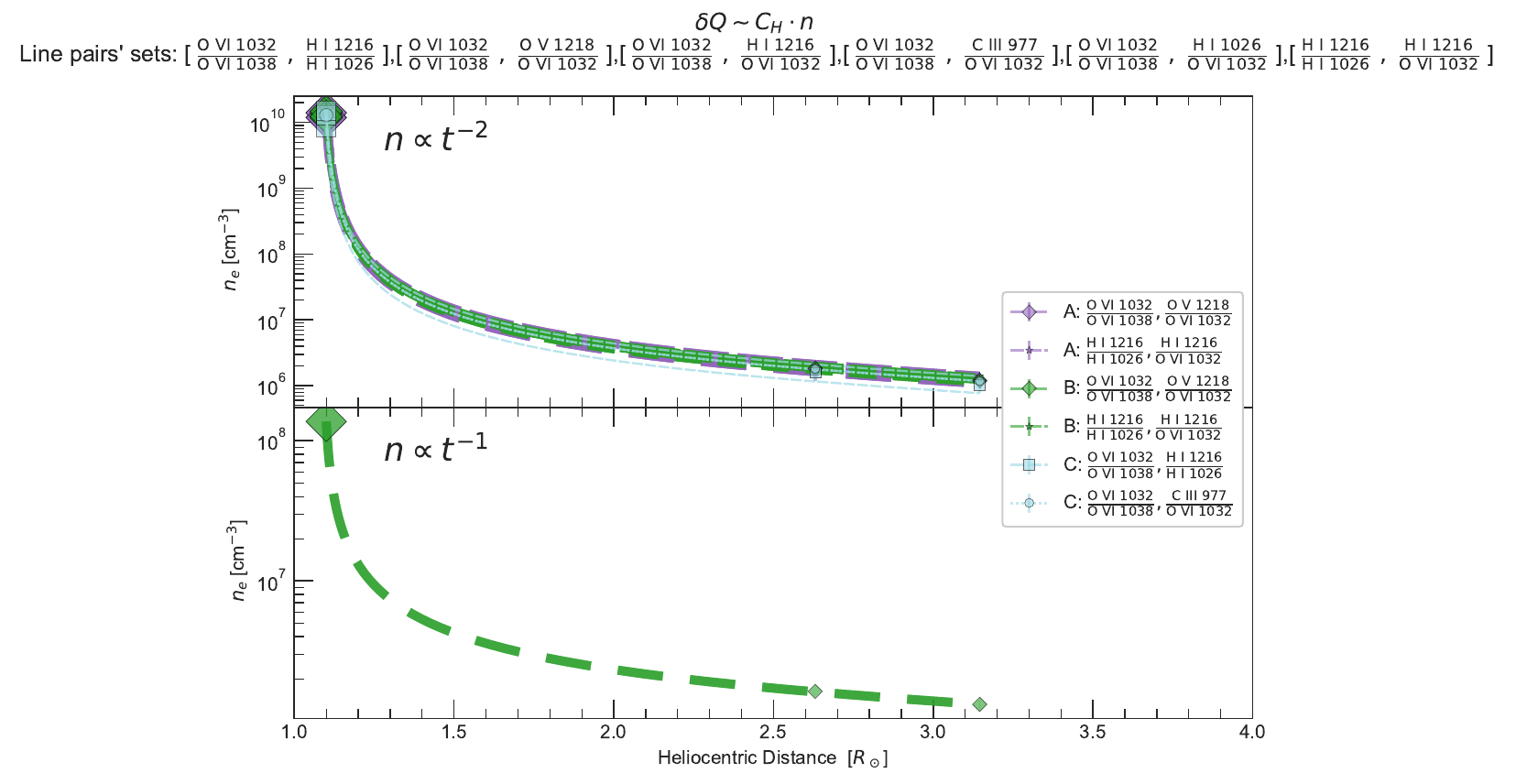}} \hfill
    \subfloat[\label{fig: temp Q-n}]{\includegraphics[page=1, width=0.49\linewidth, trim=4cm 0cm 4cm 0.65in, clip]{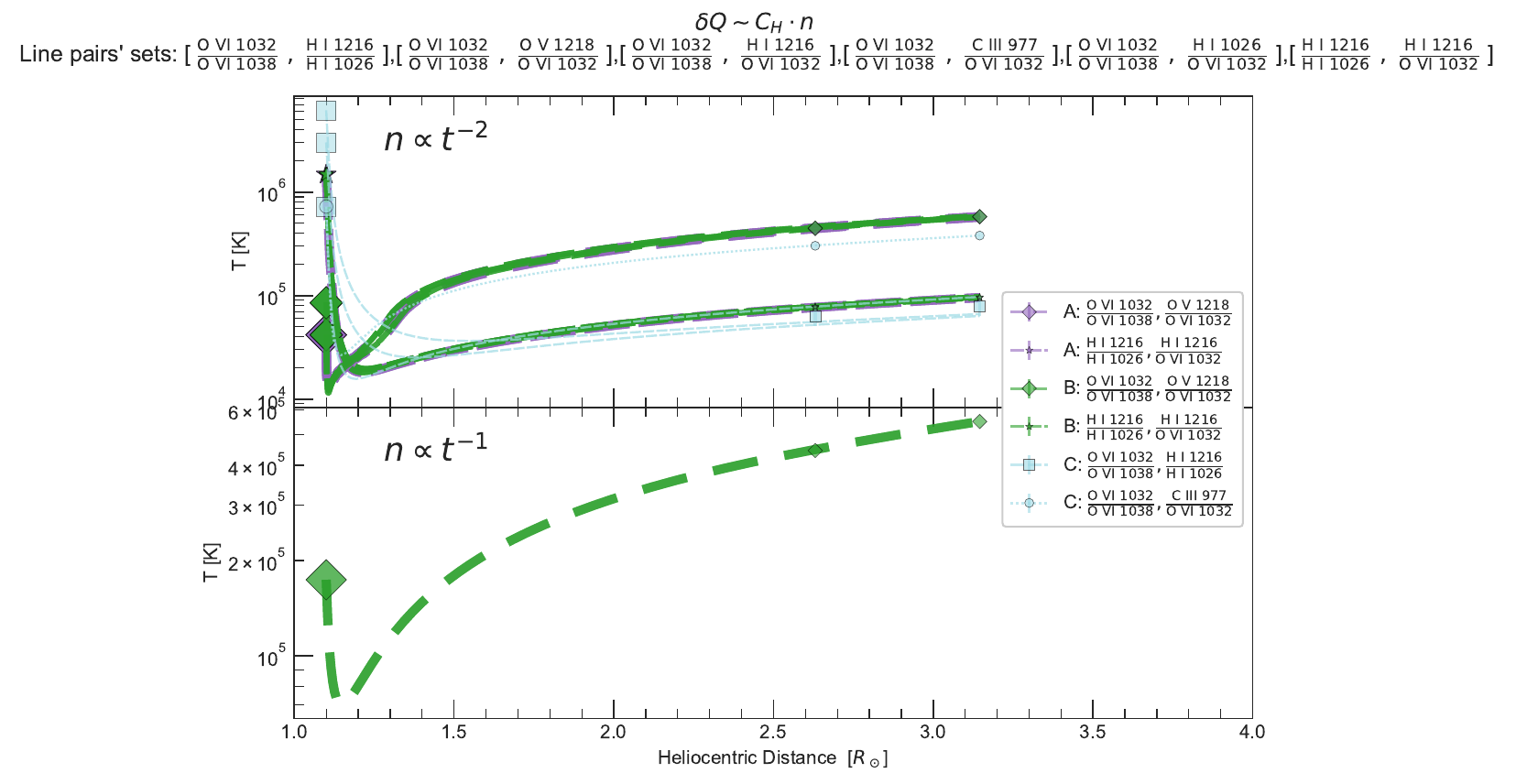}}
    
    \caption{ Tight constraints on evolution of physical conditions after using various double-ratio analyses (for the $Q_n$ heating parameterization).  }
\end{figure*}

\subsection{Non-Equilibrium Ionization Code}

In our procedure, the evolving ionization states directly affect each model's emissivities, intensity ratios, and radiative cooling.  The initial condition for each model requires that the plasma environment changes temperature on a timescale that is long enough to allow the ions' rate of ionization to balance the rate of recombination.  Within this thermodynamic timescale, the population of ionization states is independent of time and can be determined as a function of temperature.  This assumption of ionization equilibrium does not hold when astrophysical phenomena compel a plasma's thermodynamic state to change more rapidly than the ionization rate or the recombination rate.  

The ionization and recombination rates are affected by the environment's density and, even more so, the temperature.  As an example, the low-density regions distant from the dense solar surface can suppress a plasma's ability to ionize or recombine despite experiencing hot coronal temperatures.  Also, a high-speed solar wind affecting those regions can transfer the plasma quickly to other regions within timescales that are too short for the ionization and recombination processes to balance out.  Such scenarios might occur on timescales that observations do not temporally resolve; therefore, meticulous care should be taken by accounting for a net change in the population of ionization states that is predominantly due to recombination processes or predominantly due to ionization processes \citep[\egcite][]{ Raymond.1990, Rakowski.2007, Ko.2010, Bradshaw.2011, Gruesbeck.2011, Gruesbeck.2012, Landi.2012, Rivera.2019.August}.  If unknown mechanisms heat the plasma quickly while the ionization rate is much slower, an ionization equilibrium assumption for the observed ionization states would understimate the temperature.  Conversely, if the plasma is quickly cooled and observations of its ionization states are taken before slow recombination processes occur, the ionization equilibrium assumption would overestimate the temperature.  In such cases, the non-equilibrium ionization states can be determined through the formula 
\begin{equation}\label{eq: nei}
    \begin{split}
        \frac{ dn_{z} }{dt}\ = \ &  n_e n_{z-1} q_i(Z,z-1,T) \\  & - n_e n_z\big[ q_i(Z,z,T)  + \alpha_r(Z,z,T)\big] \\ & + n_e n_{z+1} \alpha_r(Z,z+1,T),
    \end{split}
\end{equation}
where $q_i$ is the ionization rate coefficient and $\alpha_r$ is the recombination rate coefficient.  This formula is incorporated into the ionization code developed by \cite{Shen.2015}.  Originally written in \textsc{fortran}\footnote{ \url{https://github.com/ionizationcalc/time_dependent_fortran} }, we use its \textsc{python}\footnote{ \url{https://github.com/PlasmaPy/PlasmaPy-NEI} } counterpart.

The ionization code solves the time-dependent equations for a parcel of gas traveling in a Lagrangian framework, in which the temporal evolution history of plasma density and temperature can be obtained. The code pre-calculated all $q_i$ and $\alpha_r$ values at a grid of temperatures and saved them into structured tables based on the atomic database Chianti v9 \citep{Dere.2019}. 
The calculations are then analytically simplified with the Eigenvalue method described by \cite{Masai.1984} and \cite{Hughes.1985} for a given temperature. To maintain temporal efficiency in the enormous computations, we apply an adaptive time step strategy (see \cite{Shen.2015} for details), and load only the tables during the simulation. We obtain these calculations for all ionization states of ten elements: H, He, C, N, O, Ne, Mg, Si, S, and Fe.  
We use the resulting ionization fractions to compute $n_{Z,z}$ for the emissivity described by Equation~\ref{eq: intensity} as well as the rate of radiative cooling described by Equation~\ref{eq: rad cooling}.

\section{MHD Model from MAS Simulation}\label{sect: mas model}

With our 1D numerical models, we can determine the physical properties of the plasma observed at two heights in the corona.  The historical evolution of the plasma's physical properties can be approximately evaluated as well; but, unfortunately, such a modelled evolution does not have spectra below 2.6~\rsun\ to act as a constraint on the evolving parameters.  This would have provided more insight on how the CME heating begins near the solar surface and continues to evolve throughout the corona.  

The historical evolution of a CME's physical properties can show how one of our heating parameterizations might be better than another heating parameterization at realistically mimicking the effects of the true CME heating mechanisms that are still unknown.  Therefore, we compare parameterizations by using a realistic 3D MHD model that provides the historical evolution of a simulated CME with similar properties to the one that we observe.  This simulation is a product of the Magnetohydrodynamic Algorithm outside a Sphere (MAS) code.

MAS models the global solar atmosphere from the top of the chromosphere to Earth and beyond, and it has been used extensively to study coronal structure \citep{1999PhPl....6.2217M,1999JGR...104.9809L,2009ApJ...690..902L,2013Sci...340.1196D,2018NatAs...2..913M}, coronal dynamics \citep{2005ApJ...625..463L,2006ApJ...642L..69L,2011ApJ...731..110L} and CMEs \citep{2003PhPl...10.1971L,2013ApJ...777...76L,2018ApJ...856...75T, Reeves.2010, Reeves.2019}. MAS solves the resistive, thermodynamic MHD equations in spherical coordinates $(r,\theta,\phi)$ on structured nonuniform meshes.  Magnetosonic waves are treated semi-implicitly, allowing us to use large time steps for the efficient computation of long-time evolution.
A simplified radial magnetic field based on observational measurements is used as the primary boundary condition.  
To drive the magnetic field evolution in MAS, the radial component of the magnetic field at the boundary can be evolved using a technique similar to that described by \citet{2013ApJ...777...76L}.

The present version of MAS employs a sophisticated thermodynamic MHD approach, where additional terms that describe energy flow in the corona and solar wind are included \citep[coronal heating, parallel thermal conduction, radiative loss, and Alfv\'e{}n wave acceleration; as fully described in  Appendix A of][]{2018ApJ...856...75T}. This treatment is essential for capturing the thermal-magnetic state of the corona and solar wind, enabling the direct comparison of a variety of forward modelled observables to real observations.

A non-equilibrium ionization module to advance the fractional charge states of minor ions according to the model of \citet{Shen.2015} was recently incorporated into MAS. The implementation is very similar to that of \citet{2019SoPh..294...13L} for a 1D solar wind code.

\section{Results and Discussion}\label{sect: results}

\newcommand{\figscale}{0.45}

The composite clumps seen at the highest coronal heights observed, 2.6 and 3.1~\rsun, are the only clumps for which we can confidently deduce two independent estimates of the velocity.  The first is a multi-height, average velocity estimate that comes from the data analysis described in \S\ref{sect: multi-height velocity}.  The second is an instantaneous velocity estimate from \ionn{O}{VI} radiative pumping analytics of the 1038~\Ang\ line as described in \S\ref{sect: emissivity}.  Both estimates provide upper and lower limits for the velocity that are further constrained by the uncertainties we assigned to the intensity ratios in Table~\ref{table: uvcs ratios}.  Additionally, as described in \S\ref{sect: plasma diagnostics}, any of our intensity ratios can serve as a diagnostic for velocity when the resonant scattering components are comparable to the collisional components.  This is common for the \rhialpha\ ratio. 

We focus on the three composite clumps emphasized in Figures \ref{fig: all spectra}~and~\ref{fig: light curve}.  The final results presented in this section suggest that all three clumps have experienced similar conditions.  After our grid of model initial conditions evolves and reaches the clumps' respective coronal heights, a range of model velocities, densities, temperatures, and ionization states is deduced for each clump.  Subsequently, we determine the historical profile of each clump from their respective models.  According to the profiles we derive, the physical parameters determined at 2.6~\rsun\ are within an order of magnitude of the parameters determined at 3.1~\rsun\ for our three clumps.  For this reason, we report the physical conditions as roughly the same for both heights in the corona.

Although the model parameters vary as a function of height, there are general characteristics of the historical profiles for density, temperature, and ionization states that are similar in all cases regardless of the heating function or expansion rate that we use.  For example, in each case, our assumption of a simple self-similar expansion (expressed in the form of Equation~\ref{eq: expand}) requires that the density profiles for ions and free electrons monotonically decrease.  An example of these observationally constrained density profiles is given in Figure~\ref{fig: density Q-B2}.

\newcommand{\gtrend}[1]{\textbf{[#1]}}
\newcommand{\trendlabel}[1]{\textit{#1}}

For temperature profiles, the minor details vary case by case; but, there are three general trends.  Examples of these three general trends can be seen in Figure~\ref{fig: temp Q-B2}.  When visualized on a logarithmic scale, the trends can be described as the following:  
\begin{itemize}
    \item[\gtrend{1}] The temperature profile begins by decreasing exponentially until it starts to plateau within 1.4~\rsun, which suggests the cooling is substantially greater than the heating immediately after the eruption but balances out later. 
    \item[\gtrend{2}] The opposite occurs.  The temperature begins with an increase and continues in a logarithmic fashion until it starts to plateau within 1.4~\rsun, which indicates a heating that is consistently greater than the cooling by a margin that is gradually decreasing over time.  
    \item[\gtrend{3}] The temperature profile exponentially decays until it reaches an inflection point within 1.4~\rsun, where it then gradually increases.  This occurs when the heating is quickly increasing but is temporarily dominated by the cooling immediately after eruption.
\end{itemize}
We refer to general trend~\gtrend{1} as \trendlabel{D-F} since its curve initially decreases but later begins to flatten out within 1.4~\rsun.  We refer to general trend~\gtrend{2} as \trendlabel{I-F} since its curve initially increases but later flattens out.  Lastly, we refer to general trend~\gtrend{3} as \trendlabel{D-I} since its curve initially decreases but later increases.

As for the evolution of ionization states, the ionization fraction profiles are not tightly constrained when using only the \rOVI\ ratio.  These profiles are heterogeneous and their corresponding broad range of temperature profiles are just as heterogeneous.  In this context, the heterogeneity is clear when temperature profiles are not limited to a specific general trend: the \trendlabel{D-F}, \trendlabel{I-F}, or \trendlabel{D-I}~trend. 
When we use the \rhialpha\ ratio or the \rLyman\ ratio, there is a strong anti-correlation between the \ionn{H}{I} ionization fraction profiles and their corresponding temperature profiles.  This is one of the reasons why the \trendlabel{D-F} trend is so prevalent for all three clumps regardless of heating function and expansion rate.  An example of our ionization fraction profiles is given by Figure~\ref{fig: ionization Q-ht}.  The cooling must be significantly greater than the heating until the temperature is low enough to yield a significant amount of \ionn{H}{I} at the clumps' respective coronal heights (cf.~$T_m$ in Table~\ref{table: spectral lines}).  This is why relatively high temperatures, around $10^6$ or $10^7$~K, are the inferred initial temperatures for many of the \ionn{H}{I} models, which often evolve to a temperature of about $10^5$~K at the final two observed coronal heights.  For the \rhialpha\ ratio, the strong temperature constraints on the ionization state of \ionn{H}{I} always narrow the range of model temperatures permitted by \ionn{O}{VI}. 

The multi-ion ratios provide diagnostics that are sensitive to ionization states.  Our results consistently indicate that the single-ion \rOVI\ ratio and single-ion \rLyman\ ratio yield a broader range of physical conditions than the constraints of the multi-ion \rhialpha\ ratio.  When we use the mulit-ion \rCIII\ ratio and assume \ionn{C}{III} ions share the same temperature, density, and velocity as \ionn{O}{VI} ions, there is very little agreement with observations.  Only the $Q_n$ heating function shows any agreement at the two observed coronal heights but only for clump~\clump{C}.  This lack of agreement suggests that our assumption that \ionn{C}{III} experiences the same conditions as \ionn{O}{VI} is not plausible for our three observed clumps.  
Using the \rOVbr\ ratio and assuming \ionn{O}{V} shares the same conditions as \ionn{O}{VI} yields models that show more agreement with observations than did the \rCIII\ ratio.  However, this agreement is seen only when assuming the $Q_n$ heating.

Regardless of the intensity ratio used, our analysis is done using three self-similar expansion indices ($\alpha_t$) distinctly.  None of our calculations using a cubic ($\alpha_t$=3) self-similar expansion rate resulted in models that agreed with the observational constraints of clumps \clump{A}, \clump{B}, or \clump{C}.  The model densities drop off excessively between the beginning of its evolution near the solar surface and the end of its evolution near 2.6 and 3.1~\rsun.  At these observed coronal heights, our model electron densities (which contribute to the radiation's collisional component) and our model ion densities (which contribute to both components of radiation) are too low to explain the clumps' observed intensities and POS sizes.  For the few models that do yield plausible densities, there is either far too much heating or far too much cooling at the beginning of the models' corresponding temperature profiles.  As a result, this excessive change in thermal energy along with our aforementioned LOS length constraint (cf.~Equation~\ref{eq: length}) have ruled out all models that utilize a cubic self-similar expansion rate for our three clumps at 2.6 and 3.1~\rsun.  For this reason, we discuss results that come from only two of our self-similar expansion rates.  

In the case of $Q_{n^2}$, only the linear expansion rate models have results that agree with the observations.  The other four heating parameterizations yield results for both the quadratic and linear expansion rates.  For a given expansion rate, all heating parameterizations suggest similar physical conditions for the observed clumps and similar energy budgets.  Therefore, we detail the results in this section only for the $Q_n$ parameterization and we elaborate on the results of the other heating parameterizations in \S\ref{sect: Q-n2}, \S\ref{sect: Q-ht}, and \S\ref{sect: Q-B}.

\newbool{CIIIincluded} \setbool{CIIIincluded}{true}
 
\newbool{noSciNot}
\setbool{noSciNot}{true}
\newbool{SciNot}
\setbool{SciNot}{false}

\subsection{Density proportional heating}\label{sect: Q-n}

Using the $Q_n$ heating, %
\ifbool{CIIIincluded}{%
there are five distinct plasma clouds modelled that agree with the observations:  \ionn{H}{I}, \ionn{O}{VI}, \ionn{H}{I} mixed with \ionn{O}{VI}, \ionn{O}{V} mixed with \ionn{O}{VI}, and (for only the quadratic expansion rate models) \ionn{C}{III} mixed with \ionn{O}{VI}.  }{%
There are four distinct plasma clouds modelled that seem to agree with the data:  \ionn{H}{I}, \ionn{O}{VI}, \ionn{H}{I} mixed with \ionn{O}{VI}, and \ionn{O}{V} mixed with \ionn{O}{VI}.  %
}%
The \ionn{H}{I} dominant material is modelled through the {\rLyman} ratio.  The \ionn{H}{I} and \ionn{O}{VI} mixture is modelled through the {\rhialpha} ratio or the {\rhibeta} ratio.  The \ionn{O}{VI} dominant material is modelled through the \rOVI\ ratio.  The \ionn{O}{V} and \ionn{O}{VI} mixture is modelled through the \rOVbr\ ratio.  %
\ifCIIIincluded%
The \ionn{C}{III} with \ionn{O}{VI} mixture is modelled through the \rCIII\ ratio.  %
\fi
For all three clumps, we find models that agree with each of these ratios, except for the \rCIII\ ratio.  There is only a very tiny region in parameter space where our models can agree with the \ionn{C}{III} observations and that agreement is only found for clump~\clump{C}.  %

We estimate the kinetic energy and gravitational potential energy of the three composite clumps, as illustrated by Figures~\ref{fig: uvcs Q-n single-ratio}~and~\ref{fig: uvcs Q-n multi-ratio}.  Note that the vertical position of each symbol within its respective error bar is not statistically more significant than the other velocity values within range of its error bar.  Each symbol is placed in the middle of its range of values for visual clarity.  In Figure~\ref{fig: uvcs Q-n single-ratio}, the \ionn{O}{VI} dominant material (plotted with diamond symbols) has the slowest velocity estimates at the height of 3.1~\rsun.  The \ionn{H}{I} with \ionn{O}{VI} mixture (via the \rhialpha\ ratio models marked by hexagons) tends to have the fastest velocities at that height.  Figures \ref{fig: uvcs Q-n multi-ratio}~and~\ref{fig: energy Q-n multi-ratio} show how for the linear expansion rate there is only one double-ratio set (\rOVI\ with \rOVbr) that has models agreeing with observations.  Overall, the double-ratio models for both expansion rates are better constrained and suggest slower velocities than the single-ratio models.  

Figures~\ref{fig: energy Q-n single-ratio}~and~\ref{fig: energy Q-n multi-ratio} summarize the cumulative heating energy amongst all models for the three clumps.  The vertical upper and lower limits correspond to a constrained range of temperature profiles.  The horizontal upper and lower limits correspond to a constrained range of kinetic energies.  Each symbol is placed in the middle (vertically and horizontally) of its range of values merely for visual clarity, and each symbol appears twice to represent each clump's observation at two coronal heights.  The cumulative heating at the higher height is, by default, always slightly greater than at the lower height since we assume the CME's heating is continuous between observations.  %
These results derive from heating rate coefficients in range of log($C_H$/erg~s$^{-1}$) $\in$ [-15.0, -12.6] for both the quadratic expansion rate models and the linear expansion rate models.
The quadratic expansion rate models suggest cumulative heating energies in range of \ifbool{noSciNot}{%
$10^{13.31 \textrm{---} 14.93}$~erg~g$^{-1}$.%
}{%
\scinot{2}{13} --- \scinot{9}{14}~erg~g$^{-1}$.}  The linear expansion rate suggests \ifbool{noSciNot}{$ 10^{12.96 \textrm{---} 14.54} $}{\scinot{9}{12} --- \scinot{3}{14}}~erg~g$^{-1}$.  Thus, our energy budget for this CME, assuming the $Q_n$ heating, suggests that the cumulative heating energy is similar to the $\sim$$10^{14}$~erg~g$^{-1}$ kinetic energy.

We now present a few examples of how the CME heating rates and energy budget may be deduced from observations of a single intensity ratio.  Amongst all of the ratios that we use, we find that the three most informative results came from using the \rLyman\ ratio, the \rOVI\ ratio, and the \rhialpha\ ratio.  

\subsubsection{$\rLyman$ ratio analysis}
All three clumps in the case of both expansion rates have velocities from 200 to 270~\unitv\ at 2.6~\rsun.  At 3.1~\rsun, the velocities are in the range 200--285~\unitv\ for the quadratic expansion rate and 200--300~\unitv\ for the linear expansion rate.  The temperature profiles exhibit the general trends~\trendlabel{D-F},~\trendlabel{I-F}, and~\trendlabel{D-I}.  Along both 2.6 and 3.1~\rsun, the temperatures range from \scinot{1}{4} to \scinot{1}{5}~K for the quadratic expansion rate case.  In the linear expansion rate case, the temperature range is from \scinot{2}{4} to \scinot{4}{6}~K.  The million Kelvin temperatures are reached here via the \trendlabel{I-F} trend, which only appears for the linear expansion rate.  Such hot temperatures are responsible for broadening the resonant scattering line profiles sufficiently to allow a broad range of models that pertain to relatively slow velocities near 200~\unitv\ and relatively fast velocities near 300~\unitv.  (The cooler temperatures favor a narrower range of velocities that are near 200~\unitv\ by narrowing the scattering line profiles.)  The hottest model temperatures also coincide with the lowest \ionn{H}{I} ionization fractions while the coldest temperatures yield the highest ionization fractions.  The temperature profiles firmly anti-correlate with the ionization fraction profiles.  This is partially due to our lower limits in temperature coinciding with the \ionn{H}{I} ion's peak formation temperature (under ionization equilibrium), $T_m \sim 3\times 10^4$~K.  The density range is from \scinot{1}{5} to \scinot{6}{6}~\unitn\ for the quadratic expansion rate and \scinot{9}{4} to \scinot{4}{6}~\unitn\ for linear expansion rate. Lastly, the range of plausible initial conditions are as follows: $n_{e,0}$ = \ifbool{noSciNot}{$ 10^{9.21 \textrm{---} 10.64} $}{\scinot{2}{9} --- \scinot{4}{10}}~\unitn, $T_{e,0}$ = \ifbool{noSciNot}{$ 10^{5.24 \textrm{---} 7.10} $}{\scinot{2}{5} --- \scinot{1}{7}}~K, and log($C_H$ / erg~s$^{-1}$) $\in$ [-15.0, -14.0] for the quadratic expansion rate; as well as, $n_{e,0}$ = \ifbool{noSciNot}{$ 10^{6.97 \textrm{---} 8.50} $}{\scinot{9}{6} --- \scinot{3}{8}}~\unitn, $T_{e,0}$ = \ifbool{noSciNot}{$ 10^{4.00 \textrm{---} 5.86} $}{\scinot{1}{4} --- \scinot{7}{5}}~K, and log($C_H$ / erg~s$^{-1}$) $\in$ [-15.0, -12.6] for the linear expansion rate.

In the case of our linear expansion rate, the models that follow the \trendlabel{I-F} trend and produce the million Kelvin temperatures are derived from the lowest initial temperature ($T_{e,0} = 10^4$~K) and greatest heating rate ($C_H = 10^{-12.6}$~erg~s$^{-1}$) in our observationally constrained models.  This anti-correlation between the minimum initial temperature and maximum heating rate only agrees with our \rLyman\ ratio constraints when the \trendlabel{I-F} trend is followed.  Under both expansion rates, the models that follow the general trends~\trendlabel{D-F}~and~\trendlabel{D-I} do not correspond to this minimum initial temperature nor this maximum heating rate.  

\subsubsection{$\rOVI$ ratio analysis}
The velocities are similar at both heights for each expansion rate: 200--265~\unitv\ for the quadratic expansion rate and 200--300~\unitv\ for the linear expansion rate.  Most of the models suggest velocities $\leq 225$~\unitv\ due to the radiative pumping.   Compared to the collisional component, the radiative pumping effect must have increased at 3.1~\rsun\ because the observed \rOVI\ intensity ratios are consistently lower at 3.1~\rsun\ than at 2.6~\rsun (cf.~Table~\ref{table: uvcs ratios}).  The velocity of the \ionn{O}{VI} material has the most influence on the strength of the radiative pumping.  Therefore, velocities that are closer to the speed of peak radiative pumping at $\sim$180~\unitv\ (due to the \ionn{C}{II} 1037~\Ang\ solar disk emission) can produce lower intensity ratios.  This velocity diagnostic becomes plagued by degeneracies as the intensity ratios get closer to 2.0.  Consequently, although it is reasonable to expect faster velocities at 2.6~\rsun\ due to its higher intensity ratios (cf. Table~\ref{table: uvcs ratios}), our resultant models do not clearly indicate that.  The intensity ratios can be affected by the lower ambient temperature and density at greater heights in the corona.

The faster velocities are only plausible in special cases where there is a balance between million-Kelvin temperatures and low densities that are less than $10^5$~\unitn.  The hot temperatures broaden the line profiles and allow the radiative pumping to be in effect for a broader range of velocities, including velocities greater than $\sim$250~\unitv.  The velocities between 250~and~300~\unitv\ imply that the 1038~\Ang\ line profile shifts away from the peak of the \ionn{C}{II} 1037~\Ang\ line profile and thus weakens the radiative pumping; but, without the thermally broadened line profile there would be no radiative pumping near 250~\unitv\ at all.  The low densities balance the ratio by diminishing the square-density dependent collisional component of the 1032~\Ang\ line much more than the density dependent resonant scattering component of the 1038~\Ang\ line that is being (weakly) pumped.  Thus, the \rOVI\ intensity ratio can remain below 2.0 in spite of the relatively weak radiative pumping at velocities between 250~and~300~\unitv.  
The degeneracies that justify these relatively fast velocities are mitigated when constraints from multiple ratios are simultaneously imposed on our models.  As a result, the double-ratio models that include the \rOVI\ ratio consistently suggest relatively slow velocities (i.e., $200 \leq v /$~\unitv$ \leq 225$).

The temperature profiles exhibit the general trends~\trendlabel{D-F},~\trendlabel{I-F}, and~\trendlabel{D-I}.  Along both 2.6 and 3.1~\rsun, the temperatures range from \scinot{3}{4} to \scinot{3}{6}~K for the quadratic expansion rate and from \scinot{2}{4} to \scinot{4}{6}~K for the linear expansion rate.  The density range is from \scinot{1}{4} to \scinot{6}{6}~\unitn\ for the quadratic expansion rate and from \scinot{3}{4} to \scinot{4}{7}~\unitn\ for the linear expansion rate.  These wide ranges of resultant physical parameters increase the chance that another single-ratio model will overlap.  Such overlap exhibited from double-ratio models will briefly be discussed later in this section.  Lastly, the range of plausible initial conditions are as follows: $n_{e,0}$ = \ifbool{noSciNot}{$ 10^{8.22 \textrm{---} 10.64} $}{\scinot{2}{8} --- \scinot{5}{10}}~\unitn, $T_{e,0}$ = \ifbool{noSciNot}{$ 10^{4.00 \textrm{---} 7.10} $}{\scinot{1}{4} --- \scinot{1}{7}}~K, and log($C_H$ / erg~s$^{-1}$) $\in$ [-14.6, -12.6] for the quadratic expansion rate; as well as, $n_{e,0}$ = \ifbool{noSciNot}{$ 10^{6.43 \textrm{---} 9.57} $}{\scinot{3}{6} --- \scinot{4}{9}}~\unitn, $T_{e,0}$ = \ifbool{noSciNot}{$ 10^{4.00 \textrm{---} 6.48} $}{\scinot{1}{4} --- \scinot{3}{6}}~K, and log($C_H$ / erg~s$^{-1}$) $\in$ [-15.0, -12.6] for the linear expansion rate.

\newbool{accel}\setbool{accel}{false}

\subsubsection{$\rhialpha$ ratio analysis}
For both expansion rates, many of the models require that the instantaneous velocity estimates increase between the heights 2.6 and 3.1~\rsun.  Some models suggest no acceleration while others can be as high as 70~m~s$^{-2}$\ifaccel , which is similar to the analysis of \cite{Landi.2010} where their maximum CME core acceleration in the \textit{SOHO} POS was 40.5~m~s$^{-2}$\fi.  
The acceleration of our models is due to this ratio's significant drop between the two coronal heights (cf. Table~\ref{table: uvcs ratios}), which occurs for all three clumps.  Many of our models attribute the drop to a decrease in \ionn{H}{I} Lyman-$\alpha$ emission (as opposed to an increase in \ionn{O}{VI} 1032~\Ang\ emission).  This implies that the resonant scattering component could have dropped substantially, which gives leeway for a greater change in velocity.  Some models however account for the drop in the intensity ratio by permitting the velocity to remain the same while the population of \ionn{H}{I} ions decreases substantially.   

The temperature profiles only follow the \trendlabel{D-F} trend.  The corresponding \ionn{H}{I} ionization fraction profiles anti-correlate with the temperature profiles.  The corresponding \ionn{O}{VI} ionization profiles do not correlate with temperature; but, for this ratio analysis, almost all of the models suggest \ionn{O}{VI} ionization state becomes frozen-in before 1.5~\rsun, and the \ionn{O}{VI} remains at its ionization fraction of $\sim$5\% onward through 2.6 and 3.1~\rsun.  Along both 2.6 and 3.1~\rsun, the temperatures \ifbool{SciNot}{are in the range $ 10^{4.57 \textrm{---} 5.08} $}{are from \scinot{4}{4} to \scinot{1}{5}}~K for the quadratic expansion rate and \ifbool{SciNot}{$ 10^{4.53 \textrm{---} 4.77} $}{from \scinot{3}{4} to \scinot{6}{4}}~K for the linear expansion rate.  This narrow range of upper and lower limit temperatures is due to the need for the presence of multiple ions (in this case \ionn{H}{I} and \ionn{O}{VI}).  The density \ifbool{SciNot}{range is $ 10^{5.89 \textrm{---} 7.15} $}{ranges from \scinot{8}{5} to \scinot{1}{7}}~\unitn\ for the quadratic expansion rate and \ifbool{SciNot}{$ 10^{6.48 \textrm{---} 6.93} $}{from \scinot{3}{6} to \scinot{8}{6}}~\unitn\ for linear expansion rate.  The densities are just as well-constrained as the temperatures due to the strong trade-off between density and temperature, which are both responsible for producing the emissivities that are necessary to match with the observed intensity ratios.  Lastly, the range of plausible initial conditions are as follows: $n_{e,0}$ = \ifbool{noSciNot}{$ 10^{9.93 \textrm{---} 11.00 } $}{\scinot{9}{9} --- \scinot{1}{11}}~\unitn, $T_{e,0}$ = \ifbool{noSciNot}{$ 10^{5.86 \textrm{---} 7.10} $}{\scinot{7}{5} --- \scinot{1}{7}}~K, and log($C_H$ / erg~s$^{-1}$) $\in$ [-14.4, -13.8] for the quadratic expansion rate; as well as,  $n_{e,0}$ = \ifbool{noSciNot}{$ 10^{8.50 \textrm{---} 8.85} $}{\scinot{3}{8} --- \scinot{7}{8}}~\unitn, $T_{e,0}$ = \ifbool{noSciNot}{$ 10^{5.55 \textrm{---} 5.86} $}{\scinot{4}{5} --- \scinot{7}{5}}~K, and log($C_H$ / erg~s$^{-1}$) $\in$ [-15.0, -14.6] for the linear expansion rate.  


\subsubsection{Common traits in ratio analyses}\label{sect: common traits}
As demonstrated, a single intensity ratio can provide a unique analysis to confirm the physical conditions of the observed CME, and this in turn constrains the initial conditions after eruption.  In addition to the three ratios discussed in detail, we also find models that agree with the observational constraints of the following ratios: \rhibeta, \rOVbr, and \rCIII.  Within the observational constraints, we find that some models work well to simultaneously explain multiple, observed intensity ratios:  \rOVI\ with \rLyman, \rOVI\ with \rhialpha, \rOVI\ with \rhibeta, \rLyman\ with \rhialpha, \rOVI\ with \rOVbr, and \rOVI\ with \rCIII.  Across all of the ratio or ratio-pair analyses we performed, the substantial agreement with observations of various spectral lines is partly due to the simplicity of the $Q_n$ heating.  This heating parameterization allows our models to sample parameter space very well and thus account for many distinct characteristics that might explain our CME's evolution. 

Due to the already tight limits on the physical conditions deriving from each multi-ion ratio, we do not find any models that agree simultaneously with two distinct multi-ion ratios (e.g., \rhialpha\ with \rhibeta).   We also note that there are no models that agree with any group of three ratios simultaneously at both of the final observation heights.  This likely could have been accomplished if we used a third single-ion ratio, such as the density-sensitive ratios of \rOV\ and \rNIII.  Due to their ionization-sensitive nature, each multi-ion ratio has a tendency to single-handedly tighten the limits on model parameters so severely that another multi-ion ratio is unlikely to match.  Regardless of each unique ratio or ratio-pair analysis, there are similarities in the deduced physical properties and energetics: velocity, temperature, density, expansion, and heating.

\textbf{Velocity:}  %
With respect to velocity, commonality can be seen in how the only (two) single-ion ratios that we use are typically the ratios that yield the slowest, instantaneous velocity estimates at 3.1~\rsun.  This is seen primarily by clumps \clump{B}~and~\clump{C} regardless of the expansion rate assumed.

\textbf{Temperature:}  %
With respect to temperature evolution, the temperature profiles that follow the \trendlabel{D-F} trend typically yield the coolest temperature estimates at the 2.6 and 3.1~\rsun\ while the profiles exhibiting the \trendlabel{I-F} trend often yield the hottest temperatures.  The inverse is true for the initial temperatures: hottest initial temperatures often correspond to the \trendlabel{D-F} trend and coolest initial temperatures often correspond to the \trendlabel{I-F} trend.  


We cannot definitively confirm that the observed material is predominantly hot coronal gas that has been cooled or predominantly cool prominence gas that has been heated.  In fact, our relatively broad but constrained range of plausible initial conditions suggests that we likely observed both types of CME material.

When considering plasma clouds that consist of \ionn{H}{I}, we find that a quadratic expansion rate limits the physical conditions sufficiently for there to be only one plausible explanation: a gradually cooled coronal gas tangled with the CME is predominantly the type of material we observed through the Lyman-$\alpha$ and Lyman-$\beta$ lines.  
The linear expansion rate however leads to more uncertainty in the models' physical conditions and thus more possible explanations.  

\cite{Akmal.2001} conducted a UVCS CME heating analysis where they found plasma clouds in the core that were somewhat separated by temperature.  In their observations, cool \ionn{C}{III} emission was evident in a small region interior to the hotter \ionn{O}{V} and \ionn{O}{VI} emission seen surrounding that small region.  It is likely that we are also observing temperature-stratified or ion-stratified behavior along the POS and LOS of our CME's core material.  Distinct regions along the LOS may be responsible for the distinct physical conditions; furthermore, we describe in \S\ref{sect: spatial confirmation} how there are individual clumps at two position angles observed within each composite clump, which means that one individual clump could be dominated by initially cool prominence material while the other individual clump (that is spatially-distant along the POS) could be dominated by initially hot coronal material.  The temperature and ion stratification could also indicate the presence of the CME's prominence-core transition region \citep[][]{Engvold.1988, Parenti.2012, Rivera.2019.August}.

\textbf{Ionization states:}  %
With respect to ionization states, the initial ionization fractions have strong correlations and anti-correlations with temperature due to our assumed initial condition of ionization equilibrium; however, our time-dependent non-equilibrium ionization calculations ensured that the evolution of the ionization fractions did not always (anti-)correlate strongly with the evolution of the respective plasma cloud's temperature.  Only the ionization fraction profiles of \ionn{H}{I} consistently showed a strong relationship with its respective plasma cloud's temperature profile.  

\textbf{Density:}  %
With respect to density, there are many models that exhibit degeneracies due to the trade-off that can occur between density and temperature in order to generate the same observed intensity ratio.  Frequently, the upper limit we determine for final temperatures corresponds to the lower limit we determine for final densities.  Fortunately, we significantly mitigate such degeneracies by performing the robust double-ratio analyses.  This trade-off occurs only to meet the required intensity ratios at the coronal heights of our observations; therefore, the trade-off is not always present in the plausible initial conditions we infer. 

\textbf{Expansion Rates:}  %
Our inferred initial densities are often influenced by the assumed expansion rate.  
The initial conditions can differ substantially between distinct expansion rate assumptions while still producing the observed intensity ratios at the final two observation heights.  Different initial conditions suggested by our models infer different historical evolutions for the CME's physical conditions.  Therefore, models using different expansion rates can be in agreement by yielding the same observed intensity ratios while disagreeing on the CME's historical evolution.  


The expansion rate's influence on the density also manifests through its simple monotonic decrease (from the aforementioned power law of Equation~\ref{eq: expand}) between the beginning of our models at 1.1~\rsun\ and the observations at 2.6 and 3.1~\rsun.  Our models suggest that when $\alpha_t$=2 the density drops by four orders of magnitude between the inferred initial density and the final density.  The density only drops by 2 orders of magnitude when $\alpha_t$=1.  This is because the observational constraints on the density must be met regardless of expansion rate.  Thus, in order for a model to reach a given density at the observation heights, its corresponding initial density must be greater for greater expansion rates.  

\textbf{Heating Energy:}
With respect to the constant heating rates, the constraints are influenced by the expansion rates.  In the three analyses detailed earlier, we report a lower limit of $C_H = 10^{-15}$~erg~s$^{-1}$ consistently when $\alpha_t=1$; but in this case, the heating becomes negligible compared to the cooling.  Any heating rate coefficient $C_H \leq 10^{-15}$~erg~s$^{-1}$ makes the heating negligible when $\alpha_t$=1.  However, this low heating rate coefficient is still significant when $\alpha_t$=2.  This is because the square-density dependent radiative cooling drops quickly for $\alpha_t$=2 and eventually becomes low enough to make the total cooling comparable to the low heating.  Due to this low heating's significance for the case of $\alpha_t$=2, the thermal energy is often allowed to be very low and thus be in disagreement with observational constraints.  In other words, the low heating rate coefficients that work well for $\alpha_t$=1 (e.g., $C_H < 10^{-15}$~erg~s$^{-1}$) are often too low for $\alpha_t$=2.  

The relationship between the heating rate, assumed expansion rate, and inferred initial conditions affects the cumulative heating energy.  As mentioned in \S\ref{sect: numerical model}, the cumulative heating energy is determined by accounting for the initial thermal energy and the continuous input of thermal energy via the heating function.  Our resultant models show that a slower expansion rate permits a lower heating rate coefficient and a lower initial temperature.  This is why the cumulative heating energies we report (in Figure~\ref{fig: Q-n}) are typically lower for the $\alpha_t$=1 models than the $\alpha_t$=2 models.  

\textbf{Degeneracies:}  %
All of the relationships exhibited amongst our various ratio analyses become more ambiguous as analyses are plagued by more degeneracies.  The single-ratio analysis using the \rOVI\ ratio suffers the most from degeneracies, and thus, although the aforementioned relationships are present, they have a minor effect on the (lack of) constraints.  However, the \rOVI\ ratio proved to be the most useful ratio in our double-ratio analyses as we mitigated the degeneracies and tightened the energy budget constraints.  The tight constraints of the double-ratio analyses are exemplified in Figures \ref{fig: energy Q-n multi-ratio}~and~\ref{fig: energy Q-n multi-ratio} for the energy budget and Figures \ref{fig: temp Q-n}~and~\ref{fig: temp Q-n} for the physical conditions.  

\begin{figure*}
\label{fig: mas visual}
    \centering
    \includegraphics[width=\linewidth, trim=0cm 0cm 0cm 0cm, clip]{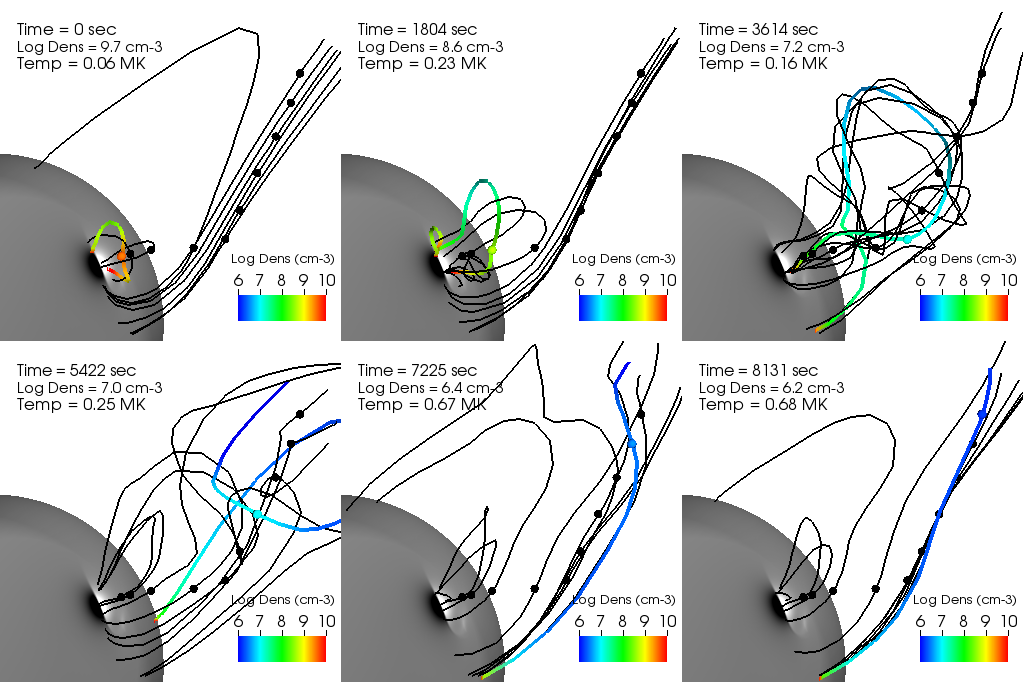}
    \caption{ 
    MAS code's simulation of a slow CME~\citep{Reeves.2019}.   The trajectory for a parcel of plasma within the dense CME core is visualized.  {\color{black}Its} local density and temperature are listed in each panel.  The only magnetic field lines that are illustrated are those for which the plasma will eventually be frozen into.  These are shown as black lines.  In each of the six panels, the one colored field line indicates where the plasma is currently located.  The color indicates the density of the environment along that field line and the large dot shows the parcel's position along that field line.  As time progresses the field lines elongate, get tangled, and extend outward into the corona.  Such morphology from the magnetic flux rope is partly responsible for the decelerations and accelerations seen in Figure~\ref{fig: mas curves trajectory}.
    }
\end{figure*}

\begin{figure}
    \label{fig: mas curves}
    \centering
    \captionsetup[subfigure]{position=top, labelfont=bf,textfont=normalfont,singlelinecheck=off,justification=raggedright} 

    \subfloat[\label{fig: mas curves trajectory}]{\includegraphics[page=1, scale=0.37, trim=1cm 1.7cm 1cm 2.2cm, clip=true]{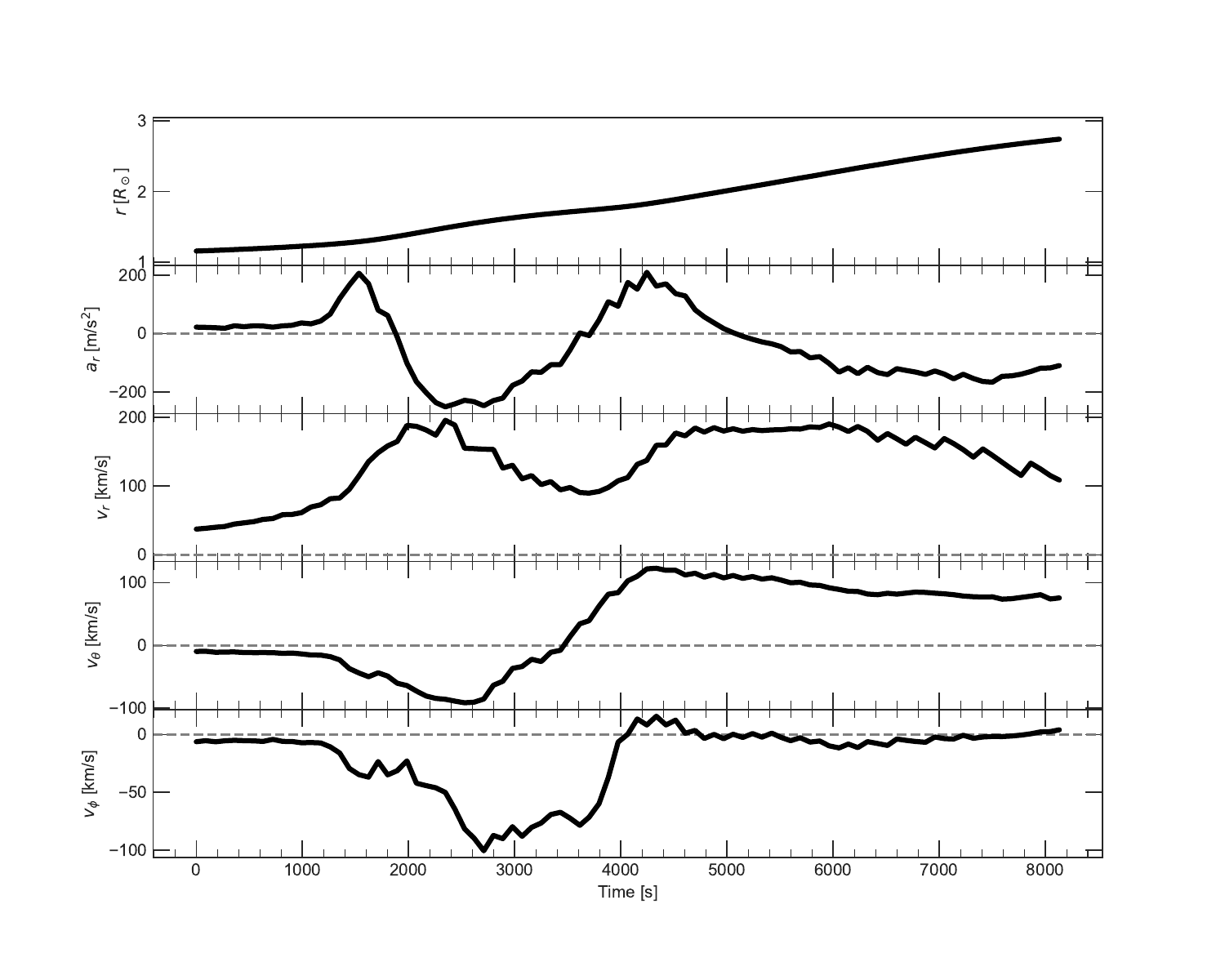}} \\  
    \subfloat[\label{fig: mas curves conditions}]{\includegraphics[page=1, scale=0.37, trim=1cm 1.7cm 1cm 1.5cm, clip=true]{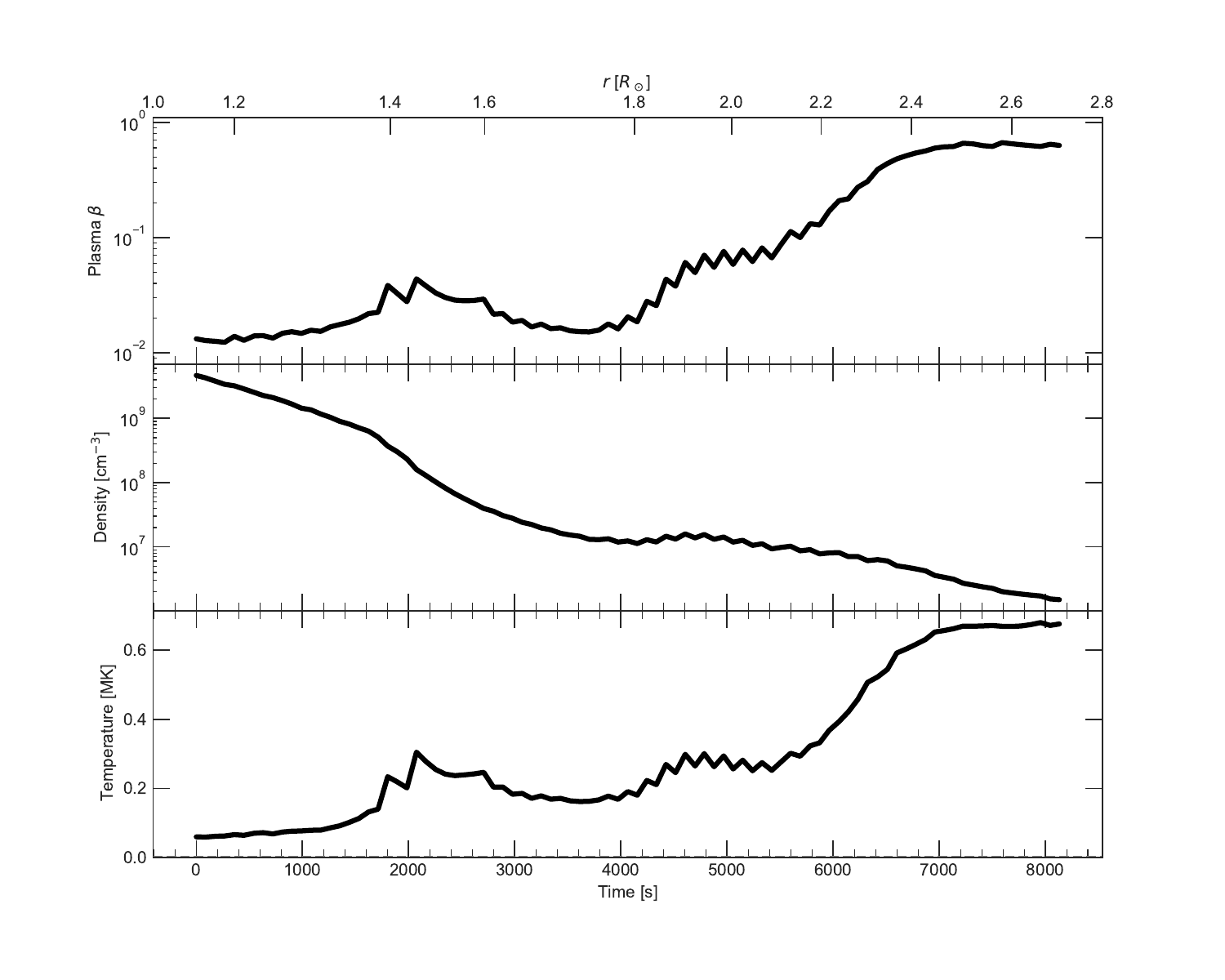}} \\  
    \subfloat[\label{fig: mas curves compare}]{\includegraphics[page=1, scale=0.37, trim=1cm 1cm 1cm 1.5cm, clip=true]{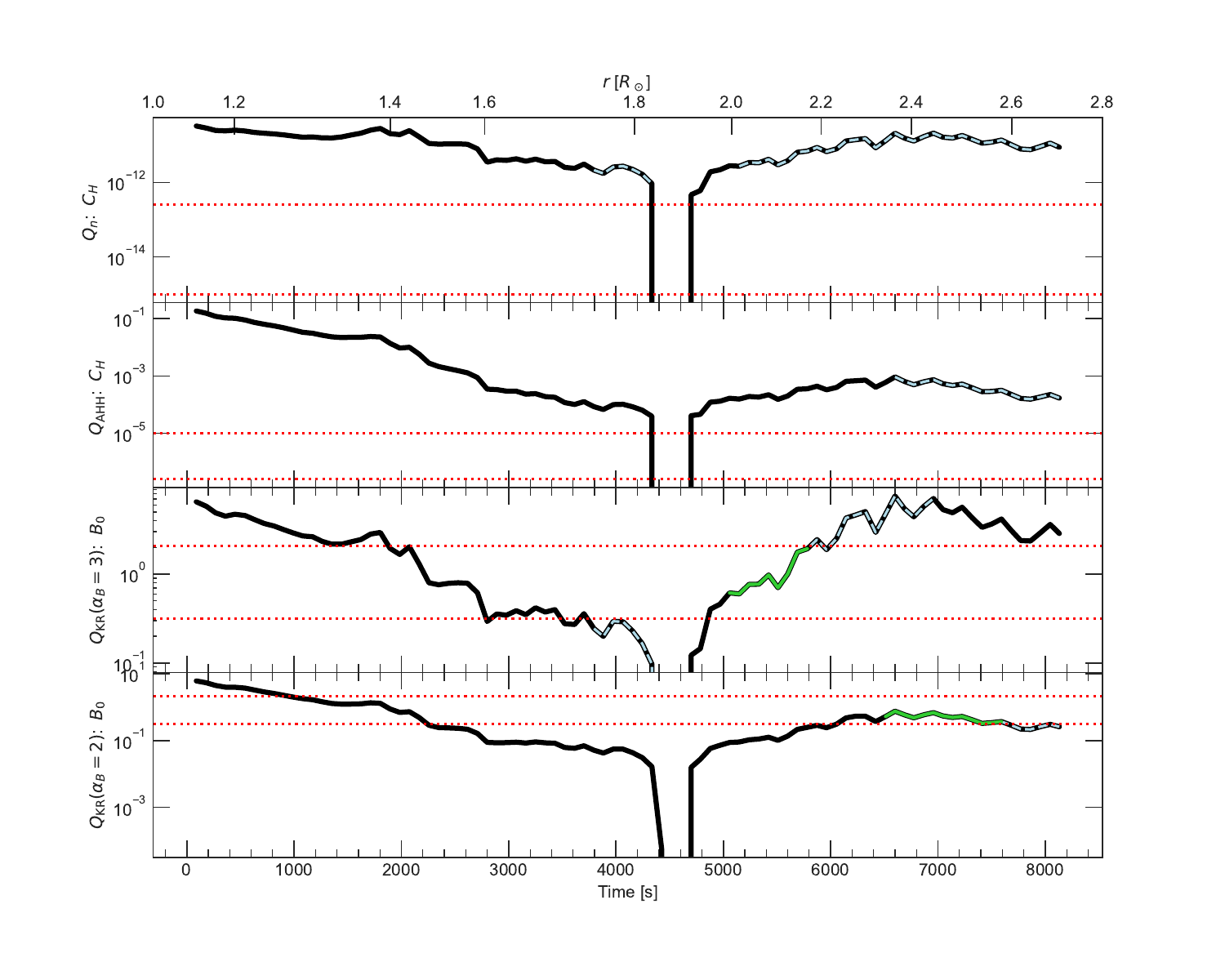}}
    
    \caption{  Simulation data is plotted for a parcel of plasma within the CME core as it travels for over 2 hours.  The \textbf{(a)} trajectory, \textbf{(b)} physical conditions, and \textbf{(c)} time-variant heating rate coefficients for the plasma are given.  See \S\ref{sect: mas insight} and Table~\ref{table: mas} for details.
    }
\end{figure}

\subsection{Insight from MAS MHD model}\label{sect: mas insight}

We use a simulated CME to compare the evolutionary effects of our heating parameterizations.  Our resultant 1D numeral models derive from the heating rate coefficients that are necessary to produce models that agree with the observations at 2.6 and 3.1~\rsun.  We can use the resultant coefficients to suggest if one heating parameterization evolves the physical properties in a way that is more realistic than another heating parameterization.  
This can be done if the realistic simulation of a CME exhibits physical conditions similar to that of the observed 1999 CME.

\cite{Reeves.2019} thoroughly describes the global behavior and energetics of the CME simulation that we use.  Within this CME, we extract an exemplary parcel of plasma and monitor its characteristics as we follow its evolution.  Its environment and localized properties are illustrated in Figure~\ref{fig: mas visual}. As can be seen, the plasma is sometimes located in a high density environment compared to the regions along its respective magnetic field line.  These densities imply that the plasma would seem very bright along one part of the field line but would seem faint if it were at another part of the field line and observed with UVCS at a given slit height. 

The trajectory of the plasma is reported in Figure~\ref{fig: mas curves trajectory}.  We only focus on its journey from $r$~=~1.2~\rsun\ to 2.7~\rsun\ since that reaches a height similar to the heights of the three observed composite clumps.  As can be seen from the simulated plasma's $v_r$, $v_\theta$, and $v_\phi$ components, it does not travel in a purely radial direction.  Moreover, its absolute speed can be as high as 220~\unitv, which is also similar to the three observed composite clumps.  There is a consistent deceleration beyond 6,000~seconds (and beyond 2.3~\rsun), which corresponds to many of the magnetic field lines being radially oriented.  In the last two panels of Figure~\ref{fig: mas visual}, at the two times beyond 7,000 seconds, the plasma is frozen-in a nearly radial magnetic field line, which allows the gravitational deceleration to strongly influence the plasma motion.  At these times, there is a $\sim$40~m~s$^{-2}$ gravitational deceleration that is largely responsible for the consistent $\sim$140~m~s$^{-2}$ deceleration reported in Figure~\ref{fig: mas curves trajectory}.  

The evolution of the plasma's localized physical conditions is reported in Figure~\ref{fig: mas curves conditions}.  As seen from the plasma $\beta$, the magnetic pressure consistently dominates early in the plasma's evolution but gradually decreases compared to the plasma pressure.  The density gradually decreases but has a momentary increase near 1.9~\rsun.  The evolution of the plasma temperature is affected by the local environment's radiation cooling, conduction, expansion, compression, ohmic heating, ambient coronal heating, and the advection of the plasma through this environment.  Also, the dominant magnetic pressure {\color{black}can} allow a portion of magnetic energy to be converted into thermal energy throughout the plasma's evolution.  The simulation's physical model, including its heating and cooling {\color{black}mechanisms}, are described in detail by \cite{Reeves.2019}.  Although this simulation was not constructed to duplicate the observed 1999 CME, {\color{black}sometimes the simulation's heating mechanisms add up to a total heating rate equivalent to our observationally constrained heating parameterization (cf. $Q$ in Equation~\ref{eq: energy}).  The observations do not directly provide constraints for each of the simulation's heating mechanisms (e.g., ohmic heating), but the heating rates can be similar between the physical 3D MHD model and our parameterized 1D heating models.  Also,} we find that at certain times and heights, the simulated plasma within the dense CME core exhibits a density and temperature that are similar to that of the observed clumps.

To compare heating parameterizations, we focus only on moments in the simulated plasma's journey when the density and temperature agree with the observed clumps.  We treat the simulated density and temperature as data inputs for our 1D modelling procedure (cf.~\S\ref{sect: numerical model}) and re-evaluate the coefficient $C_H$ for each heating parameterization.  Now, we make the heating rate coefficient vary with time.  We consider the length of time in which it agrees with the observationally constrained $C_H$ coefficients.  

To choose an expansion rate for our calculations, we considered the simulated plasma's density profile.  It does not follow a smooth power law but the density profile does drop by at least three orders of magnitude, which is done consistently by our constrained quadratic expansion rate models but is never done by our linear expansion rate models.  This suggests that an expansion index of $\alpha_t$~=~2 for our power law is more realistic than $\alpha_t$~=~1 when describing a plasma within the CME core. 

The evolution of each time-variant heating rate coefficient is presented in Figure~\ref{fig: mas curves compare} as a black solid line.  The red dotted lines indicate the upper and lower limits of the time-invariant coefficient that described the observed clumps.  Since we have ruled out the linear expansion rate, only the limits derived from quadratic expansion rate models are plotted as red dotted lines.  The coefficient for the $Q_{n^2}$ parameterization is not considered since only the linear expansion rate models could describe the observed clumps (cf.~\S\ref{sect: Q-n2}).  This implies that the $Q_{n^2}$ parameterization is not as realistic as the rest of our five parameterizations, which all attempt to imitate the rate of heating caused by the CME heating mechanisms.

\capstartfalse
\begin{table*}
    \begin{longtable}{c|ccc|cc}
    
        \multicolumn{1}{r}{ {\scshape \tablename\ \thetable}: } &  \multicolumn{5}{l}{Using 3D MHD CME evolution to compare }
        \label{table: mas} 
        
        \\\multicolumn{1}{l}{  } & \multicolumn{5}{l}{ realistic heating rates and parameterizations } \\\bottomrule\bottomrule
          & \multicolumn{3}{c|}{ 1D model limits for } &
        \multicolumn{2}{c}{ 3D evolution matches } \\
          & \multicolumn{3}{c|}{ observed plasma (cf. \S\ref{sect: numerical model}) } &
        \multicolumn{2}{c}{ observed limits (cf. \S\ref{sect: mas insight}) }
        \\
        Heating  &  Log $n_e$ & Log $T_e$ & Log $C_H$ &  $(\%)_{n,T}$ & $(\%)_{C_H,(n,T)}$   \\\hline
        $Q_n$ & [4.176, 7.146] & [4.167, 6.399] & [-15, -12.6] &  43 & 0 \\
        $Q_{\rm{AHH}}$ & [5.529, 6.788] & [4.370, 6.621] & [-6.6, -5.0] & 19 & 0 \\
        $Q_{\rm{KR}}(\alpha_B=3)$ & [5.852, 7.146] & [4.088, 5.808] & [-2.400, -0.760]$^*$ &  30 & 30 \\
        $Q_{\rm{KR}}(\alpha_B=2)$ & [5.495, 6.788] & [4.139, 6.100] & [-2.400, -0.760]$^*$ & 19 & 71 \\\hline
        \multicolumn{6}{ p{0.70\linewidth} }{ \textbf{Notes.} 
        The heating rate coefficient $C_H$ is re-evaluated as a time-variant parameter while the density and temperature profiles from the MHD CME simulation act as data inputs for the modelling procedure described in \S\ref{sect: numerical model}.  
        }\\\multicolumn{6}{ p{0.70\linewidth} }{ Only the 1D model limits derived from the quadratic expansion rate models are considered, which excludes the $Q_{n^2}$ parameterization (cf.~\S\ref{sect: Q-n2}).  %
        }\\\multicolumn{6}{ p{0.70\linewidth} }{ The $(\%)_{n,T}$ is the fraction of time covered by the dashed turquoise lines (cf. Figure~\ref{fig: mas curves compare}) compared to the 2.3 hours covered by the black solid lines in Figure~\ref{fig: mas curves}.  The solid green line is overplotted onto the turquoise dashed line.  The $(\%)_{C_H,(n,T)}$ is the fraction of time covered by the green line compared to the total turquoise line.  See Figure~\ref{fig: mas curves compare} and \S\ref{sect: mas insight} for details. 
        }\\\multicolumn{6}{ p{0.70\linewidth} }{ $^*$This corresponds to log($B_0$/G)~$\in$~[-0.5, 0.32], which ranges from 0.3 to 2.1~Gauss.    } 
    \end{longtable}
\end{table*}
\capstarttrue

For the four remaining parameterizations, each time-variant coefficient drops drastically near 1.9~\rsun.  Within this region, the coefficients become negative.  This is due to the increase in density (i.e., compression) at the time of 4,300~seconds which is also one of the times at which the plasma's radial acceleration $a_r$ peaks and begins to drop (cf. Figure~\ref{fig: mas curves trajectory}).  The temperature increases at this moment also.  This is a result of the cooling from adiabatic expansion being reversed substantially while the plasma experiences a high temperature via advection.  At this moment, the heating rates from the simulated plasma's advection and adiabatic compression are at their greatest; however, this is counteracted by the thermal conduction becoming a cooling term as it carries thermal energy away from the local environment near 1.9~\rsun.  For our 1D modelling, the time-variant coefficients account for this cooling by becoming negative and converting our heating term into a cooling term, which is a systematic response to how the increase in density near 1.9~\rsun\ converts our expansion cooling term into a {\color{black}compression} heating term (cf.~Equation~\ref{eq: energy expansion}).  

The turquoise dashed line in Figure~\ref{fig: mas curves compare} indicates the period of time when the plasma simultaneously has a density and temperature that agree with the observed clumps.  Since these two physical conditions match with observations, this is a period of time for which the time-variant $C_H$ coefficient is likely to match with the observationally constrained time-invariant $C_H$ coefficient.  When this matching of $C_H$ also occurs, the moment is marked by a solid green line, which obscures a portion of the turquoise line.  These moments are reported in Table~\ref{table: mas}.  The $(\%)_{n,T}$ is the portion of time when the density and temperature simultaneously match with the observed clumps (i.e., the turquoise line) compared to the total 2.3~hours of data extracted from the MAS simulation (i.e., the black line).  The $(\%)_{C_H,(n,T)}$ is the portion of time when the density, temperature, and time-variant $C_H$ all simultaneously match with the observed clumps (i.e, the green line) compared to the total time in which only the physical conditions of density and temperature match (i.e., the turquoise line). 

The $(\%)_{C_H,(n,T)}$ signifies how well the heating parameterization can produce a realistic $C_H$ coefficient within the period of time that the physical conditions of density and temperature are realistic. The realistic limits for these three parameters are defined by the observed clumps' results and are summarized in Table~\ref{table: mas}.  The simulated plasma exhibits the same density and temperature as the observed clumps for 43\% of the time when the observed clumps are analyzed with the $Q_n$ parameterization; however, the re-evaluated $C_H$ is never the same as the observed clumps within this portion of time (i.e., 0\% of this time).  In contrast, the simulated plasma exhibits the same density and temperature as the observed clumps for 19\% of the time when the $Q_{\rm{KR}}(\alpha_B=2)$ parameterization is used; and, within this time interval, the $C_H$ coefficient matches 71\% of the time.  This suggests that the rate of heating given by our $Q_{\rm{KR}}(\alpha_B=2)$ parameterization can realistically describe the heating of a plasma for a longer portion of time than our other heating parameterizations.  However, this does not imply that the parameterization is an accurate description of a plasma's heating mechanisms.

\section{Summary and Conclusions}\label{sect: summary}

We have presented an analysis detailing the physical properties and energetics for the core material of a CME.  This CME event occurred in 1999 and was observed by \textit{SOHO}'s EIT, LASCO, and UVCS instruments.  We proved that there were structures within the CME's core that crossed the (single) slit of UVCS once at 2.6~\rsun\ and once again at 3.1~\rsun.  Three different approaches were used to confirm this serendipitous result.  For the clumps of plasma observed, we revealed patterns of behavior in their positioning along the slit and in the shape of their light curves.  The third form of confirmation came from the agreement between their average velocity estimates from multi-height observations and their instantaneous velocity estimates from the \ionn{O}{VI} doublet intensity ratios. 

To better understand the CME heating problem, we used 1D numerical models to evaluate the internal thermal energy of the plasma as a function of height.  We assumed the plasma is being continuously heated and we investigated five different parameterizations to represent the unknown CME heating mechanisms.  We monitored the evolution of the model plasma's physical conditions, which included the temperature, density, and ionization states.  The evolutionary profiles for these conditions extended from 1.1 to 3.1~\rsun.  We monitored the ionization states of \ionn{H}{I}, \ionn{C}{III}, \ionn{O}{V}, and \ionn{O}{VI} by using non-equilibrium ionization calculations.  We required that these model ions produce emission that gives the same intensity ratios that UVCS observed at 2.6 and 3.1~\rsun. 

The intensity ratios allowed us to exploit the Doppler dimming effect and diagnose the instantaneous physical conditions of the observed clumps of plasma within the CME core.  The evolutionary profiles were constrained by the observed intensity ratios, which in turn constrained the initial conditions of the CME material.  We found evidence of initially cool but gradually heated prominence material as well as initially hot but gradually cooled coronal material being present within the observed clumps of plasma.  We also found that the cumulative heating energy is comparable to the kinetic energy and gravitational potential energy, which signifies how important the heating processes are during the eruption and evolution of the CME.

We monitored the evolution of a realistic MHD simulation of plasma being heated within the dense CME core in order to determine which heating parameterizations provide the most realistic heating rates.  We found that a quadratic self-similar expansion rate is more realistic than a linear self-similar expansion rate.  Models derived from the quadratic expansion rate suggest that our magnetic heating parameterization is the most realistic parameterization when its magnetic field expansion is predominantly two-dimensional instead of three-dimensional. 

Our robust analyses could have been improved if our observational constraints came from three heights in the corona instead of just two or if the two heights in the corona were more than 0.5~\rsun\ apart from each other.  In either case, a longer baseline of the plasma's historical behavior would have been observed, which would have tightened our range of plausible evolutionary profiles---including our inferred initial conditions of the CME.  For coronagraph spectrometers, the observations of the same CME structures at merely two heights is actually a fortuitous achievement.  Historically, the lack of such observations is due partly to the single-slit aperture of these instruments; and, even more so, it is due to the unpredictable nature of a CME's initial location, time of eruption, and velocity.  This problem is exacerbated for observations of diffuse and dynamic features in the CME core, which are difficult to track from one height to another.

Coronagraph spectrometers have been acquiring ultraviolet spectroscopic measurements of the extended corona ($d_H$=1.5--10\rsun) since 1979, and yet, the type of fortuitous multi-height observations that we examined in our analysis is still seldom acquired \citep[\egcite][]{Ko.2005}.  The first coronagraph spectrometer acquired measurements during its three suborbital flights (in 1979, 1980, and 1982) on the Nike boosted Black Brant V sounding rockets \citep[\egcite][]{Kohl.1980}.  Later, the Ultraviolet Coronal Spectrometer (UVCS) instrument on board \textit{SPARTAN 201} acquired measurements during four of NASA's Space Transportation System (STS) missions (in 1993, 1994, 1995, and 1998) \citep[\egcite][]{Strachan.1994, Kohl.1994}.  The \textit{SOHO}/UVCS instrument was launched in 1995 as an improved version of \textit{SPARTAN}/UVCS.  Unfortunately, all three of these space-based ultraviolet coronagraph spectrometers are no longer operational.

Now, the new era of coronagraph spectrometers will have more than one slit aperture.  In this way, the type of multi-height CME spectra analyzed in this paper can be achieved more frequently.  The unpredictable nature of CMEs may remain but the multiple slits will monitor different heights in the corona simultaneously along the same position angle.  Therefore, if a three-part CME is observed by one slit at one height then all three parts can be observed again by the next slit at the next height.  We expect to see CME observations like this from the following multi-slit coronagraph spectrometer missions: the UltraViolet Spectro-Coronagraph (UVSC) Pathfinder instrument \citep{Strachan.2017} is scheduled to launch in 2021, and the Large Optimized Coronagraphs for KeY Emission line Research (LOCKYER) instrument \citep{Ko.2016, Laming.2019} is currently being designed.

\acknowledgements

The authors gratefully acknowledge helpful conversations with Cooper Downs and Yeimy Rivera.  This research is supported by the NASA grant NNH18ZD001N given to the Smithsonian Astrophysical Observatory and supported by the Scholarly Studies program of the Smithsonian Institution.  SOHO is a mission from the joint collaboration of ESA and NASA.  The LASCO CME catalog is generated and maintained at the CDAW Data Center by NASA and The Catholic University of America in cooperation with the Naval Research Laboratory.  This work has benefitted from the use of NASA's Astrophysics Data System.

\textit{Facilities:} \textit{SOHO} (EIT, LASCO, UVCS)

\appendix 

\section{Square-density proportional heating}\label{sect: Q-n2}

Using the $Q_{n^2}$ heating, there are three distinct plasma clouds modelled that agree with the observations: \ionn{H}{I}, \ionn{O}{VI}, and \ionn{H}{I} mixed with \ionn{O}{VI}.  The same observational constraints applied to the $Q_n$ heating analysis are also applied here.  As a result, the physical conditions we derive for the CME when using this $Q_{n^2}$ heating are similar to that of our $Q_n$ heating results.  However, this is only true for our linear expansion rate.  This $Q_{n^2}$ heating compels our models to have an excessive amount of heating near the solar surface when we use the quadratic expansion rate.  As explained in \S\ref{sect: common traits}, there is a correlation between the expansion rates and our inferred initial densities.  The inferred initial densities are systematically greater for the faster expansion rate and, consequently, the square-density dependence of the $Q_{n^2}$ function drives the thermal energy to excessively high temperatures.  Conversely, lower initial densities lead to excessively low final densities that cannot explain our observed intensity ratios.

\begin{figure*}\label{fig: Q-n2}
    \centering
    
    \captionsetup[subfigure]{position=top, labelfont=bf,textfont=normalfont,singlelinecheck=off,justification=raggedright} 
    
    \subfloat[\label{fig: uvcs Q-n2 single-ratio}]{\includegraphics[page=1, width=0.49\linewidth, trim=0cm 0cm 0cm 0cm, clip=true]{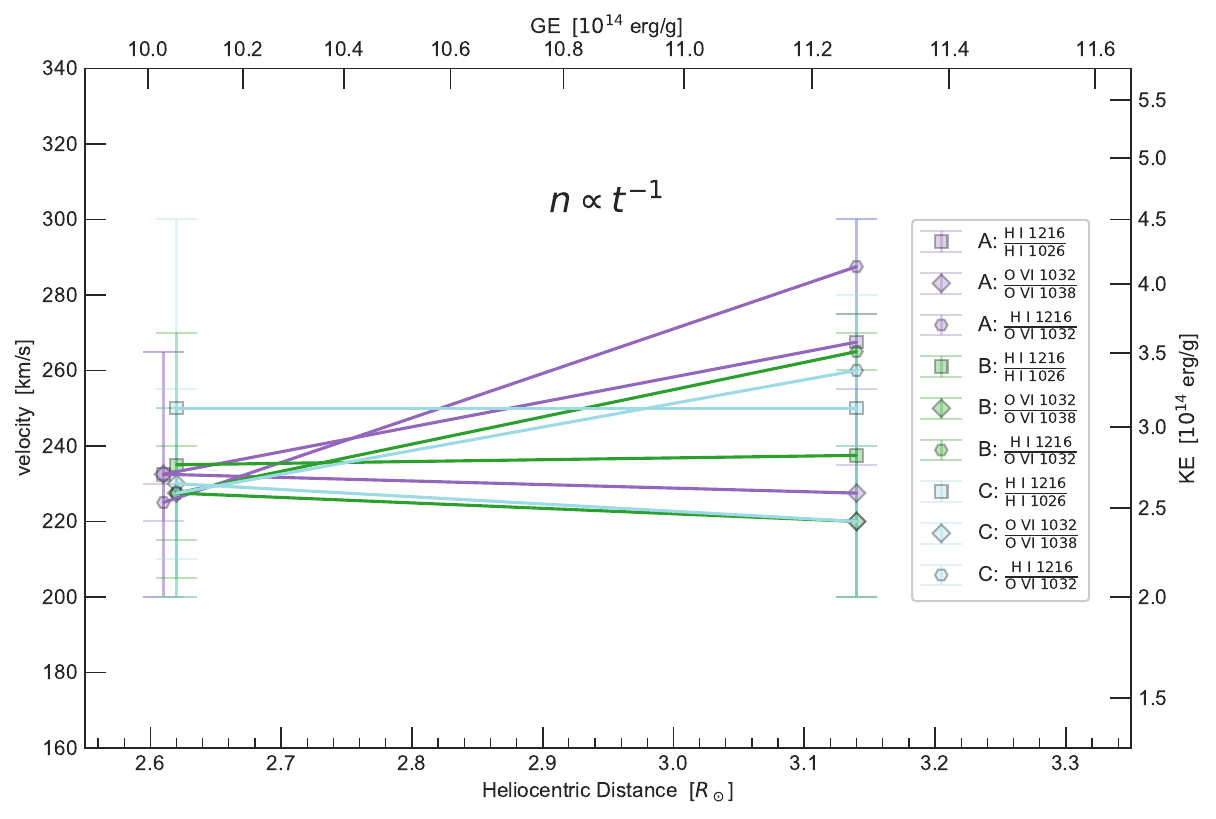}} \hfill  
    \subfloat[\label{fig: energy Q-n2 single-ratio}]{\includegraphics[page=1, width=0.49\linewidth, trim=0cm 0.1in 0cm 0.6in, clip=true]{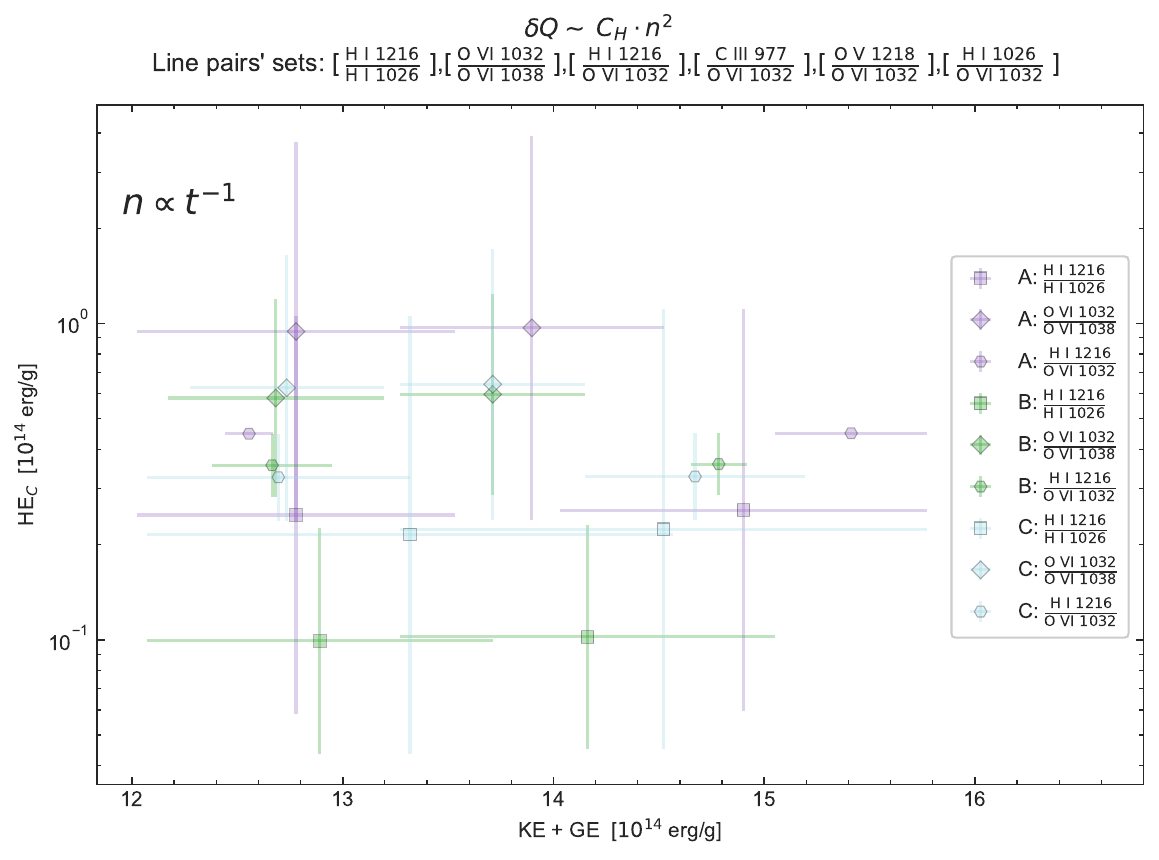}}
    
    \caption{ Observationally constrained models using the $Q_{n^2}$ heating parameterization.  }
\end{figure*}

The energy budget under this heating parameterization is summarized in Figure~\ref{fig: Q-n2}.  The kinetic and potential energies for the three clumps are given in Figure~\ref{fig: uvcs Q-n2 single-ratio}.  At the height of 3.1~\rsun, our models describe the \ionn{O}{VI} dominant material as having the slowest velocity estimates, just as in the $Q_n$ results.  The cumulative heating energies are given in Figure~\ref{fig: energy Q-n2 single-ratio}.  We find this to be in the range of $10^{12.64 \textrm{---} 14.59}$~erg~g$^{-1}$, which is similar to the $Q_n$ heating results for its linear expansion rate models.  The corresponding heating rate coefficients are in the range log($C_H$~/~erg~cm$^3$~s$^{-1}$) $\in$~[-22.0,~-19.8].

For each of the three single-ratio analyses detailed in \S\ref{sect: Q-n}, the characteristics exhibited when using the $Q_n$ heating function are similar to the characteristics exhibited when using the $Q_{n^2}$ heating function.  
For a given expansion rate, the common traits seen across all of our ratio analyses are also present with this heating function.  However, the relationships that correlate or anti-correlate with the choice of expansion rate cannot be reaffirmed due to the lack of models that agree with observations when a quadratic expansion rate is assumed.  Also, none of our resultant models match to give a double-ratio analysis with this heating function.  


\begin{figure*}\label{fig: Q-ht}
    \centering
    
    \captionsetup[subfigure]{position=top, labelfont=bf,textfont=normalfont,singlelinecheck=off,justification=raggedright} 
    
    \subfloat[\label{fig: uvcs Q-ht single-ratio}]{\includegraphics[page=1, width=0.49\linewidth, trim=0cm 0cm 0cm 0cm, clip=true]{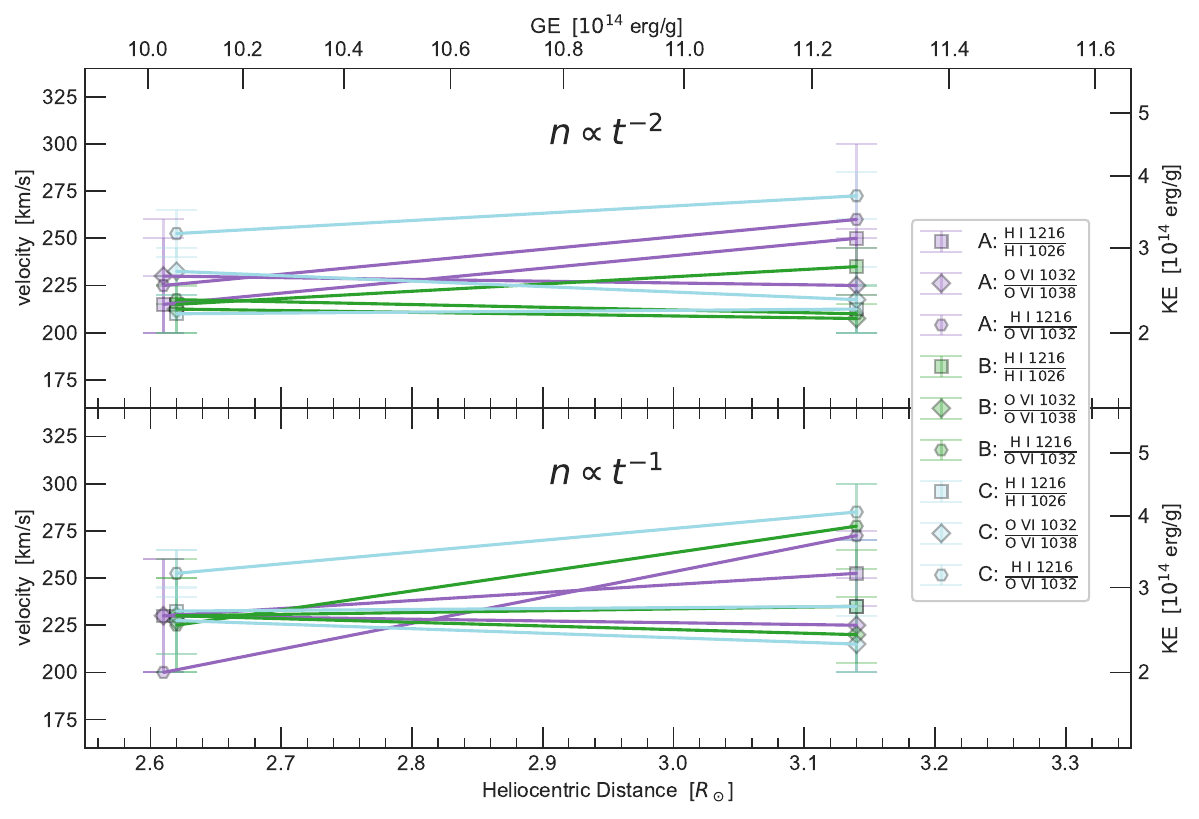}} \hfill  
    \subfloat[\label{fig: energy Q-ht single-ratio}]{\includegraphics[page=1, width=0.49\linewidth, trim=0cm 0.1in 0cm 0.6in, clip=true]{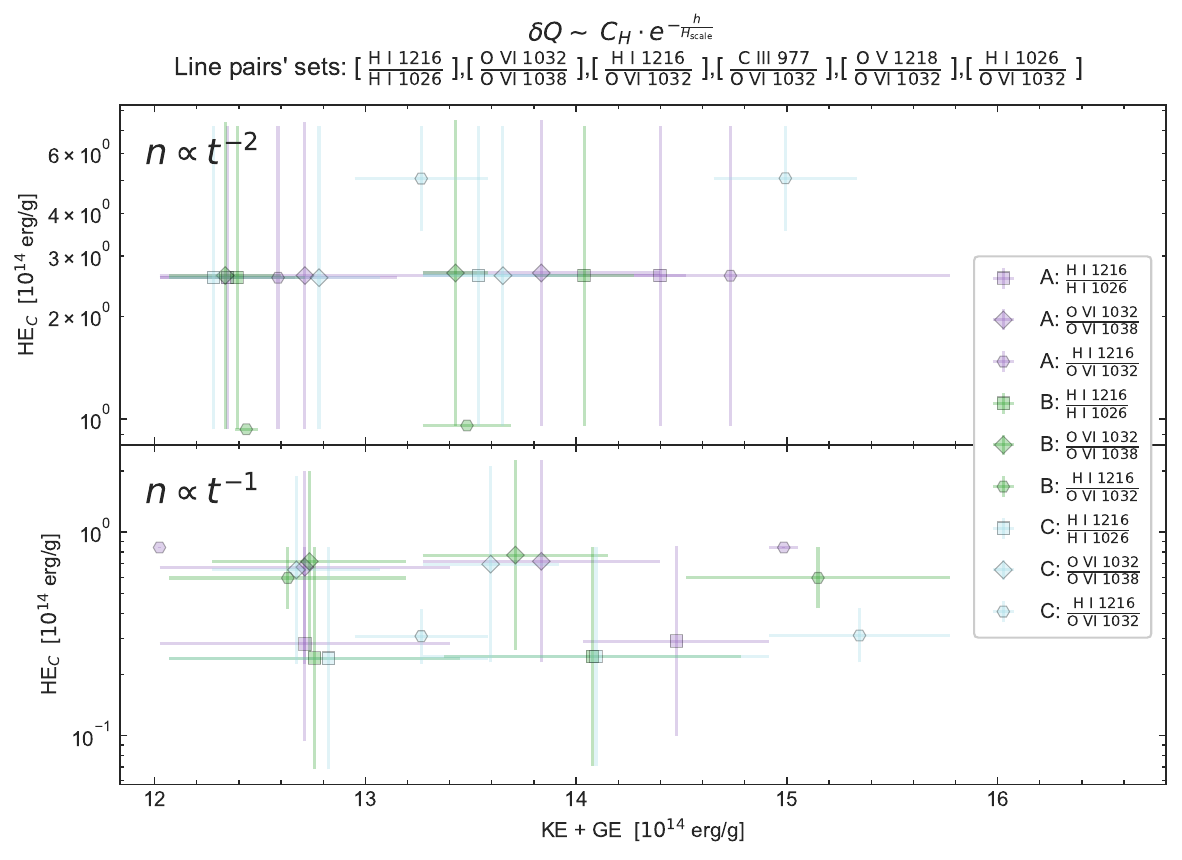}}
    
    \subfloat[\label{fig: uvcs Q-ht multi-ratio}]{\includegraphics[page=1, width=0.49\linewidth, trim=0cm 0cm 0cm 0cm, clip=true]{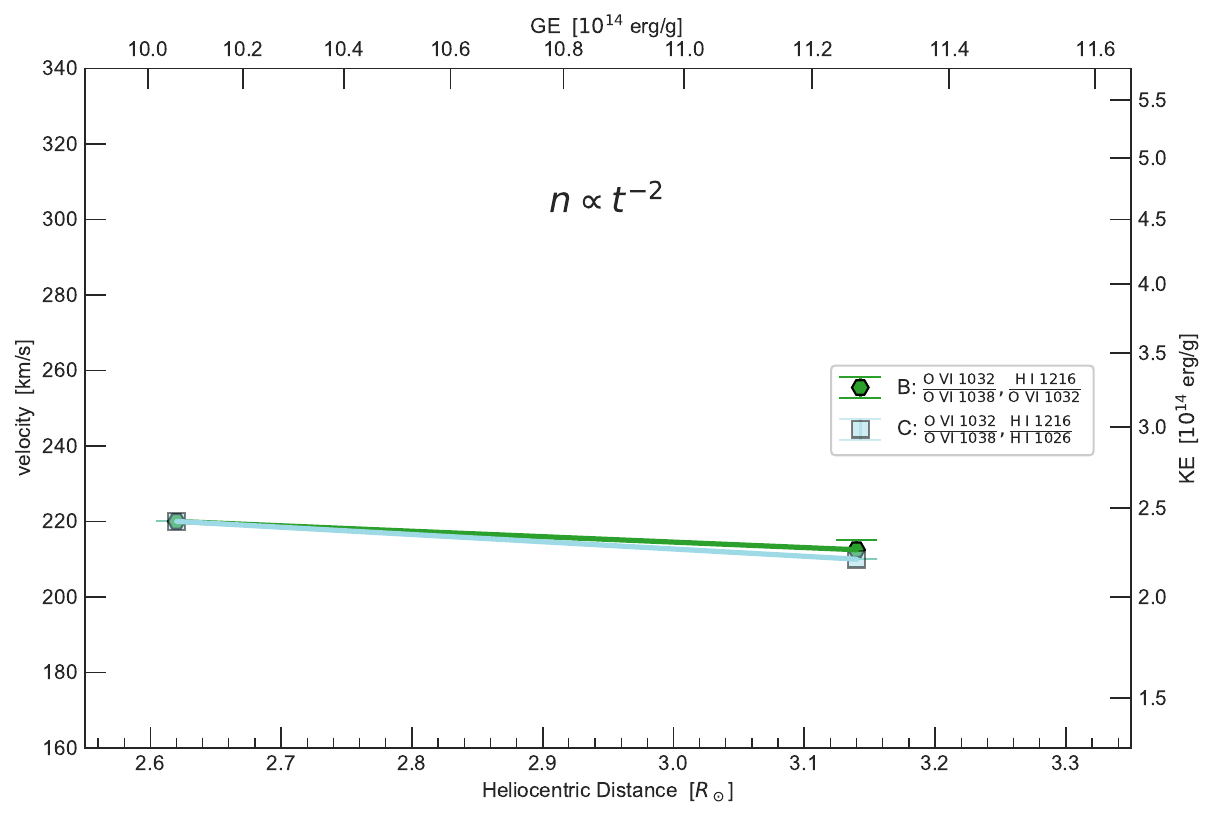}} \hfill  
    \subfloat[\label{fig: energy Q-ht multi-ratio}]{\includegraphics[page=1, width=0.49\linewidth, trim=4cm 0.1in 4.0cm 0.6in, clip=true]{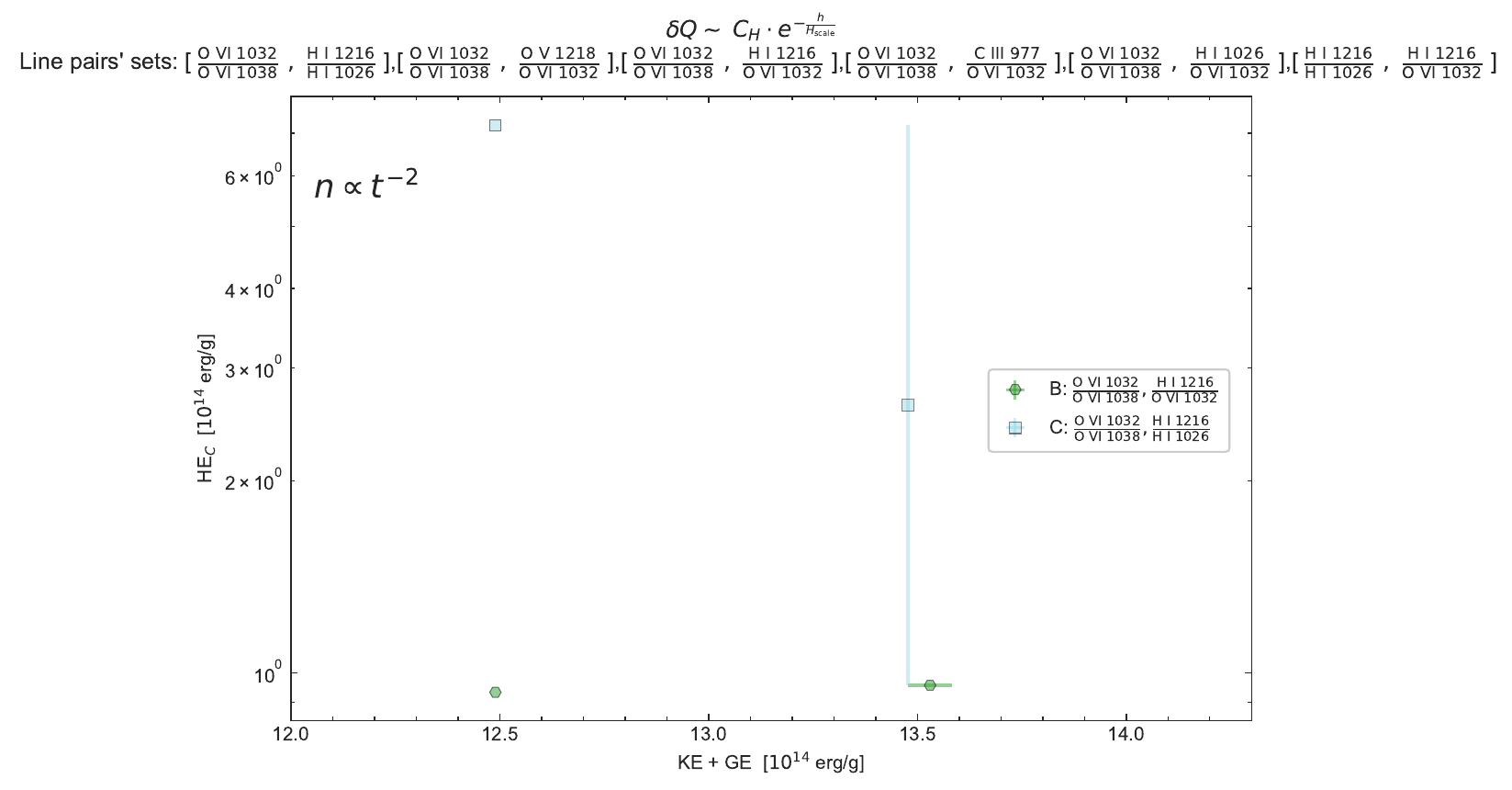}}
    
    \caption{ Observationally constrained models using the $Q_{\rm{AHH}}$ heating parameterization.  }
\end{figure*}


\section{Wave Heating Parameterization}\label{sect: Q-ht}

Using the $Q_{\rm{AHH}}$ heating, there are three distinct plasma clouds modelled that agree with the observations: \ionn{H}{I}, \ionn{O}{VI}, and \ionn{H}{I} mixed with \ionn{O}{VI}.  As with the $Q_{n^2}$ heating results, the physical conditions we derive for these plasma clouds are similar to that of our $Q_n$ heating results due to our use of the same observational constraints.   The evolution of these physical conditions varies between heating parameterizations, but the energy budgets remain similar regardless of the heating parameterizations.  Also, the common traits seen in various ratio analyses for this heating parameterization exhibit the same relationships that we detailed in \S\ref{sect: common traits}.

The kinetic and potential energies are given in Figures \ref{fig: uvcs Q-ht single-ratio}~and~\ref{fig: uvcs Q-ht multi-ratio}.  Just as in the $Q_n$ and $Q_{n^2}$ heating results, the \ionn{O}{VI} dominant material has the slowest velocities at 3.1~\rsun.  For our double-ratio analyses, we find very few models that agree with observations, and among these models, agreement is found only with clumps \clump{B}~and~\clump{C}.  The cumulative heating energies, given in Figures \ref{fig: energy Q-ht single-ratio}~and~\ref{fig: energy Q-ht multi-ratio}, are within the range $10^{13.97 \textrm{---} 14.86}$~erg~g$^{-1}$ for the quadratic expansion rate and $10^{12.84 \textrm{---} 14.36}$~erg~g$^{-1}$ for the linear expansion rate.  

The corresponding heating rate coefficients are in the range log($C_H$ / erg~cm$^{-3}$~s$^{-1}$) $\in$ [-6.6, -5.0] for the quadratic expansion rate and log($C_H$ / erg~cm$^{-3}$~s$^{-1}$) $\in$ [-11.0, -5.8] for the linear expansion rate.  The heating rate's lower limit of $C_H = 10^{-11.0}$~erg~cm$^{-3}$~s$^{-1}$ gives negligible heating (compared to the cooling) under a linear expansion rate.  Thus, within our observational constraints, all heating rates of $C_H \leq 10^{-11.0}$~erg~cm$^{-3}$~s$^{-1}$ suggest that a model with no heating is sufficient to explain the physical conditions when assuming a linear expansion rate.  We presented a similar circumstance in our $Q_n$ heating results (cf. \S\ref{sect: common traits}).  A model with negligible heating is more likely to be valid when the cooling is more steady due to slower expansion rates.  The total cooling has a significant contribution from the square-density dependent radiative cooling that drops slowly under slow expansion rates.  This steady cooling with no heating creates only small changes in the evolution of the material's physical conditions.  Such a model is valid only when the initial density and initial temperature were already close to meeting our observational constraints at 2.6~and~3.1~\rsun.    

For comparison, we have used the same heating function ($Q_{\rm{AHH}}$) and scale height ($\mathcal{H}$) as \cite{Allen.1998} used in their thermal energy equations as they modelled the electron temperature ($T_e$) for the fast solar wind.  They found that a heating rate coefficient 2.5~$\times$~$10^{-7}$ (or $10^{-6.6}$)  erg~cm$^{-3}$~s$^{-1}$ sufficed to have their models agree with observations.  This is within the upper and lower limits of our heating rate coefficient ($C_H$) for both the quadratic expansion rate and the linear expansion rate models.  This supports the notion that some of our models correspond to the coronal material (as opposed to prominence material) within regions of the CME core along the LOS and POS.


\begin{figure*}\label{fig: Q-B3}
    \centering
    \captionsetup[subfigure]{position=top, labelfont=bf,textfont=normalfont,singlelinecheck=off,justification=raggedright} 
    \subfloat[\label{fig: uvcs Q-B3 single-ratio}]{\includegraphics[page=1, width=0.49\linewidth, trim=0cm 0cm 0cm 0cm, clip=true]{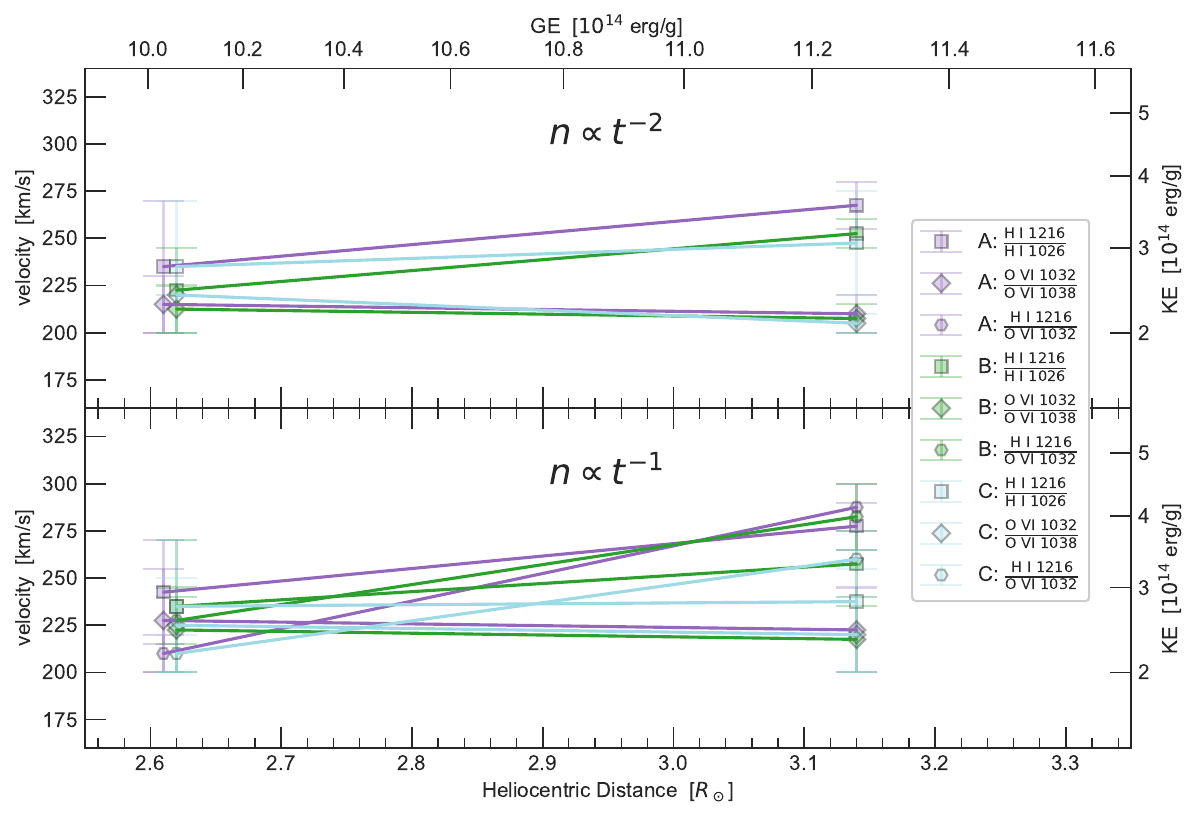}} \hfill  
    \subfloat[\label{fig: energy Q-B3 single-ratio}]{\includegraphics[page=1, width=0.49\linewidth, trim=0cm 0.1in 0cm 0.65in, clip=true]{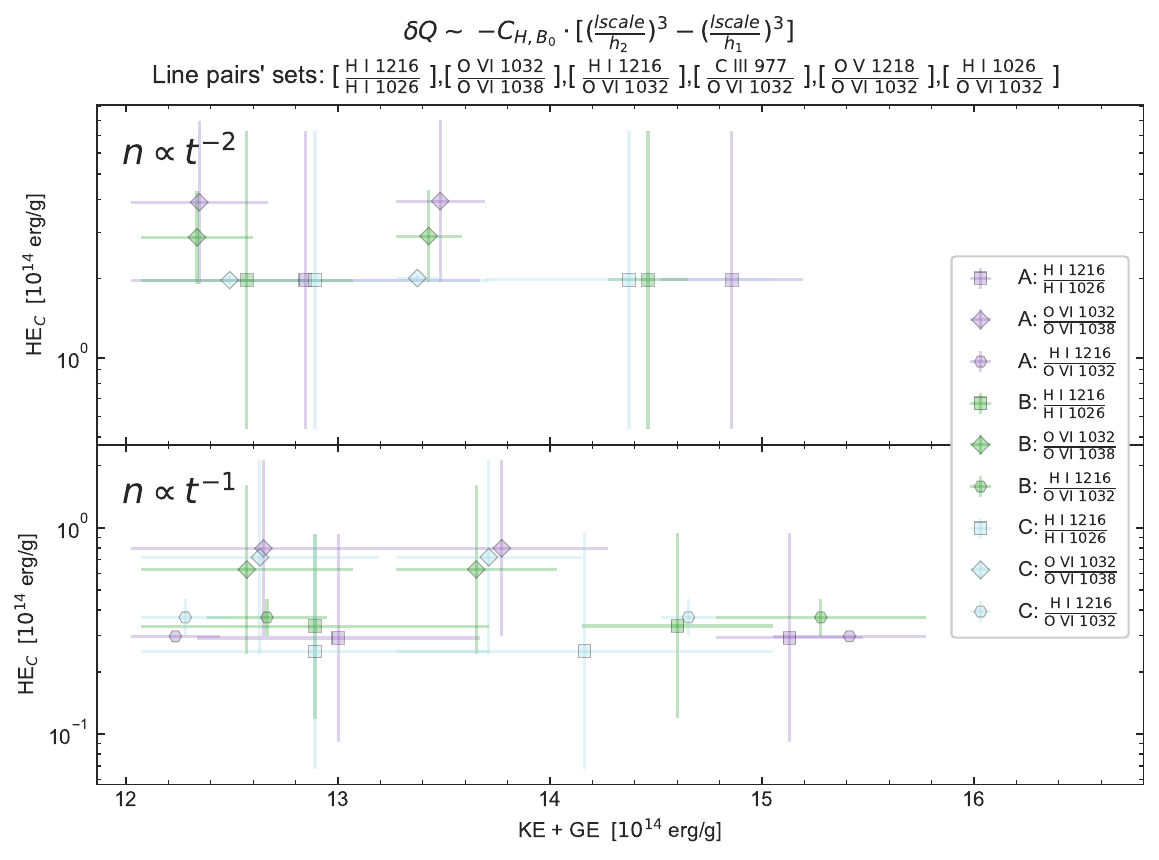}}
    \caption{ Observationally constrained models using the the $Q_{\rm{KR}}$ heating function with $\alpha_{\rm{B}}=3$.  }
\end{figure*}

\begin{figure*}\label{fig: Q-B2}
    \centering
    \captionsetup[subfigure]{position=top, labelfont=bf,textfont=normalfont,singlelinecheck=off,justification=raggedright} 
    
    \subfloat[\label{fig: uvcs Q-B2 single-ratio}]{\includegraphics[page=1, width=0.49\linewidth, trim=0cm 0cm 0cm 0cm, clip=true]{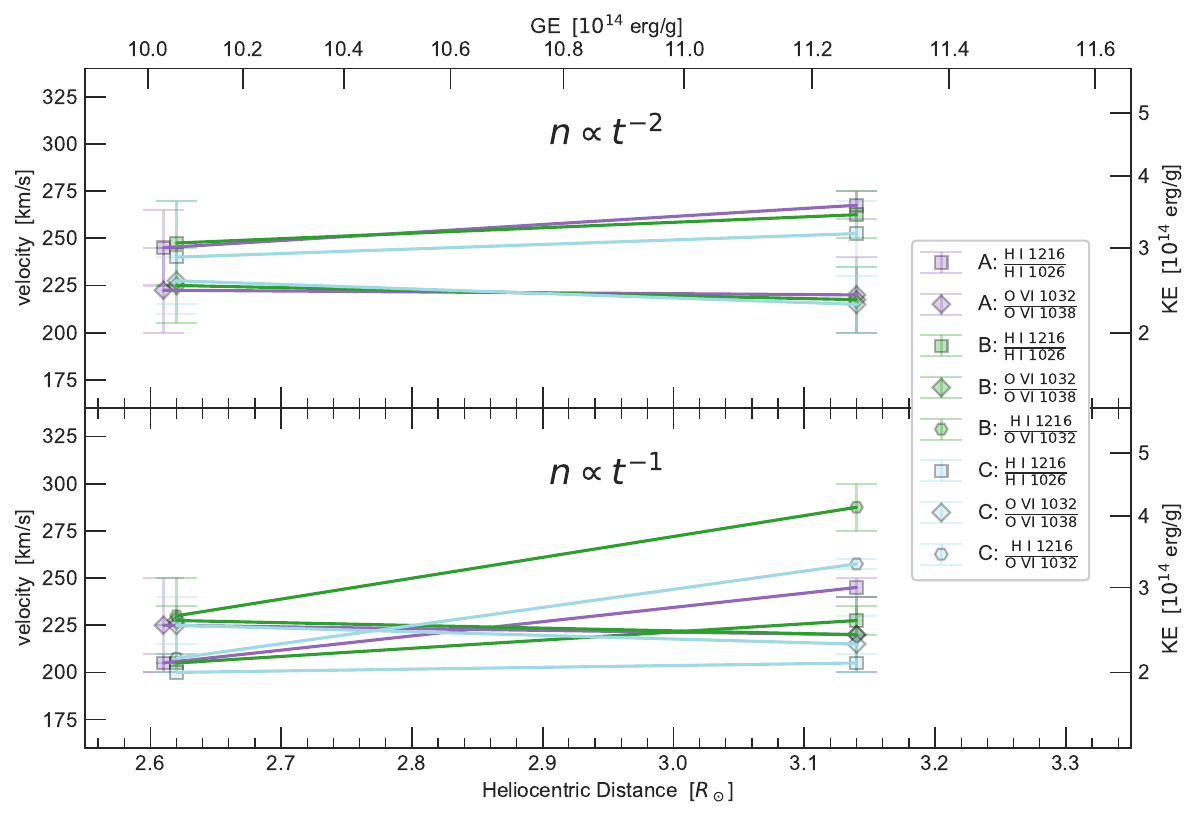}} \hfill  
    \subfloat[\label{fig: energy Q-B2 single-ratio}]{\includegraphics[page=1, width=0.49\linewidth, trim=0cm 0.1in 0cm 0.65in, clip=true]{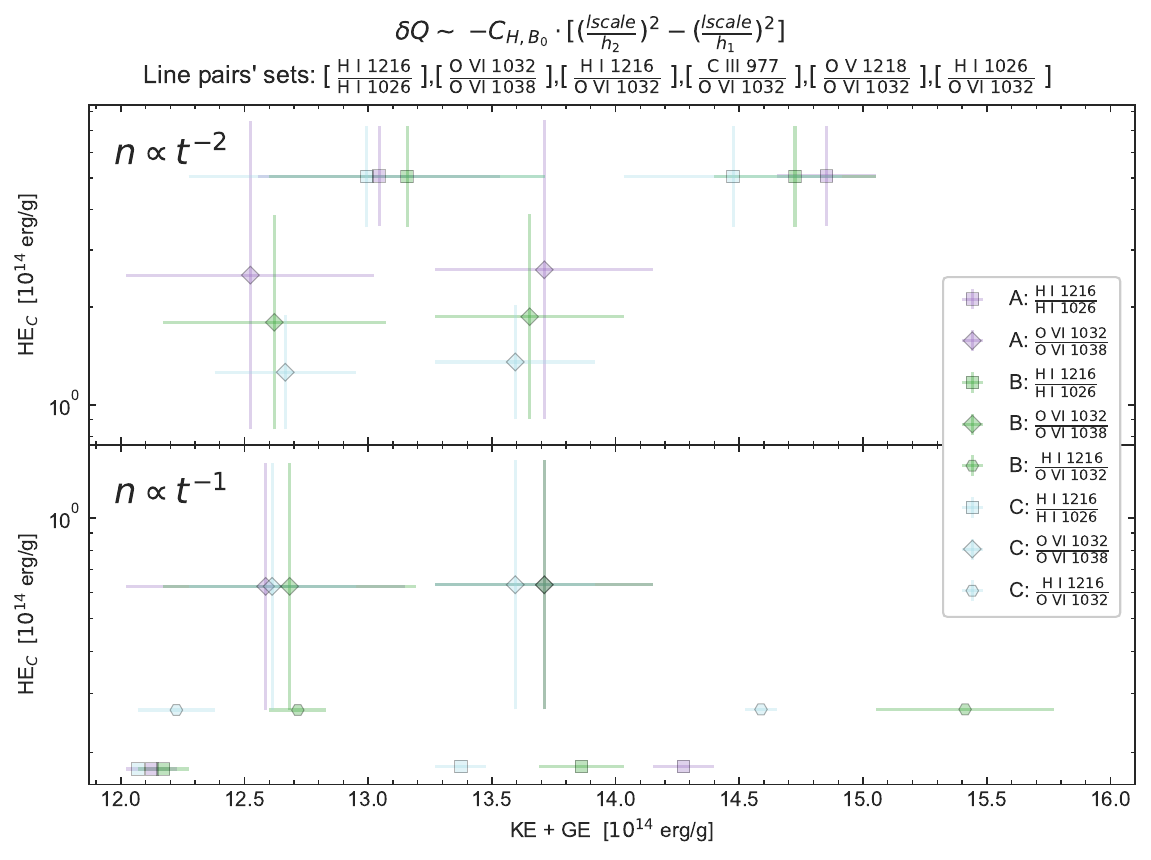}}
    
    \caption{ Observationally constrained models using the $Q_{\rm{KR}}$ heating function with $\alpha_{\rm{B}}=2$.  }
\end{figure*}

\section{Magnetic Heating Parameterization}\label{sect: Q-B}

Using the $Q_{\rm{KR}}$ heating, there are three distinct plasma clouds modelled that agree with the observations: \ionn{H}{I}, \ionn{O}{VI}, and \ionn{H}{I} mixed with \ionn{O}{VI}.  None of our models for the \ionn{H}{I} with \ionn{O}{VI} mixture agree with observations when a quadratic expansion rate is assumed.  The physical conditions we derive for these plasma clouds are similar to the results obtained when using the $Q_n$, $Q_{n^2}$, and $Q_{\rm{AHH}}$ heating functions.  This is the case for both the three-dimensional magnetic field expansion ($\alpha_B=3$) and the two-dimensional magnetic field expansion ($\alpha_B=2$).  For each choice of $\alpha_B$, the common traits seen in various ratio analyses for this $Q_{\rm{KR}}$ heating exhibit the same relationships that we detailed in \S\ref{sect: common traits}.

The energy budget in the case of $\alpha_B=3$ is summarized in Figure~\ref{fig: Q-B3}.  In our kinetic energy estimates, the \ionn{O}{VI} dominant material has the slowest velocity at 3.1~\rsun. The cumulative heating energies of the three plasma clouds we model are in the range $10^{13.73 \textrm{---} 14.90}$~erg~g$^{-1}$ for a quadratic expansion rate and $10^{12.83 \textrm{---} 14.33}$~erg~g$^{-1}$ for a linear expansion rate.  The cumulative heating energy is influenced by our choice of an initial magnetic energy that can contribute to the heating.  We considered magnetic field strengths within the range \ifbool{noSciNot}{log($B_0$/G) $\in [-0.50, 4.0]$}{$B_0 = 10^{-0.5\ \textrm{---}\ 4}$~G}.
The magnetic field strengths that correspond to the cumulative heating results \ifbool{noSciNot}{are within the range log($B_0$/G) $\in [-0.50, 0.32]$\iffalse $10^{-0.50\ \textrm{---}\ 0.32}$~G \fi}{range from 0.32 to 2.08~G} for the quadratic expansion rate and \ifbool{noSciNot}{within the range log($B_0$/G) $\in [-0.50, 0.73]$\iffalse $10^{-0.50\ \textrm{---}\ 0.73}$~G\fi}{from 0.32 to 5.34~G} for the linear expansion rate.  The lower limits of $10^{-0.50}$~G are due to our cutoff for plausible magnetic field strengths.  
These initial conditions correspond to ratios of plasma pressure to magnetic pressure in the range of (initial) plasma-beta $\beta_0 \in [0.26, \scinot{3}{3}]$ for the quadratic expansion rate and $\beta_0 \in [\scinot{2}{-5}, 40.8]$ for the linear expansion rate. 



The energy budget in the case of $\alpha_B=2$ is summarized in Figure~\ref{fig: Q-B2}.  As with all other heating parameterizations, our models suggest that the \ionn{O}{VI} dominant material has the lowest kinetic energy at 3.1~\rsun\ among the plasma clouds we consider.  The cumulative heating energies we find are in the range $10^{13.93 \textrm{---} 14.88}$~erg~g$^{-1}$ for the quadratic expansion rate and $10^{13.25 \textrm{---} 14.17}$~erg~g$^{-1}$ for the linear expansion rate.  Their corresponding magnetic field strengths are within the range \ifbool{noSciNot}{log($B_0$/G) $\in [-0.50, 0.32]$\iffalse within the range $10^{-0.5\ \textrm{---}\ 0.32}$~G \fi}{from 0.32 to 2.08~G} for the quadratic expansion rate and \ifbool{noSciNot}{log($B_0$/G) $\in [-0.50, 0.73]$\iffalse within the range $10^{-0.5\ \textrm{---}\ 0.73}$~G \fi}{from 0.32 to 5.3~G} for the linear expansion rate.  The corresponding initial plasma-beta values are in the range $\beta_0 \in [0.75, \scinot{2}{4}]$ for the quadratic expansion rate and $\beta_0 \in [\scinot{4}{-5}, 190]$ for the linear expansion rate.

Between $\alpha_B=3$ and $\alpha_B=2$, the limits for $B_0$ are the same although the limits for $\beta_0$ differ.  This attests to how the observational constraints from our intensity ratios influence the acceptable initial plasma pressure much more than the initial magnetic pressure.  This is perhaps a consequence of using intensity ratios that come from emissivities that are directly affected by the plasma density and temperature.




\begin{thebibliography}{}
\expandafter\ifx\csname natexlab\endcsname\relax\def\natexlab#1{#1}\fi

\bibitem[{{Akmal} {et~al.}(2001){Akmal}, {Raymond}, {Vourlidas}, {Thompson},
  {Ciaravella}, {Ko}, {Uzzo}, \& {Wu}}]{Akmal.2001}
{Akmal}, A., {Raymond}, J.~C., {Vourlidas}, A., {et~al.} 2001, \apj, 553, 922

\bibitem[{{Allen} {et~al.}(1998){Allen}, {Habbal}, \& {Hu}}]{Allen.1998}
{Allen}, L.~A., {Habbal}, S.~R., \& {Hu}, Y.~Q. 1998, \jgr, 103, 6551

\bibitem[{{Allen} {et~al.}(2000){Allen}, {Habbal}, \& {Li}}]{Allen.2000}
{Allen}, L.~A., {Habbal}, S.~R., \& {Li}, X. 2000, \jgr, 105, 23123

\bibitem[{{Aschwanden}(2017)}]{Aschwanden.2017}
{Aschwanden}, M.~J. 2017, \apj, 847, 27

\bibitem[{{Aschwanden} {et~al.}(2014){Aschwanden}, {Xu}, \&
  {Jing}}]{Aschwanden.2014}
{Aschwanden}, M.~J., {Xu}, Y., \& {Jing}, J. 2014, \apj, 797, 50

\bibitem[{{Bemporad} {et~al.}(2006){Bemporad}, {Poletto}, {Suess}, {Ko},
  {Schwadron}, {Elliott}, \& {Raymond}}]{Bemporad.2006}
{Bemporad}, A., {Poletto}, G., {Suess}, S.~T., {et~al.} 2006, \apj, 638, 1110

\bibitem[{Berger \& Field(1984)}]{Berger.1984}
Berger, M.~A., \& Field, G.~B. 1984, Journal of Fluid Mechanics, 147, 133–148

\bibitem[{{Bradshaw} \& {Klimchuk}(2011)}]{Bradshaw.2011}
{Bradshaw}, S.~J., \& {Klimchuk}, J.~A. 2011, \apjs, 194, 26

\bibitem[{{Brueckner} {et~al.}(1995){Brueckner}, {Howard}, {Koomen},
  {Korendyke}, {Michels}, {Moses}, {Socker}, {Dere}, {Lamy}, {Llebaria},
  {Bout}, {Schwenn}, {Simnett}, {Bedford}, \& {Eyles}}]{Brueckner.1995}
{Brueckner}, G.~E., {Howard}, R.~A., {Koomen}, M.~J., {et~al.} 1995, \solphys,
  162, 357

\bibitem[{{Ciaravella} {et~al.}(2001){Ciaravella}, {Raymond}, {Reale},
  {Strachan}, \& {Peres}}]{Ciaravella.2001}
{Ciaravella}, A., {Raymond}, J.~C., {Reale}, F., {Strachan}, L., \& {Peres}, G.
  2001, \apj, 557, 351

\bibitem[{{Ciaravella} {et~al.}(2003){Ciaravella}, {Raymond}, {van
  Ballegooijen}, {Strachan}, {Vourlidas}, {Li}, {Chen}, \&
  {Panasyuk}}]{Ciaravella.2003}
{Ciaravella}, A., {Raymond}, J.~C., {van Ballegooijen}, A., {et~al.} 2003,
  \apj, 597, 1118

\bibitem[{{Cranmer} {et~al.}(2008){Cranmer}, {Panasyuk}, \&
  {Kohl}}]{Cranmer.2008}
{Cranmer}, S.~R., {Panasyuk}, A.~V., \& {Kohl}, J.~L. 2008, \apj, 678, 1480

\bibitem[{{Davies} {et~al.}(2020){Davies}, {Forsyth}, {Good}, \&
  {Kilpua}}]{Davies.2020}
{Davies}, E.~E., {Forsyth}, R.~J., {Good}, S.~W., \& {Kilpua}, E. K.~J. 2020,
  \solphys, 295, 157

\bibitem[{{Delaboudini{\`e}re} {et~al.}(1995){Delaboudini{\`e}re}, {Artzner},
  {Brunaud}, {Gabriel}, {Hochedez}, {Millier}, {Song}, {Au}, {Dere}, {Howard},
  {Kreplin}, {Michels}, {Moses}, {Defise}, {Jamar}, {Rochus}, {Chauvineau},
  {Marioge}, {Catura}, {Lemen}, {Shing}, {Stern}, {Gurman}, {Neupert},
  {Maucherat}, {Clette}, {Cugnon}, \& {van Dessel}}]{Delaboudiniere.1995}
{Delaboudini{\`e}re}, J.~P., {Artzner}, G.~E., {Brunaud}, J., {et~al.} 1995,
  \solphys, 162, 291

\bibitem[{{Dere} {et~al.}(2019){Dere}, {Del Zanna}, {Young}, {Landi}, \&
  {Sutherland}}]{Dere.2019}
{Dere}, K.~P., {Del Zanna}, G., {Young}, P.~R., {Landi}, E., \& {Sutherland},
  R.~S. 2019, \apjs, 241, 22

\bibitem[{{Dere} {et~al.}(1997){Dere}, {Landi}, {Mason}, {Monsignori Fossi}, \&
  {Young}}]{Dere.1997}
{Dere}, K.~P., {Landi}, E., {Mason}, H.~E., {Monsignori Fossi}, B.~C., \&
  {Young}, P.~R. 1997, \aaps, 125, 149

\bibitem[{{Downs} {et~al.}(2013){Downs}, {Linker}, {Miki{\'c}}, {Riley},
  {Schrijver}, \& {Saint-Hilaire}}]{2013Sci...340.1196D}
{Downs}, C., {Linker}, J.~A., {Miki{\'c}}, Z., {et~al.} 2013, Science, 340,
  1196

\bibitem[{{Emslie} {et~al.}(2005){Emslie}, {Dennis}, {Holman}, \&
  {Hudson}}]{Emslie.2005}
{Emslie}, A.~G., {Dennis}, B.~R., {Holman}, G.~D., \& {Hudson}, H.~S. 2005,
  Journal of Geophysical Research (Space Physics), 110, A11103

\bibitem[{{Emslie} {et~al.}(2004){Emslie}, {Kucharek}, {Dennis}, {Gopalswamy},
  {Holman}, {Share}, {Vourlidas}, {Forbes}, {Gallagher}, {Mason}, {Metcalf},
  {Mewaldt}, {Murphy}, {Schwartz}, \& {Zurbuchen}}]{Emslie.2004}
{Emslie}, A.~G., {Kucharek}, H., {Dennis}, B.~R., {et~al.} 2004, Journal of
  Geophysical Research (Space Physics), 109, A10104

\bibitem[{{Engvold}(1988)}]{Engvold.1988}
{Engvold}, O. 1988, in Solar and Stellar Coronal Structure and Dynamics, ed.
  R.~C. {Altrock}, 151--169

\bibitem[{{Feldman} {et~al.}(1992){Feldman}, {Mandelbaum}, {Seely}, {Doschek},
  \& {Gursky}}]{Feldman.1992}
{Feldman}, U., {Mandelbaum}, P., {Seely}, J.~F., {Doschek}, G.~A., \& {Gursky},
  H. 1992, \apjs, 81, 387

\bibitem[{{Feng} {et~al.}(2010){Feng}, {Yang}, {Xiang}, {Wu}, {Zhou}, \&
  {Zhong}}]{Feng.2010}
{Feng}, X., {Yang}, L., {Xiang}, C., {et~al.} 2010, \apj, 723, 300

\bibitem[{{Filippov} \& {Koutchmy}(2002)}]{Filippov.2002}
{Filippov}, B., \& {Koutchmy}, S. 2002, \solphys, 208, 283

\bibitem[{{Gardner} {et~al.}(2000){Gardner}, {Atkins}, {Fineschi}, {Smith},
  {Kohl}, {Maccari}, \& {Romoli}}]{Gardner.2000}
{Gardner}, L.~D., {Atkins}, N., {Fineschi}, S., {et~al.} 2000, in Society of
  Photo-Optical Instrumentation Engineers (SPIE) Conference Series, Vol. 4139,
  Instrumentation for UV/EUV Astronomy and Solar Missions, ed. S.~{Fineschi},
  C.~M. {Korendyke}, O.~H. {Siegmund}, \& B.~E. {Woodgate}, 362--369

\bibitem[{{Gardner} {et~al.}(1996){Gardner}, {Kohl}, {Daigneau}, {Dennis},
  {Fineschi}, {Michels}, {Nystrom}, {Panasyuk}, {Raymond}, {Reisenfeld},
  {Smith}, {Strachan}, {Suleiman}, {Noci}, {Romoli}, {Ciaravella},
  {Modigliani}, {Huber}, {Antonucci}, {Benna}, {Giordano}, {Tondello},
  {Nicolosi}, {Naletto}, {Pernechele}, {Spadaro}, {Siegmund}, {Allegra},
  {Carosso}, \& {Jhabvala}}]{Gardner.1996}
{Gardner}, L.~D., {Kohl}, J.~L., {Daigneau}, P.~S., {et~al.} 1996, in Society
  of Photo-Optical Instrumentation Engineers (SPIE) Conference Series, Vol.
  2831, Ultraviolet Atmospheric and Space Remote Sensing: Methods and
  Instrumentation, ed. R.~E. {Huffman} \& C.~G. {Stergis}, 2--24

\bibitem[{{Gardner} {et~al.}(2002){Gardner}, {Smith}, {Kohl}, {Atkins},
  {Ciaravella}, {Miralles}, {Panasyuk}, {Raymond}, {Strachan}, {Suleiman},
  {Romoli}, \& {Fineschi}}]{Gardner.2002}
{Gardner}, L.~D., {Smith}, P.~L., {Kohl}, J.~L., {et~al.} 2002, ISSI Scientific
  Reports Series, 2, 161

\bibitem[{{Gilly} \& {Cranmer}(2020)}]{Gilly.2020}
{Gilly}, C.~R., \& {Cranmer}, S.~R. 2020, \apj, 901, 150

\bibitem[{{Gopalswamy} {et~al.}(2009){Gopalswamy}, {Yashiro}, {Michalek},
  {Stenborg}, {Vourlidas}, {Freeland}, \& {Howard}}]{Gopalswamy.2009}
{Gopalswamy}, N., {Yashiro}, S., {Michalek}, G., {et~al.} 2009, Earth Moon and
  Planets, 104, 295

\bibitem[{{Gosling} {et~al.}(1974){Gosling}, {Hildner}, {MacQueen}, {Munro},
  {Poland}, \& {Ross}}]{Gosling.1974}
{Gosling}, J.~T., {Hildner}, E., {MacQueen}, R.~M., {et~al.} 1974, \jgr, 79,
  4581

\bibitem[{{Gruesbeck} {et~al.}(2012){Gruesbeck}, {Lepri}, \&
  {Zurbuchen}}]{Gruesbeck.2012}
{Gruesbeck}, J.~R., {Lepri}, S.~T., \& {Zurbuchen}, T.~H. 2012, \apj, 760, 141

\bibitem[{{Gruesbeck} {et~al.}(2011){Gruesbeck}, {Lepri}, {Zurbuchen}, \&
  {Antiochos}}]{Gruesbeck.2011}
{Gruesbeck}, J.~R., {Lepri}, S.~T., {Zurbuchen}, T.~H., \& {Antiochos}, S.~K.
  2011, \apj, 730, 103

\bibitem[{{Hansen} {et~al.}(1971){Hansen}, {Garcia}, {Grognard}, \&
  {Sheridan}}]{Hansen.1971}
{Hansen}, R.~T., {Garcia}, C.~J., {Grognard}, R.~J.~M., \& {Sheridan}, K.~V.
  1971, \pasa, 2, 57

\bibitem[{{Howard} \& {Vourlidas}(2018)}]{Howard.2018}
{Howard}, R.~A., \& {Vourlidas}, A. 2018, \solphys, 293, 55

\bibitem[{{Hughes} \& {Helfand}(1985)}]{Hughes.1985}
{Hughes}, J.~P., \& {Helfand}, D.~J. 1985, \apj, 291, 544

\bibitem[{{Hyder} \& {Lites}(1970)}]{Hyder.1970}
{Hyder}, C.~L., \& {Lites}, B.~W. 1970, \solphys, 14, 147

\bibitem[{{Illing} \& {Hundhausen}(1985)}]{Illing.1985}
{Illing}, R.~M.~E., \& {Hundhausen}, A.~J. 1985, \jgr, 90, 275

\bibitem[{{Kahler}(1994)}]{Kahler.1994}
{Kahler}, S. 1994, \apj, 428, 837

\bibitem[{{Ko} {et~al.}(2010){Ko}, {Raymond}, {Vr{\v{s}}nak}, \&
  {Vuji{\'c}}}]{Ko.2010}
{Ko}, Y.-K., {Raymond}, J.~C., {Vr{\v{s}}nak}, B., \& {Vuji{\'c}}, E. 2010,
  \apj, 722, 625

\bibitem[{{Ko} {et~al.}(2016){Ko}, {Vourlidas}, {Korendyke}, \&
  {Laming}}]{Ko.2016}
{Ko}, Y.~K., {Vourlidas}, A., {Korendyke}, C., \& {Laming}, J.~M. 2016, in AGU
  Fall Meeting Abstracts, SH43B--2569

\bibitem[{{Ko} {et~al.}(2005){Ko}, {Raymond}, {Gibson}, {Alexander},
  {Strachan}, {Holzer}, {Gilbert}, {Cyr}, {Thompson}, {Pike}, {Mason},
  {Burkepile}, {Thompson}, \& {Fletcher}}]{Ko.2005}
{Ko}, Y.~K., {Raymond}, J.~C., {Gibson}, S.~E., {et~al.} 2005, \apj, 623, 519

\bibitem[{{Kohl} {et~al.}(1994){Kohl}, {Gardner}, {Strachan}, \&
  {Hassler}}]{Kohl.1994}
{Kohl}, J.~L., {Gardner}, L.~D., {Strachan}, L., \& {Hassler}, D.~M. 1994,
  \ssr, 70, 253

\bibitem[{{Kohl} {et~al.}(2006){Kohl}, {Noci}, {Cranmer}, \&
  {Raymond}}]{Kohl.2006}
{Kohl}, J.~L., {Noci}, G., {Cranmer}, S.~R., \& {Raymond}, J.~C. 2006, \aapr,
  13, 31

\bibitem[{{Kohl} {et~al.}(1978){Kohl}, {Reeves}, \& {Kirkham}}]{Kohl.1978}
{Kohl}, J.~L., {Reeves}, E.~M., \& {Kirkham}, B. 1978, in New Instrumentation
  for Space Astronomy, ed. K.~A. {van der Hucht} \& G.~{Vaiana}, 91--94

\bibitem[{{Kohl} {et~al.}(1980){Kohl}, {Weiser}, {Withbroe}, {Noyes},
  {Parkinson}, {Reeves}, {Munro}, \& {MacQueen}}]{Kohl.1980}
{Kohl}, J.~L., {Weiser}, H., {Withbroe}, G.~L., {et~al.} 1980, \apjl, 241, L117

\bibitem[{{Kohl} \& {Withbroe}(1982)}]{Kohl.1982}
{Kohl}, J.~L., \& {Withbroe}, G.~L. 1982, \apj, 256, 263

\bibitem[{{Kohl} {et~al.}(1995){Kohl}, {Esser}, {Gardner}, {Habbal},
  {Daigneau}, {Dennis}, {Nystrom}, {Panasyuk}, {Raymond}, {Smith}, {Strachan},
  {van Ballegooijen}, {Noci}, {Fineschi}, {Romoli}, {Ciaravella}, {Modigliani},
  {Huber}, {Antonucci}, {Benna}, {Giordano}, {Tondello}, {Nicolosi}, {Naletto},
  {Pernechele}, {Spadaro}, {Poletto}, {Livi}, {von der L{\"u}he}, {Geiss},
  {Timothy}, {Gloeckler}, {Allegra}, {Basile}, {Brusa}, {Wood}, {Siegmund},
  {Fowler}, {Fisher}, \& {Jhabvala}}]{Kohl.1995}
{Kohl}, J.~L., {Esser}, R., {Gardner}, L.~D., {et~al.} 1995, \solphys, 162, 313

\bibitem[{{Kumar} \& {Rust}(1996)}]{Kumar.1996}
{Kumar}, A., \& {Rust}, D.~M. 1996, \jgr, 101, 15667

\bibitem[{{Laming} {et~al.}(2013){Laming}, {Moses}, {Ko}, {Ng}, {Rakowski}, \&
  {Tylka}}]{Laming.2013}
{Laming}, J.~M., {Moses}, J.~D., {Ko}, Y.-K., {et~al.} 2013, \apj, 770, 73

\bibitem[{{Laming} \& {Vourlidas}(2019)}]{Laming.2019}
{Laming}, J.~M., \& {Vourlidas}, A. 2019, in AGU Fall Meeting Abstracts, Vol.
  2019, SH31B--15

\bibitem[{{Landi} {et~al.}(2012){Landi}, {Alexander}, {Gruesbeck}, {Gilbert},
  {Lepri}, {Manchester}, \& {Zurbuchen}}]{Landi.2012}
{Landi}, E., {Alexander}, R.~L., {Gruesbeck}, J.~R., {et~al.} 2012, \apj, 744,
  100

\bibitem[{{Landi} {et~al.}(2010){Landi}, {Raymond}, {Miralles}, \&
  {Hara}}]{Landi.2010}
{Landi}, E., {Raymond}, J.~C., {Miralles}, M.~P., \& {Hara}, H. 2010, \apj,
  711, 75

\bibitem[{{Lee} {et~al.}(2009){Lee}, {Raymond}, {Ko}, \& {Kim}}]{Lee.2009}
{Lee}, J.~Y., {Raymond}, J.~C., {Ko}, Y.~K., \& {Kim}, K.~S. 2009, \apj, 692,
  1271

\bibitem[{{Lee} {et~al.}(2017){Lee}, {Raymond}, {Reeves}, {Moon}, \&
  {Kim}}]{Lee.2017}
{Lee}, J.-Y., {Raymond}, J.~C., {Reeves}, K.~K., {Moon}, Y.-J., \& {Kim}, K.-S.
  2017, \apj, 844, 3

\bibitem[{{Lepri} {et~al.}(2001){Lepri}, {Zurbuchen}, {Fisk}, {Richardson},
  {Cane}, \& {Gloeckler}}]{Lepri.2001}
{Lepri}, S.~T., {Zurbuchen}, T.~H., {Fisk}, L.~A., {et~al.} 2001, \jgr, 106,
  29231

\bibitem[{{Leroy} {et~al.}(1983){Leroy}, {Bommier}, \&
  {Sahal-Brechot}}]{Leroy.1983}
{Leroy}, J.~L., {Bommier}, V., \& {Sahal-Brechot}, S. 1983, \solphys, 83, 135

\bibitem[{{Liewer} {et~al.}(2009){Liewer}, {de Jong}, {Hall}, {Howard},
  {Thompson}, {Culhane}, {Bone}, \& {van Driel-Gesztelyi}}]{Liewer.2009}
{Liewer}, P.~C., {de Jong}, E.~M., {Hall}, J.~R., {et~al.} 2009, \solphys, 256,
  57

\bibitem[{{Linker} {et~al.}(2011){Linker}, {Lionello}, {Miki{\'c}}, {Titov}, \&
  {Antiochos}}]{2011ApJ...731..110L}
{Linker}, J.~A., {Lionello}, R., {Miki{\'c}}, Z., {Titov}, V.~S., \&
  {Antiochos}, S.~K. 2011, \apj, 731, 110

\bibitem[{{Linker} {et~al.}(2003){Linker}, {Miki{\'c}}, {Lionello}, {Riley},
  {Amari}, \& {Odstrcil}}]{2003PhPl...10.1971L}
{Linker}, J.~A., {Miki{\'c}}, Z., {Lionello}, R., {et~al.} 2003, Physics of
  Plasmas, 10, 1971

\bibitem[{{Linker} {et~al.}(1999){Linker}, {Miki{\'c}}, {Biesecker}, {Forsyth},
  {Gibson}, {Lazarus}, {Lecinski}, {Riley}, {Szabo}, \&
  {Thompson}}]{1999JGR...104.9809L}
{Linker}, J.~A., {Miki{\'c}}, Z., {Biesecker}, D.~A., {et~al.} 1999, \jgr, 104,
  9809

\bibitem[{{Lionello} {et~al.}(2019){Lionello}, {Downs}, {Linker}, {Miki{\'c}},
  {Raymond}, {Shen}, \& {Velli}}]{2019SoPh..294...13L}
{Lionello}, R., {Downs}, C., {Linker}, J.~A., {et~al.} 2019, \solphys, 294, 13

\bibitem[{{Lionello} {et~al.}(2013){Lionello}, {Downs}, {Linker},
  {T{\"o}r{\"o}k}, {Riley}, \& {Miki{\'c}}}]{2013ApJ...777...76L}
---. 2013, \apj, 777, 76

\bibitem[{{Lionello} {et~al.}(2009{\natexlab{a}}){Lionello}, {Linker}, \&
  {Miki{\'c}}}]{Lionello.2009}
{Lionello}, R., {Linker}, J.~A., \& {Miki{\'c}}, Z. 2009{\natexlab{a}}, \apj,
  690, 902

\bibitem[{{Lionello} {et~al.}(2009{\natexlab{b}}){Lionello}, {Linker}, \&
  {Miki{\'c}}}]{2009ApJ...690..902L}
---. 2009{\natexlab{b}}, \apj, 690, 902

\bibitem[{{Lionello} {et~al.}(2006){Lionello}, {Linker}, {Miki{\'c}}, \&
  {Riley}}]{2006ApJ...642L..69L}
{Lionello}, R., {Linker}, J.~A., {Miki{\'c}}, Z., \& {Riley}, P. 2006, \apjl,
  642, L69

\bibitem[{{Lionello} {et~al.}(2005){Lionello}, {Riley}, {Linker}, \&
  {Miki{\'c}}}]{2005ApJ...625..463L}
{Lionello}, R., {Riley}, P., {Linker}, J.~A., \& {Miki{\'c}}, Z. 2005, \apj,
  625, 463

\bibitem[{{Ma} {et~al.}(2011){Ma}, {Raymond}, {Golub}, {Lin}, {Chen}, {Grigis},
  {Testa}, \& {Long}}]{Ma.2011}
{Ma}, S., {Raymond}, J.~C., {Golub}, L., {et~al.} 2011, \apj, 738, 160

\bibitem[{{Masai}(1984)}]{Masai.1984}
{Masai}, K. 1984, \apss, 98, 367

\bibitem[{{Miki{\'c}} {et~al.}(2018){Miki{\'c}}, {}, {Downs}, {Linker},
  {Caplan}, {Mackay}, {Upton}, {Riley}, {Lionello}, {T{\"o}r{\"o}k}, {Titov},
  {Wijaya}, {Druckm{\"u}ller}, {Pasachoff}, \& {Carlos}}]{2018NatAs...2..913M}
{Miki{\'c}}, {}, Z., {Downs}, C., {et~al.} 2018, Nature Astronomy, 2, 913

\bibitem[{{Miki{\'c}} {et~al.}(1999){Miki{\'c}}, {Linker}, {Schnack},
  {Lionello}, \& {Tarditi}}]{1999PhPl....6.2217M}
{Miki{\'c}}, Z., {Linker}, J.~A., {Schnack}, D.~D., {Lionello}, R., \&
  {Tarditi}, A. 1999, Physics of Plasmas, 6, 2217

\bibitem[{{Murphy} {et~al.}(2011){Murphy}, {Raymond}, \&
  {Korreck}}]{Murphy.2011}
{Murphy}, N.~A., {Raymond}, J.~C., \& {Korreck}, K.~E. 2011, \apj, 735, 17

\bibitem[{{Noci} {et~al.}(1987){Noci}, {Kohl}, \& {Withbroe}}]{Noci.1987}
{Noci}, G., {Kohl}, J.~L., \& {Withbroe}, G.~L. 1987, \apj, 315, 706

\bibitem[{{Olsen} {et~al.}(1994){Olsen}, {Leer}, \& {Holzer}}]{Olsen.1994}
{Olsen}, E.~L., {Leer}, E., \& {Holzer}, T.~E. 1994, \apj, 420, 913

\bibitem[{{Parenti} {et~al.}(2012){Parenti}, {Schmieder}, {Heinzel}, \&
  {Golub}}]{Parenti.2012}
{Parenti}, S., {Schmieder}, B., {Heinzel}, P., \& {Golub}, L. 2012, \apj, 754,
  66

\bibitem[{{Pernechele} {et~al.}(1997){Pernechele}, {Naletto}, {Nicolosi},
  {Tondello}, {Fineschi}, {Romoli}, {Noci}, {Spadaro}, \&
  {Kohl}}]{Pernechele.1997}
{Pernechele}, C., {Naletto}, G., {Nicolosi}, P., {et~al.} 1997, \ao, 36, 813

\bibitem[{{Rakowski} {et~al.}(2007){Rakowski}, {Laming}, \&
  {Lepri}}]{Rakowski.2007}
{Rakowski}, C.~E., {Laming}, J.~M., \& {Lepri}, S.~T. 2007, \apj, 667, 602

\bibitem[{{Raymond}(1990)}]{Raymond.1990}
{Raymond}, J.~C. 1990, \apj, 365, 387

\bibitem[{{Raymond} \& {Ciaravella}(2004)}]{Raymond.2004}
{Raymond}, J.~C., \& {Ciaravella}, A. 2004, \apjl, 606, L159

\bibitem[{{Reeves} {et~al.}(2010){Reeves}, {Linker}, {Miki{\'c}}, \&
  {Forbes}}]{Reeves.2010}
{Reeves}, K.~K., {Linker}, J.~A., {Miki{\'c}}, Z., \& {Forbes}, T.~G. 2010,
  \apj, 721, 1547

\bibitem[{{Reeves} {et~al.}(2015){Reeves}, {McCauley}, \& {Tian}}]{Reeves.2015}
{Reeves}, K.~K., {McCauley}, P.~I., \& {Tian}, H. 2015, \apj, 807, 7

\bibitem[{{Reeves} {et~al.}(2019){Reeves}, {T{\"o}r{\"o}k}, {Miki{\'c}},
  {Linker}, \& {Murphy}}]{Reeves.2019}
{Reeves}, K.~K., {T{\"o}r{\"o}k}, T., {Miki{\'c}}, Z., {Linker}, J., \&
  {Murphy}, N.~A. 2019, \apj, 887, 103

\bibitem[{{Rivera} {et~al.}(2019{\natexlab{a}}){Rivera}, {Landi}, \&
  {Lepri}}]{Rivera.2019.August}
{Rivera}, Y.~J., {Landi}, E., \& {Lepri}, S.~T. 2019{\natexlab{a}}, \apjs, 243,
  34

\bibitem[{{Rivera} {et~al.}(2019{\natexlab{b}}){Rivera}, {Landi}, {Lepri}, \&
  {Gilbert}}]{Rivera.2019.April}
{Rivera}, Y.~J., {Landi}, E., {Lepri}, S.~T., \& {Gilbert}, J.~A.
  2019{\natexlab{b}}, \apj, 874, 164

\bibitem[{{Rottman} {et~al.}(2001){Rottman}, {Woods}, {Snow}, \&
  {DeToma}}]{Rottman.2001}
{Rottman}, G., {Woods}, T., {Snow}, M., \& {DeToma}, G. 2001, Advances in Space
  Research, 27, 1927

\bibitem[{{Scolini} {et~al.}(2020){Scolini}, {Chan{\'e}}, {Temmer}, {Kilpua},
  {Dissauer}, {Veronig}, {Palmerio}, {Pomoell}, {Dumbovi{\'c}}, {Guo},
  {Rodriguez}, \& {Poedts}}]{Scolini.2020}
{Scolini}, C., {Chan{\'e}}, E., {Temmer}, M., {et~al.} 2020, \apjs, 247, 21

\bibitem[{{Shen} {et~al.}(2015){Shen}, {Raymond}, {Murphy}, \&
  {Lin}}]{Shen.2015}
{Shen}, C., {Raymond}, J.~C., {Murphy}, N.~A., \& {Lin}, J. 2015, Astronomy and
  Computing, 12, 1

\bibitem[{{Solanki} {et~al.}(2003){Solanki}, {Lagg}, {Woch}, {Krupp}, \&
  {Collados}}]{Solanki.2003}
{Solanki}, S.~K., {Lagg}, A., {Woch}, J., {Krupp}, N., \& {Collados}, M. 2003,
  \nat, 425, 692

\bibitem[{{Strachan} {et~al.}(1994){Strachan}, {Gardner}, {Hassler}, \&
  {Kohl}}]{Strachan.1994}
{Strachan}, L., {Gardner}, L.~D., {Hassler}, D.~M., \& {Kohl}, J.~L. 1994,
  \ssr, 70, 263

\bibitem[{{Strachan} {et~al.}(2012){Strachan}, {Panasyuk}, {Kohl}, \&
  {Lamy}}]{Strachan.2012}
{Strachan}, L., {Panasyuk}, A.~V., {Kohl}, J.~L., \& {Lamy}, P. 2012, \apj,
  745, 51

\bibitem[{{Strachan} {et~al.}(2017){Strachan}, {Laming}, {Ko}, {Tun Beltran},
  {Korendyke}, {Brown}, {Socker}, {Galysh}, {Finne}, {Eisenhower}, {Brechbiel},
  {Noya}, {Provornikova}, \& {Gardner}}]{Strachan.2017}
{Strachan}, L., {Laming}, J.~M., {Ko}, Y.-K., {et~al.} 2017, in AAS/Solar
  Physics Division Meeting, Vol.~48, AAS/Solar Physics Division Abstracts \#48,
  110.07

\bibitem[{Taylor(1974)}]{Taylor.1974}
Taylor, J.~B. 1974, Phys. Rev. Lett., 33, 1139

\bibitem[{{T{\"o}r{\"o}k} {et~al.}(2018){T{\"o}r{\"o}k}, {Downs}, {Linker},
  {Lionello}, {Titov}, {Miki{\'c}}, {Riley}, {Caplan}, \&
  {Wijaya}}]{2018ApJ...856...75T}
{T{\"o}r{\"o}k}, T., {Downs}, C., {Linker}, J.~A., {et~al.} 2018, \apj, 856, 75

\bibitem[{{Tousey} {et~al.}(1973){Tousey}, {Bartoe}, {Bohlin}, {Brueckner},
  {Purcell}, {Scherrer}, {Sheeley}, {Schumacher}, \&
  {Vanhoosier}}]{Tousey.1973}
{Tousey}, R., {Bartoe}, J.~D.~F., {Bohlin}, J.~D., {et~al.} 1973, \solphys, 33,
  265

\bibitem[{{Usmanov} \& {Dryer}(1995)}]{Usmanov.1995}
{Usmanov}, A.~V., \& {Dryer}, M. 1995, \solphys, 159, 347

\bibitem[{{Vernazza} \& {Reeves}(1978)}]{Vernazza.1978}
{Vernazza}, J.~E., \& {Reeves}, E.~M. 1978, \apjs, 37, 485

\bibitem[{{Vourlidas} {et~al.}(2010){Vourlidas}, {Howard}, {Esfandiari},
  {Patsourakos}, {Yashiro}, \& {Michalek}}]{Vourlidas.2010}
{Vourlidas}, A., {Howard}, R.~A., {Esfandiari}, E., {et~al.} 2010, \apj, 722,
  1522

\bibitem[{{Withbroe}(1988)}]{Withbroe.1988}
{Withbroe}, G.~L. 1988, \apj, 325, 442

\end{thebibliography}

\end{document}